\documentclass[longauth]{aa}  

\usepackage{graphicx}
\usepackage[varg]{txfonts}
\usepackage{natbib}
\bibpunct{(}{)}{;}{a}{}{,} 
\usepackage[colorlinks=true,allcolors=blue]{hyperref}
\usepackage{caption}
\usepackage{subcaption}
\usepackage{amsmath}
\usepackage{euclid}
\usepackage{multirow}

\newcommand{\eqdef}{\mathrel{\mathop:}=
} 

\graphicspath{{graphs/}}

\begin{document}

\renewcommand{\sectionautorefname}{Sect.}
\renewcommand{\subsectionautorefname}{Sect.}
\renewcommand{\subsubsectionautorefname}{Sect.}
\renewcommand{\figureautorefname}{Fig.}
\def\equationautorefname~#1\null{%
  Eq.~(#1)\null
}

 \title{\Euclid: Nonparametric point spread function field recovery through interpolation on a graph Laplacian\thanks{This paper is published on behalf of the
Euclid Consortium.}}

   \titlerunning{Non-parametric PSF field recovery}
   
   \authorrunning{M.A.~Schmitz et al.}

\author{M.A.~Schmitz$^{1}$, J.-L.~Starck$^{1}$, F.~Ngole Mboula$^{2}$, N.~Auricchio$^{3}$, J.~Brinchmann$^{4,5}$, R.I.~Vito Capobianco$^{6}$, R.~Cl\'{e}dassou$^{7}$, L.~Conversi$^{8}$, L.~Corcione$^{6}$, N.~Fourmanoit$^{9}$, M.~Frailis$^{10}$, B.~Garilli$^{11}$, F.~Hormuth$^{12}$, D.~Hu$^{13}$, H.~Israel$^{14}$, S.~Kermiche$^{9}$, T. D.~Kitching$^{13}$, B.~Kubik$^{15}$, M.~Kunz$^{16}$, S.~Ligori$^{6}$, P.B.~Lilje$^{17}$, I.~Lloro$^{18,19}$, O.~Mansutti$^{10}$, O.~Marggraf$^{20}$, R.J.~Massey$^{21}$, F.~Pasian$^{10}$, V.~Pettorino$^{1}$, F.~Raison$^{22}$, J.D.~Rhodes$^{23}$, M.~Roncarelli$^{3,24}$, R.P.~Saglia$^{14,22}$, P.~Schneider$^{20}$, S.~Serrano$^{18,25}$, A.N.~Taylor$^{26}$, R.~Toledo-Moreo$^{27}$, L.~Valenziano$^{3}$, C.~Vuerli$^{10}$, J.~Zoubian$^{9}$}

\institute{$^{1}$ AIM, CEA, CNRS, Universit\'{e} Paris-Saclay, Universit\'{e} Paris Diderot, Sorbonne Paris Cit\'{e}, F-91191 Gif-sur-Yvette, France\\
        $^{2}$ Institut LIST, CEA, Universit\'{e} Paris-Saclay, F-91191 Gif-Sur-Yvette Cedex, France\\
        $^{3}$ INAF-Osservatorio di Astrofisica e Scienza dello Spazio di Bologna, Via Piero Gobetti 93/3, I-40129 Bologna, Italy\\
        $^{4}$ Instituto de Astrof\'isica e Ci\^encias do Espa\c{c}o, Universidade do Porto, CAUP, Rua das Estrelas, PT4150-762 Porto, Portugal\\
        $^{5}$ Leiden Observatory, Leiden University, Niels Bohrweg 2, 2333 CA Leiden, The Netherlands\\
        $^{6}$ INAF-Osservatorio Astrofisico di Torino, Via Osservatorio 20, I-10025 Pino Torinese (TO), Italy\\
        $^{7}$ Centre National d'Etudes Spatiales, Toulouse, France\\
        $^{8}$ ESAC/ESA, Camino Bajo del Castillo, s/n., Urb. Villafranca del Castillo, 28692 Villanueva de la Ca\~nada, Madrid, Spain\\
        $^{9}$ Aix-Marseille Univ, CNRS/IN2P3, CPPM, Marseille, France\\
        $^{10}$ INAF-Osservatorio Astronomico di Trieste, Via G. B. Tiepolo 11, I-34131 Trieste, Italy\\
        $^{11}$ INAF-IASF Milano, Via Alfonso Corti 12, I-20133 Milano, Italy\\
        $^{12}$ von Hoerner \& Sulger GmbH, Schlo{\ss}Platz 8, D-68723 Schwetzingen, Germany\\
        $^{13}$ Mullard Space Science Laboratory, University College London, Holmbury St Mary, Dorking, Surrey RH5 6NT, UK\\
        $^{14}$ Universit\"ats-Sternwarte M\"unchen, Fakult\"at f\"ur Physik, Ludwig-Maximilians-Universit\"at M\"unchen, Scheinerstrasse 1, 81679 M\"unchen, Germany\\
        $^{15}$ Institut de Physique Nucl\'eaire de Lyon, 4, rue Enrico Fermi, 69622, Villeurbanne cedex, France\\
        $^{16}$ Universit\'e de Gen\`eve, D\'epartement de Physique Th\'eorique and Centre for Astroparticle Physics, 24 quai Ernest-Ansermet, CH-1211 Gen\`eve 4, Switzerland\\
        $^{17}$ Institute of Theoretical Astrophysics, University of Oslo, P.O. Box 1029 Blindern, N-0315 Oslo, Norway\\
        $^{18}$ Institute of Space Sciences (ICE, CSIC), Campus UAB, Carrer de Can Magrans, s/n, 08193 Barcelona, Spain\\
        $^{19}$ Institut d'Estudis Espacials de Catalunya (IEEC), 08034 Barcelona, Spain\\
        $^{20}$ Argelander-Institut f\"ur Astronomie, Universit\"at Bonn, Auf dem H\"ugel 71, 53121 Bonn, Germany\\
        $^{21}$ Centre for Extragalactic Astronomy, Department of Physics, Durham University, South Road, Durham, DH1 3LE, UK\\
        $^{22}$ Max Planck Institute for Extraterrestrial Physics, Giessenbachstr. 1, D-85748 Garching, Germany\\
        $^{23}$ Jet Propulsion Laboratory, California Institute of Technology, 4800 Oak Grove Drive, Pasadena, CA, 91109, USA\\
        $^{24}$ Dipartimento di Fisica e Astronomia, Universit\'a di Bologna, Via Gobetti 93/2, I-40129 Bologna, Italy\\
        $^{25}$ Institute of Space Sciences (IEEC-CSIC), c/Can Magrans s/n, 08193 Cerdanyola del Vall\'es, Barcelona, Spain\\
        $^{26}$ Institute for Astronomy, University of Edinburgh, Royal Observatory, Blackford Hill, Edinburgh EH9 3HJ, UK\\
        $^{27}$ Depto. de Electr\'onica y Tecnolog\'ia de Computadoras Universidad Polit\'ecnica de Cartagena, 30202, Cartagena, Spain\\
        \email{morgan.schmitz@astro.princeton.edu}}

   \date{Received ..}

 \abstract{Future weak lensing surveys, such as the \Euclid mission, will attempt to measure the shapes of billions of galaxies in order to derive cosmological information. These surveys will attain very low levels of statistical error, and systematic errors must be extremely well controlled. 
In particular, the point spread function (PSF) must be estimated using stars in the field, and recovered with high accuracy.}{The aims of this paper are twofold. Firstly, we took steps toward a nonparametric method to address the issue of recovering the PSF field, namely that of finding the correct PSF at the position of any galaxy in the field, applicable to \Euclid. Our approach relies solely on the data, as opposed to parametric methods that make use of our knowledge of the instrument. Secondly, we studied the impact of imperfect PSF models on the shape measurement of galaxies themselves, and whether common assumptions about this impact hold true in an \Euclid scenario.}{We extended the recently proposed resolved components analysis approach, which performs super-resolution on a field of under-sampled observations of a spatially varying, image-valued function. We added a spatial interpolation component to the method, making it a true $2$-dimensional PSF model.  We compared our approach to \texttt{PSFEx}, then quantified the impact of PSF recovery errors on galaxy shape measurements through image simulations.}{Our approach yields an improvement over \texttt{PSFEx} in terms of the PSF model and on observed galaxy shape errors, though it is at present far from reaching the required \Euclid accuracy. We also find that the usual formalism used for the propagation of PSF model errors to weak lensing quantities no longer holds in the case of an \Euclid-like PSF. In particular, different shape measurement approaches can react differently to the same PSF modeling errors.}{}

\maketitle


\section{Introduction}\label{sec:intro}

 As light from background galaxies travels through the universe, it gets deflected due to variations of the gravitational potential. 
In the vast majority of cases, distortions due to gravitational lensing are of very small amplitude, and are largely dominated by the observed objects' intrinsic ellipticities: this is the weak lensing regime. By observing a great number of sources, one can however retrieve the lensing signal, probe the large-scale structure of the Universe and derive information about the matter distribution. This makes weak lensing a very interesting cosmological probe~\citep[see e.g.,][]{kilbinger2015}.
   
Galaxy shape measurement for weak lensing has been carried out in several past surveys such as CFHTLenS~\citep{miller2013}. Several ground-based surveys are currently ongoing as well, and have already led to tighter cosmological constraints. These include the Kilo-Degree Survey~\citep[KiDS, ][]{kuijken2015}, the Dark Energy Survey~\citep[DES, ][]{jarvis2016}, and the HyperSuprime Cam~\citep[HSC, ][]{mandelbaum2018}. In the future, Stage IV surveys will allow the cosmological information extracted from the weak lensing signal to achieve unprecedented accuracies, and include both ground-based observations with the Rubin Observatory's Legacy Survey of Space and Time~\citep[LSST, ][]{LSST2009}, and space telescopes such as the Wide-Field Infrared Survey Telescope~\citep[WFIRST, ][]{green2012}, and \Euclid~\citep{laureijs2011}, which is the focus of the present work.
   
In order to fully exploit the potential of these surveys, the level of systematic errors must be kept below that of statistical uncertainty. In the case of \Euclid, where the number of measured objects will be extremely high, this leads to drastic requirements on the various sources of systematic errors. The point spread function (PSF) can induce important systematic effects, since the PSF distorts object shapes, which could lead to very strong bias in ellipticity measurements if not correctly accounted for~\citep{paulin2008,massey2012,cropper2013}.

Two approaches are possible to estimate the PSF. In the \textit{parametric} approach, a PSF model is derived using the known information about the instrument (and the observed sources), typically yielding a simulator that can recreate, based on a set of parameters, the instrument's PSF at any position in the field. These parameters are then chosen by fitting observed stars in the field to yield an accurate PSF model. These methods thus fall in the forward modeling category. An example of this is the TinyTim software~\citep{krist1995} for the Hubble Space Telescope.
The second, or \textit{nonparametric}, approach is based on data only, using unresolved stars in the field as direct measurements of the PSF and estimating the PSF at galaxy positions from  these measurements.
The \texttt{PSFEx} software~\citep{bertin2011} is a typical example of such an approach, and has been successfully applied to real data in the context of weak lensing shape measurement~\citep[for instance in the DES survey:][]{jarvis2016, zuntz2018}. The PSF modeling approach used in CFHTLenS~\citep{miller2013} and KiDS~\citep{kuijken2015} similarly falls in this category, though unlike \texttt{PSFEx}, it allows for discontinuities in the PSF variations at the boundaries between charge-coupled devices (CCD), and can thus be fit on the whole field at once rather than on each detector individually. 

The latter class of methods are ultimately limited by the amount of information that can be recovered from available data. In particular, the maximum accuracy they can achieve is directly limited by the number of observed stars. In the case of \Euclid, the requirement on the multiplicative shear bias is that it should be lower than $2\times 10^{-3}$, which in turn leads to stringent requirements on the PSF model accuracy: the root mean square (RMS) error on each ellipticity component $\left(e_i^\mathrm{PSF}\right)_{i\in\{1,2\}}$ should be lower than $5\times10^{-5}$, and that on the relative size $\delta R^2_\mathrm{PSF}/R^2_\mathrm{PSF}$ (as defined from quadrupole moments) lower than $5\times10^{-4}$. Achieving this accuracy is further complicated by stars suffering from under-sampling, and our experiments indeed show the nonparametric approach proposed in this work is not yet at the level to fulfill these requirements. Because of these considerations, a forward modeling approach of the Euclid visible instrument (VIS) PSF capable of achieving these requirements is being developed (Duncan et al., in prep.), and is already at a more mature state of implementation. Nonetheless, also having a nonparametric PSF model available remains advantageous, as the combination of the two approaches could lead to a PSF model that outperforms either individually. This however requires developing a nonparametric PSF model that can handle all of Euclid VIS' specificities, while striving to be as accurate as possible given its intrinsic limitations. The present work offers a solution to take steps toward a full nonparametric PSF modeling applicable to \Euclid. 

To that end, we considered a simplified setting that includes some of the complexity arising in modeling the VIS PSF (including under-sampling). Binary stars can impact the PSF model if they are not removed from those objects used to fit the PSF model. Since previous work~\citep{kuntzer2016,kuntzer2017} deals with the identification of such objects, we assumed in the present work they had already been removed (and, more generally, that our star catalogs were empty from contamination). Other aspects that also need to be handled but that are left for future work are:

\begin{itemize}
    \item chromatic variations of the PSF~\citep{cypriano2010,eriksen2018};
    \item effects caused by the satellite's Attitude and Orbit Control Subsystem (AOCS) and guiding errors;
    \item detector effects such as charge transfer inefficiency (CTI) and the brighter-fatter effect~\citep[for which][recently proposed a model]{coulton2018};
    \item manufacturing and polishing errors.
\end{itemize}
The latter can induce variations of the PSF that occur on very small spatial scales. While these are not included in the simulated VIS PSF used in the present work, as discussed below, the proposed method can, by construction, handle these high-spatial frequency variations (with the strong caveat that observations need to fall within the area of variation for our model to account for it). Handling detector effects within the PSF model might prove  hard, as they are flux dependent and not convolutional (though they could potentially be corrected for prior to fitting the PSF model). Lastly, the work we carried out for the present work was done at a single point in time, with a low number of observed stars. The telescope will in truth vary with time, which means the PSF modeling should be performed on each exposure separately (or on a set of exposures taken closely enough in time that the temporal variation can be neglected). However, another approach is to include the temporal variation within the nonparametric model itself, and fit it either to several exposures simultaneously, or ``online'', meaning updating the model with each new available exposure. Not only could this improve the quality of the PSF model, it might also help mitigate two serious limitations of the nonparametric method: its quality depending on the number of stars available, and the aforementioned need to observe one precisely at the position of high-spatial frequency variations (which should be constant with time).

Many nonparametric methods have been proposed to model the PSF from observed stars.~\citet{gentile2013} reviewed, 
in the context of GREAT3~\citep{mandelbaum2014,mandelbaum2015}, several traditional interpolation approaches to deal with spatial variations of the PSF. Other, more recent methods rely on 
optimal transport~\citep{ngole2017} or deep learning~\citep{kuntzer2018thesis,herbel2018}. 
\texttt{PSFEx} remains the most widely used method and is, to the best of our knowledge, 
the only one to deal with both the super-resolution and the spatial variation steps at the same time.~\cite{mandelbaum2018} found the \texttt{PSFEx}-based model of the HSC PSF to perform poorly when seeing becomes better than a certain value, close to the threshold at which \texttt{PSFEx} automatically switches to the super-resolution mode~\citep{bosch2017}, and could indicate issues with \texttt{PSFEx}'s handling of super-resolution (and the need for other nonparametric methods to deal with this problem).
Super-resolution is a well-studied problem in image processing, where sparsity-based methods~\citep{starck2015} have been shown to perform extremely well.~\cite{ngole2015} showed this to hold true in the particular case of PSFs. However, contrary to the typical setting of the super-resolution problem where the object of interest is observed several times with slight shifts~\citep{rowe2011}, in the case of \Euclid, we instead have several under-sampled observations of the PSF \textit{at different positions} in the field of view (FOV).~\citet{ngole2016} recently introduced resolved components analysis (RCA), a method specifically designed to handle such a problem, but estimating the PSFs only at star positions.

An early study of the impact of PSF modeling errors was carried out by~\citet{hoekstra2004}, where the PSF was modeled solely through its anisotropy.~\citet{paulin2008} introduced a mathematical description of PSF errors and their impact 
on galaxy shape measurements, which was further explored in~\citet{massey2012}. 
This formalism has been widely used in the context of weak lensing to set requirements  on future surveys~\citep{cropper2013}, such as the minimum number of stars in the field required to achieve a given accuracy~\citep{paulin2009}, or to validate the PSF quality on actual data~\citep{rowe2010,jarvis2016}. These studies rely on the use of unweighted quadrupole moments. The addition of weighting functions to avoid divergent moments leads to mixing with higher order moments. This is a well-studied issue in the case of galaxy shape measurement~\citep[see e.g.,][for the particular case of color gradients]{semboloni2013}. In the case of the PSF errors, however, the assumption that unweighted moments can be used is still widely made. This was reasonable in the case of ground surveys, where the PSF has a simpler profile. In the \Euclid case, however, the PSF profile will have divergent moments. While this is not a concern when considering galaxy shape measurements, as the considered galaxy's light profile then effectively acts like a weighting function, it raises the question of whether the usual expression for the propagation of PSF errors still holds, since the quantities involved are the unweighted PSF moments. If that were not the case, it could be more difficult to disentangle the shape measurement 
errors due to an imperfect PSF model from other effects such as bias due to the method used to derive the galaxy shapes. 
Bias on shape measurements were investigated in several papers ~\citep{hoekstra2015, hoekstra2017, pujol2017},  but under the assumption that the PSF was perfectly known.

In this paper, we expand the RCA method by capturing spatial variations of the PSF through a set of PSF graphs. We can thus estimate the PSF at any arbitrary position in the field, while preserving all the properties of the RCA-recovered PSF field. This leads to a new approach that can deal, similarly to \texttt{PSFEx},
 with both super-resolution and spatial variations, simultaneously. The python library is freely available\footnote{\url{https://github.com/CosmoStat/rca}}.
Neither of these models prove sufficient, in their current state, to achieve \Euclid requirements. This is likely mostly driven by the number of available stars. For a purely nonparametric PSF modeling approach to reach the accuracy required for \Euclid weak lensing, more stars than those available within a single CCD of a science exposure will thus need to be used. This could either rely only the instrument's temporal stability, or a modeling of the PSF's variations with time.

We used the opportunity provided by the comparison of two imperfect PSF models in a \Euclid-like setting to explore the impact of PSF errors on galaxy shape measurement. In particular, by propagating the errors of both PSF models through different shape measurement methods, we examined whether the assumption that these two issues can be treated separately still holds for \Euclid.  
 
 The rest of the paper is organised as follows: \autoref{intro_psffield} describes the formalism of the PSF recovery field problem we adopted; 
 \autoref{sec:RCA} gives a quick overview of the RCA method; and \autoref{sec:frca} presents the new PSF field recovery method. In \autoref{sec:xp}, we apply both \texttt{PSFEx} and our approach to recover simplified \Euclid-like PSFs and compare the resulting models. We then use them for galaxy shape measurement and study the impact of PSF modeling errors in \autoref{sec:gaals}. We conclude and offer some perspectives in \autoref{sec:conclusion}.

\section{Modelling the PSF field from stars}
\label{intro_psffield}
This section introduces the formalism used in the present work for the PSF field estimation problem, and the approach taken by \texttt{PSFEx} to solve it.

\subsection{Notations}\label{sec:notations}
Let us first describe the problem at hand. Let $\mathcal{H}(u)$ denote the (unknown) PSF that we wish to estimate; $\mathcal{H}$ is a function of $u=(x, y)$, a 2-dimensional vector containing the position within the image. It is the full, untruncated PSF intensity profile, and thus outputs a continuous image at any position $u$. Here and throughout this paper, all such positions are given in "image" coordinates (i.e., within the instrument's CCDs), since the position of objects on the sky has no influence on the PSF they are affected by. Similarly, here we consider $\mathcal{H}$ to be a single PSF that aggregates all effects (e.g., diffraction, imperfect optical elements, jitter of the telescope). In particular, we do not consider the intermediary, relative position of incoming light rays from a given object on each individual optical component. We also consider the spatial variations of the PSF to be slow enough that the entirety of an object whose center lies at position $u$ is affected by the same $\mathcal{H}(u)$.

Assuming we observe a set of $n_\mathrm{stars}$ stars across the FOV, at positions $\mathcal{U}_\mathrm{stars}\eqdef\left(u_1,\dots,u_{n_{\mathrm{stars}}}\right)$, each star $i$ gives us a noisy, under-sampled observation of $\mathcal{H}$:
\begin{align}
        Y_i = \mathcal{F}\left(\mathcal{H}(u_i)\right) + N_i\;,
\end{align}
where $N_i$ is a noise vector and $\mathcal{F}$ is the degradation operator, meaning the effect of the realization on the instrument's CCDs. In our case, we separate its effects in two distinct operators, 
\begin{align}
        \mathcal{F}=F_\mathrm{d} \circ\mathcal{F}_\mathrm{s}\;.
\end{align}
$\mathcal{F}_\mathrm{s}$ is a discrete sampling into a finite number of pixels, which turns each continuous image $\mathcal{H}(u)$ into a truncated $p\times p$ image sampled at our target pixel size. $F_\mathrm{d}$ is composed of a sub-pixel shift, and a further down-sampling matrix $M$ (i.e., the pixel response of our instrument) that can lead to under-sampling. Denoting by $D$ the down-sampling factor caused by $M$, the available observations $Y_i$ are thus $Dp\times Dp$ images. In the following, we treat them as flattened vectors of size $D^2p^2$.

The problem at hand is composed of the two following parts. Firstly, from observations $Y\eqdef(Y_1,\dots,Y_{n_{\mathrm{stars}}})$, an estimator $\hat{H}$ of the true PSF $\mathcal{H}$ must be built at corresponding positions $\mathcal{U}_\mathrm{stars}$. Secondly, the PSF must be recovered at the galaxy positions, $\mathcal{U}_\mathrm{gal} \neq \mathcal{U}_\mathrm{stars}$.
In our present case of under-sampled observations, while still discretized, we want our PSF model $\hat{H}$ to be sampled on a finer grid than observations $(Y_i)_i$, that is, to counter the effect of $F_\mathrm{d}$.

\subsection{\texttt{PSFEx}}\label{sec:psfex}
Before introducing our proposed approach to solve this PSF field reconstruction problem, we give a quick overview of the \texttt{PSFEx} method~\citep{bertin2011} that we use in our experiments for comparison purposes.
In its default configuration~\cite[and the one typically used in weak lensing surveys, e.g.,][]{zuntz2018}, \texttt{PSFEx} uses the stars in the field to fit a model directly in the pixel domain. Users can specify any \texttt{Source Extractor}~\citep{bertin1996} parameter to be used, as well as the maximum polynomial degree $d$ allowed for their corresponding components. These parameters are usually chosen to be position parameters $u=(x,y)$, leading to PSF reconstructions of the form

\begin{align}\label{eq:polynomial}
    \hat{H}^\mathrm{PSFEx}(u) = \sum_{\substack{p,q\ge0\\p+q\le d}} x^py^qS_{pq}\;.
\end{align}
The reconstructed PSFs at the positions of the stars $\mathcal{U}_\mathrm{stars}$ can thus be rewritten as follows:

\begin{align}
\forall i\in\{1,\dots,n_\mathrm{stars}\}, \,\hat{H}_i^\mathrm{PSFEx} \eqdef \hat{H}^\mathrm{PSFEx}(u_i) = SA_i\;,
\end{align}
which in turn allows us to recast the \texttt{PSFEx} model as one of matrix factorization, that is, as a way of finding two matrices $S$ and $A$ so that $Y\approx F_\mathrm{d} (SA)$. In this case, the matrix $A$ is then chosen to be

\begin{align}
    \forall i, \, A_i = \left(x_i^py_i^q\right)_{p+q\le d}\;.
\end{align}
The components that make up $S$ are obtained through the minimization of a function of the form

\begin{align}\label{eq:expb}
\min_{\Delta S} \chi^2(\Delta S) +\|T\Delta S\|_\mathrm{F}^2\;,
\end{align}
where $S \eqdef S_0 + \Delta S$, $S_0$ being a first guess obtained from a median image of the shifted observations, $T$ is chosen to be a scalar weighting, and the $\chi^2$ data fidelity term is
                                                                                                                     
\begin{align}
    \chi^2(\Delta S) = \sum_{i=1}^{n_\mathrm{stars}} \left\|\frac{Y_i- F_\mathrm{d}\left([S_0+\Delta S]A_i\right)}{\hat\sigma_i}\right\|_2^2\;,
\end{align}
where $\hat \sigma_i^2$ contains the estimated per-pixel variances. The $\|T\Delta S\|_\mathrm{F}^2$ term is a regularization. Often referred to as Tikhonov regularization, it favors certain solutions among all those possible (in the present case of a scalar $T$, those with a smaller $l_2$ norm). Here, we include the flux normalization, sub-pixel shifting, and potential down-sampling (if super resolution is required) operators in $F_\mathrm{d}$. Shifting the PSF models to the same grid as those of the observed stars is performed, both within \texttt{PSFEx} and for our proposed approach in the upcoming section, through the use of a Lanczos interpolant.

\section{Resolved components analysis}\label{sec:RCA}
This section summarizes the RCA approach to super resolution introduced by \cite{ngole2016}. Special attention is given, in \autoref{sec:rcagraph}, to its spatial regularization scheme, and how it can be formulated using tools from graph theory.

\subsection{Overview}\label{sec:RCAov}
\begin{figure}
        \centering
        \includegraphics[width=.5\textwidth]{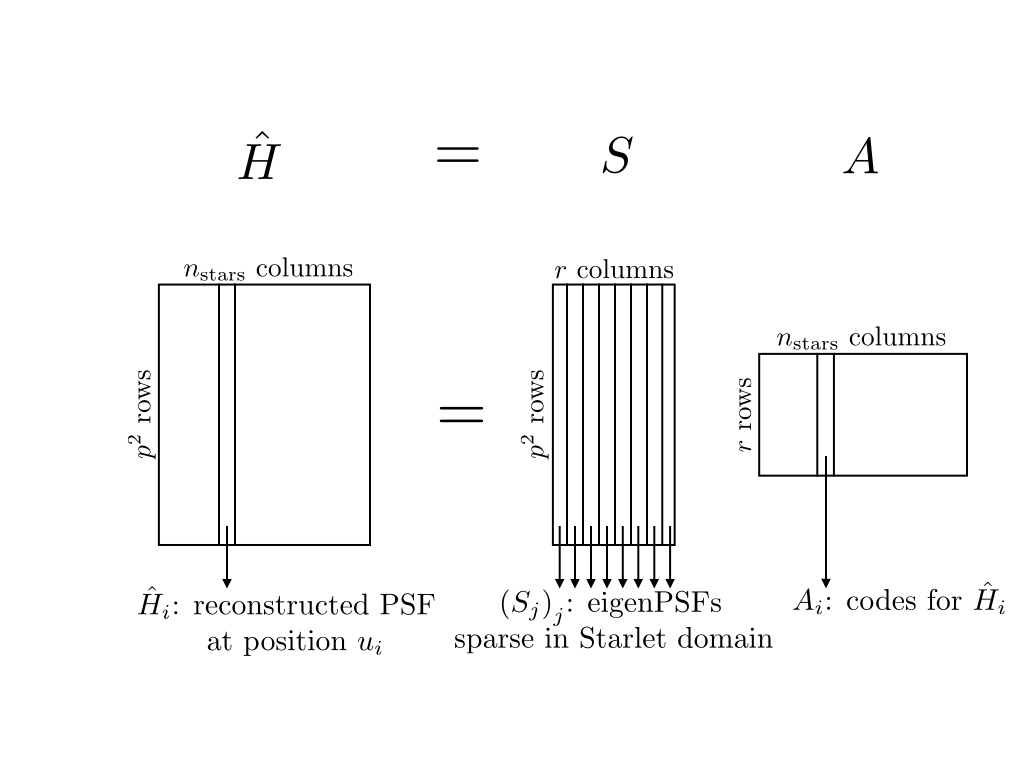}
        \caption{RCA's matrix factorization.}
        \label{fig:matfac}
    \end{figure}

        The RCA method, like many others (including \texttt{PSFEx}, as shown in \autoref{sec:psfex}), achieves super-resolution through matrix factorization. The PSF at the position $u_i$ of each star is reconstructed through a linear combination of a set of eigenPSFs, $S_j$:
    \begin{align}
        \hat{H}_i^\mathrm{RCA} \eqdef \hat{H}^\mathrm{RCA}(u_i) = \sum_{j=1}^r S_jA_{ij} = SA_i\;,
    \end{align}
    where each eigen PSF $S_j$ is an image of the same size as the PSF images. Introducing the set of all reconstructed PSFs at star positions, $\hat{H} = \left(\hat{H}_1,\dots,\hat{H}_{n_\mathrm{stars}}\right)$, we thus have the matrix formulation illustrated in \autoref{fig:matfac}.
    
    Because our data is under-sampled, a strong degeneracy needs to be broken: infinitely many finely sampled PSFs would manage to reproduce the observed under-sampled stars. In RCA, this degeneracy is broken by enforcing the following constraints on both $S$ and $A$, chosen to reflect known properties of the PSF field:    
    \begin{enumerate}
        \item \emph{Low rank}: the PSF variations across the field should be capturable through a small number of eigen PSFs $r$. This can be enforced by choosing $S$ to be of dimension $p^2\times r$, with $r \ll p^2$.
        \item \emph{Sparsity}: the PSF should have a sparse representation in an appropriate basis, which can be enforced through a sparsity constraint on the eigen PSFs.
        \item \emph{Positivity}: the final PSF model should contain no negative pixel values.
        \item \emph{Spatial constraints}: variations of the PSF across the field are highly structured, and the smaller the difference between two PSFs' positions $u_i,u_j$, the smaller the difference between their representations $\hat{H}_i, \hat{H}_j$ should be.
    \end{enumerate}
    The last of these constraints is achieved through a further factorization of $A$ itself. This step is described in the following subsection.

\subsection{Spatial regularization on the graph}\label{sec:rcagraph}
The spatial variations of the PSF across the FOV is highly structured, with both smooth variations that take place across the whole field, and some much more localized changes that only affect PSFs in a small part of it.  If we had access to evenly spaced samples, this would amount to variations occurring at different (spatial) frequencies. We could then capture these variations by making each of our eigen PSFs contain information related to a given spatial frequency. Our sampling of the PSF is, however, dependent on the position of stars in the field, over which we have no control. 

In RCA, we overcome this hurdle through the introduction of graph harmonics: each row $A_k$ of $A$, which contains the weights given to all observed star positions for eigen PSF $k$, is associated with a graph. For $k\in\{1,\dots,r\}$, $P_{e_k,a_k}$ denotes the Laplacian (up to a constant multiplication on the diagonal, see \autoref{appdx:graphprimer}) of the graph associated with $A_k$ (and thus to the $k$th eigen PSF). We define it as

\begin{align}\label{eq:graphlap}
    \nonumber \left(P_{e_k,a_k}\right)_{ij} &\eqdef \frac{-1}{\|u_i-u_j\|_2^{e_k}}\,\,\,\,\,\,\,\,\,\,\,\,\mbox{if $i\ne j$}\;,\\
    \left(P_{e_k,a_k}\right)_{ii} &\eqdef \sum_{\substack{j=1\\j\ne i}}^{n_\mathrm{stars}}\frac{a_k}{\|u_i-u_j\|_2^{e_k}}\;.
\end{align}
In other words, each of our $r$ PSF graphs are fully connected graphs with the edge between positions $u_i$ and $u_j$ given a weight of $1/\|u_i-u_j\|_2^{e_k}$.

By carefully choosing the parameters of our set of graphs, $\left(e_k,a_k\right)_{k\in\{1,\dots,n_\mathrm{stars}\}}$, we make each of them sensitive to different ranges of distances, which leads to the harmonic interpretation. See~\citet[particularly Sect. 5.2, 5.5.3, and Appendix A]{ngole2016} for more details, as well as a scheme to select appropriate $\left(e_k,a_k\right)_k$ from the data.

\begin{figure}
    \centering
    \includegraphics[width=.5\textwidth]{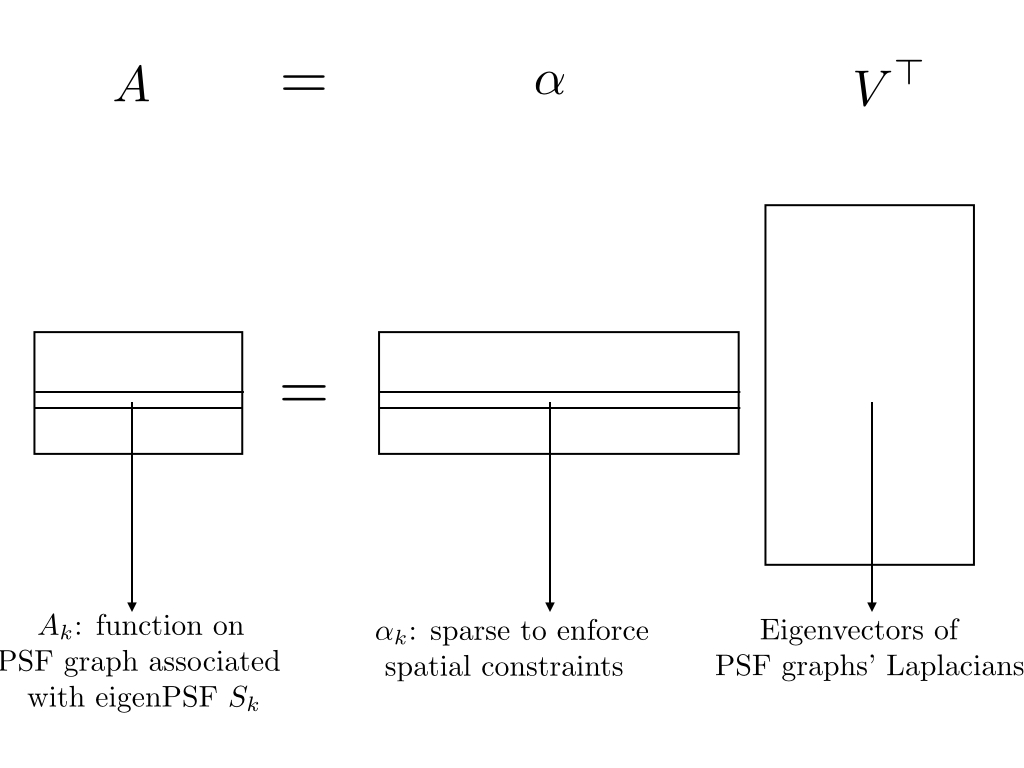}
    \caption{Matrices involved in RCA's spatial constraints.}
    \label{fig:graphfac}
\end{figure}

We enforce the link between $A$'s rows and their corresponding graph through the addition of a constraint on the former. Namely, we want to preserve the graph's geometry through $A$ so that the amplitude of coefficients associated with a certain eigen PSF varies with the right spatial harmonics. We achieve this in the following way: since $P_{e_k,a_k}$ is real and symmetric, we decompose it as

\begin{align}
    P_{e_k,a_k} \eqdef V_{e_k,a_k} \Sigma_{e_k,a_k} V_{e_k,a_k}^\top\;,
\end{align}
where $\Sigma_{e_k,a_k}$ is diagonal. $V\eqdef\left(V_{e_1,a_1},\dots,V_{e_r,a_r}\right)$ the matrix made up of the eigen vectors associated with each of our $r$ PSF graphs. Our spatial constraint can now be expressed by further factorizing $A$ by $V^\top$, and forcing the resulting coefficients $\alpha$ to be sparse. This is illustrated in \autoref{fig:graphfac}. For a quick introduction to the necessary graph theory concepts (and more insight into the construction of this spatial regularization), see \autoref{appdx:graphprimer}.

As mentioned in the introduction, manufacturing and polishing defects in the VIS instrument will inevitably lead to very localized, but strong variations of the PSF at some (fixed) positions. While these are not yet included in the simulated PSFs we use in \autoref{sec:xp}, they should be naturally handled by our proposed approach with the addition of extra eigen PSFs. Each of these additions would diminish the role of constraint 1 (low rank), but the added graph (and corresponding eigen PSF) would capture only those very localized changes in the PSF. However, all this can only be accomplished if some of the observed stars do fall within the area where these variations occur. As discussed in the introduction, this caveat could be alleviated by adding a temporal component to our model, and fitting it on stars extracted from several different exposures.

\subsection{Optimization problem}\label{sec:opti}
  Combining the factorizations illustrated in Figs.~\ref{fig:matfac} and~\ref{fig:graphfac}, reconstruction of the PSF field at the star positions through RCA amounts to solving the following problem:

  \begin{multline}\label{eq:RCApb}
  \shoveright{\min_{S,\alpha} \Bigg(\frac{1}{2} \|Y - F_\mathrm{d}(S\alpha V^\top)\|^2_\mathrm{F}} \\ + \sum_{i=1}^{r} \|w_i \odot \Phi s_i\|_1 + \iota_+(S\alpha V^\top) + \iota_\Omega(\alpha)\Bigg)\;,
  \end{multline}
  where $(w_i)_i$ are weights, $\odot$ denotes the Hadamard (or entry-wise) product, $\Phi$ is a transform through which our eigen PSFs should have a sparse representation~\cite[in our case, $\Phi$ will always be the Starlet transform,][]{starck2011}, $\iota_+$ is the positivity indicator function, that is, 
  
  \begin{align}
        \iota_+:X \mapsto \begin{cases}
                0 &\mbox{if no entry of $X$ is strictly negative},\\
                +\infty &\mbox{otherwise.}
        \end{cases}
  \end{align}
  Similarly, $\iota_\Omega$ is $0$ if $\alpha\in\Omega$ and $+\infty$ otherwise, and $\Omega$ is a sparsity enforcing set:
  
  \begin{align}
  \Omega \eqdef \left\{\alpha, \forall i\in\left\{1,\dots,r\right\}, \|(\alpha^\top)_i\|_0 \le \eta_i \right\}\;,
  \end{align}
  where $\|.\|_0$ is the ``norm'' that returns the number of nonzero entries of a vector. Here, $\alpha$ belongs to $\Omega$ if each of its row $i$ has at most $\eta_i$ nonzero entries.
  
  Breaking down \autoref{eq:RCApb} into its four summands, we can get a sense of how solving it would yield a PSF model that fits the observations while satisfying the list of constraints we introduced at the end of \autoref{sec:RCAov}. Indeed, the first term is the data fidelity term, which ensures we recover the observed star images after applying the correct under-sampling operator. The second term promotes the sparsity of our eigen PSFs, thus satisfying constraint 2. The third term ensures our PSF model only contains positive pixel values, enforcing constraint 3. The fourth term forces the learned $\alpha$ to be sparse, in turn satisfying constraint 4 as detailed in \autoref{sec:rcagraph}. Lastly, as mentioned above, constraint 1 is achieved by setting a low enough number of eigenPSFs $r$.
  
  Finding the eigenPSFs and their associated coefficients for each star amounts to solving \autoref{eq:RCApb}. This can be done through alternated minimization, that is, by solving in turn for $S$ then for $\alpha$ iteratively. Each minimization is performed through the use of \textit{proximal} methods. Examples of such algorithms adequate to our set up (where we have several constraints) include the generalized forward backward splitting~\citep{raguet2013} and that proposed by~\cite{condat2013}.
  For more details on solving the optimization problem, as well as how parameters $(e_k,a_k)_k$, $r$, $(w_i)_i,$ and $(\eta_i)_i$ are set, we refer the reader to~\cite{ngole2016}.

\section{PSF field recovery from graph harmonics}\label{sec:frca}
We now turn to the problem of interpolating our PSF model from the positions of stars, $\mathcal{U}_\mathrm{stars}$, to that of galaxies, $\mathcal{U}_\mathrm{gal}$. 

\subsection{Spatial interpolation of the PSF}\label{sec:spatialinterp}
Several standard methods exist to perform spatial interpolation, that is, to estimate the (unknown) value of some function $f$ at a new position $u=(x,y)$ given its measurements at several other positions: $\left(f(u_k)\right)_k$. See~\cite{gentile2013} for a review of such methods in the particular framework of PSF spatial interpolation. The most natural (and the one used by \texttt{PSFEx}) is probably the use of a polynomial function of positions:

\begin{align}\label{eq:polyinterp}
\hat{f}(u) = \sum_{\substack{i,j\ge0\\i+j\le d}} x^iy^jQ_{ij}\;,
\end{align}
where the maximum polynomial degree $d$ is a user-selected parameter, and the $(Q_{ij})_{i,j}$ are chosen so that $\hat{f}(u_k)  \approx f(u_k)$ at every position where $f$ was observed (in our case, $u_k\in\mathcal{U}_\mathrm{stars}$). We note that the particular set-up of \texttt{PSFEx} shown in \autoref{eq:polynomial} can be recovered when choosing $f\eqdef \hat H^\mathrm{PSFEx}$ and $Q_{ij}$ the \texttt{PSFEx}-learned, image-sized components.

An alternative to the polynomial approach is the use of radial basis functions~\cite[RBF, ][]{buhmann2003}. An RBF is a kernel $\varphi$ that only depends on the distance between two points. The polynomial formulation of \autoref{eq:polyinterp} can then be replaced by

\begin{align}\label{eq:RBFinterp}
\hat{f}(u) = \sum_{i=1}^{n_\mathrm{neighbors}} \, Q_i\varphi(\|u-u_i\|)\;,
\end{align}
where the positions in the sum correspond to the closest neighbors of $u$, and the $(Q_i)_i$ are, once again, chosen so that $\hat f$ coincides with the observed $f$ at all sampled positions. Broadly speaking, the idea behind these schemes is that the closer a position $u_i$ is, the more its observed value $f(u_i)$ should contribute to the estimated $\hat f(u)$, and RBF interpolation can be thought of as a generalization of inverse distance weighting schemes. Note that an assumption underlying the use of RBFs is that the PSF's amount of similarity to its neighbors in $f$ is isotropic, meaning the same in every direction.

Because of its simplicity and good performance exhibited in~\cite{gentile2013}, we chose to use RBF interpolation in the present work, and selected the commonly used thin plate RBF kernel~\cite[see][Sect. 3.2, for a quick discussion of its physical interpretation]{ngole2017}. In what follows, we always set $n_\mathrm{neighbors}$ to 15.

\subsection{Spatial regularity using RCA graph harmonics} 

Aside from the choice of the spatial interpolator discussed in the previous subsection, one must also decide \textit{which} function $f$ to interpolate. In our case, where the PSFs are images of $p^2$ pixels, the simplest approach would be to consider each of these pixels as a scalar function and interpolate it independently from the others. While simple, this approach is extremely sensitive to single-pixel fluctuations, which are not unexpected in our data-driven estimations of the PSF, for instance if some noise-related artefacts remain. 

As mentioned, instead of using $p^2$ different $f$ scalar functions, \texttt{PSFEx} instead considers $f$ to be $R^{p^2}$-valued. By construction, it performs a polynomial interpolation of its learned components. Spatial interpolation can also be carried within any chosen basis of representation -- a typical example being the use of principal components analysis (PCA), wherein the inputs are first decomposed, and the spatial interpolation is carried over the coefficients associated with the first few principal components.

\begin{figure}
    \centering 
    \includegraphics[width=.5\textwidth]{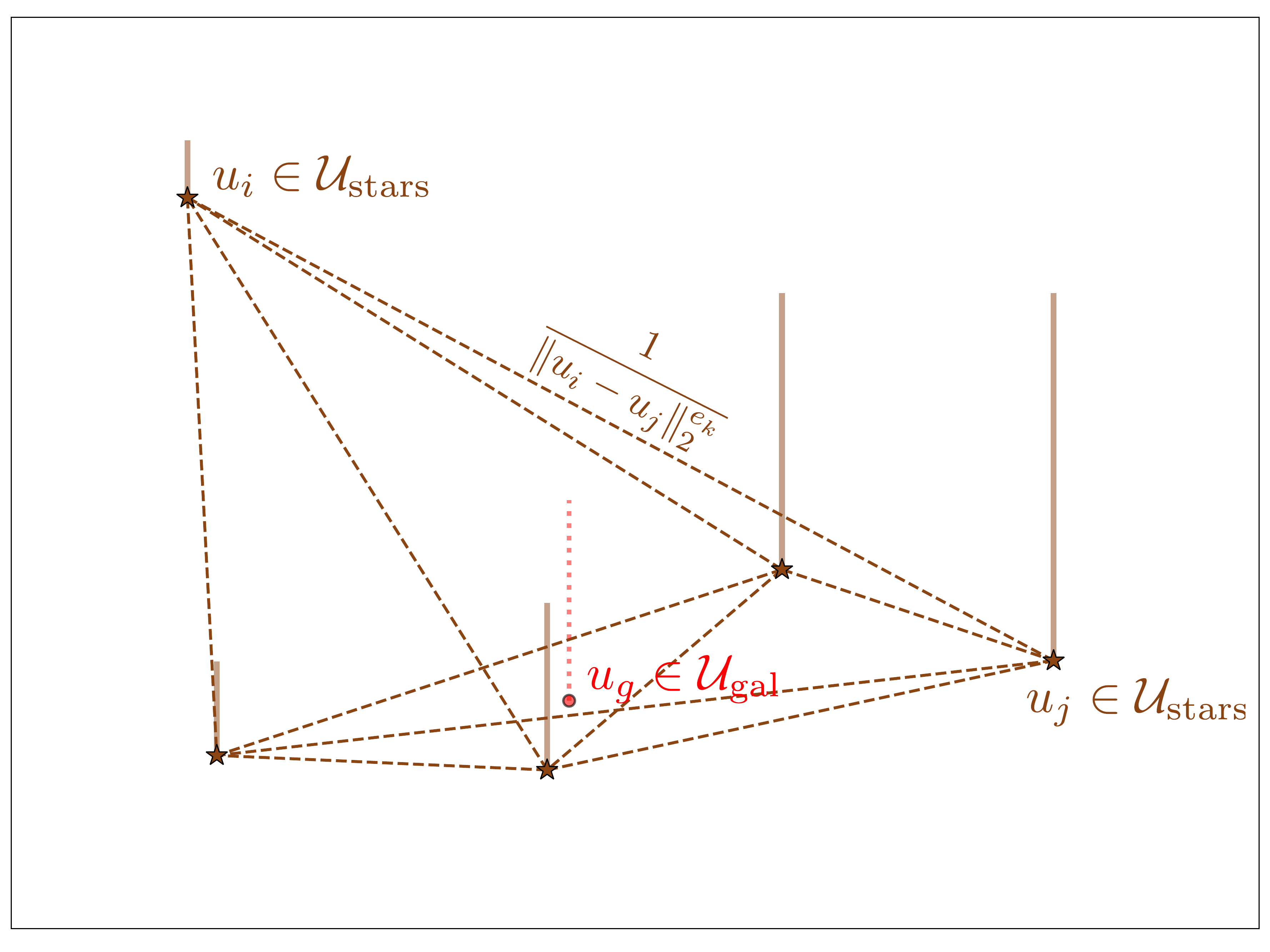}
    \caption{Graphical representation of the PSF graph associated with eigen PSF $S_k$. The height of the vertical bar at each position $u_i$ corresponds to the amplitude of coefficient $A_{ki}$.}
    \label{fig:PSFgraph}
\end{figure}

Our proposed approach is to perform this spatial interpolation step within the graph harmonics framework of RCA. We showed in \autoref{sec:rcagraph} that the rows of matrix $A$ are functions on a set of graphs, each containing the spatial information related to one particular eigen PSF. This is illustrated in \autoref{fig:PSFgraph}: the coefficients related to eigenPSF $S_k$ encode the spatial variations for a given range of distances. By performing the spatial interpolation in each of the $r$ rows of the RCA-learned $A$ matrix, we are moving along each of the corresponding PSF graphs. For any new position $u$, we can then reconstruct a new set of coefficients $A_u\in\mathbb{R}^p$ through $r$ RBF applications as in \autoref{eq:RBFinterp}, and reconstruct the PSF as
\begin{align}
    \hat H(u) \eqdef SA_u\;.
\end{align}
This amounts to adding a new point on the PSF graphs, as shown in red in \autoref{fig:PSFgraph}. Since $S$ was learned from the observed stars and $A_u$ preserves the graph harmonics, this step ensures the constraints we highlighted at the end of \autoref{sec:RCAov} are still applied to the new PSF at the galaxy positions. In particular, the spatial constraints are preserved thanks to the PSF graphs.

We note that an additional advantage to this approach lies in the fact that the most computation-intensive steps are performed during the reconstruction of the PSF field through RCA (\autoref{sec:RCA}). In an \Euclid-like framework where star images are under-sampled, if we were to use RCA to perform the necessary super-resolution step, the dictionary $S$ and the graph harmonics encoded in $A$ would already be computed. The proposed method can thus execute the spatial interpolation step in a particularly appropriate representation at no additional computational cost save for that of fitting the RBF weights. Conversely, if one wanted to use any other representation, even one as simple as PCA would require some extra computation (spectral value decomposition, in this case).

\section{Comparison of PSF models}\label{sec:xp}
In this section, we apply \texttt{PSFEx} and the proposed approach to simulated stars.

\subsection{Data set}\label{sec:psf}
\begin{figure*}
        \centering
    \begin{subfigure}{0.49\linewidth}
        \centering
        \includegraphics[width=.8\textwidth]{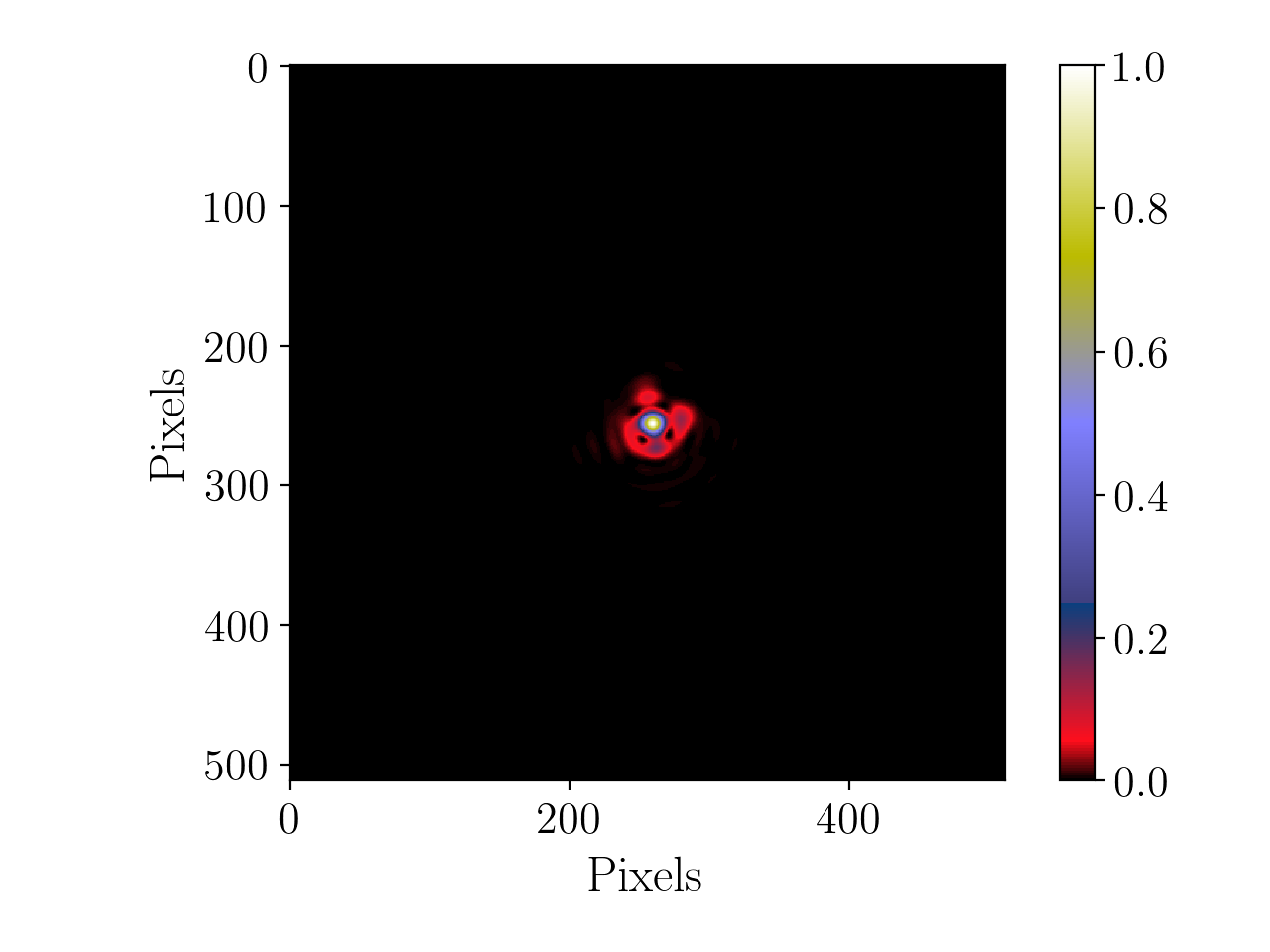}
    \end{subfigure}
    \begin{subfigure}{0.49\linewidth}
        \centering
        \includegraphics[width=.8\textwidth]{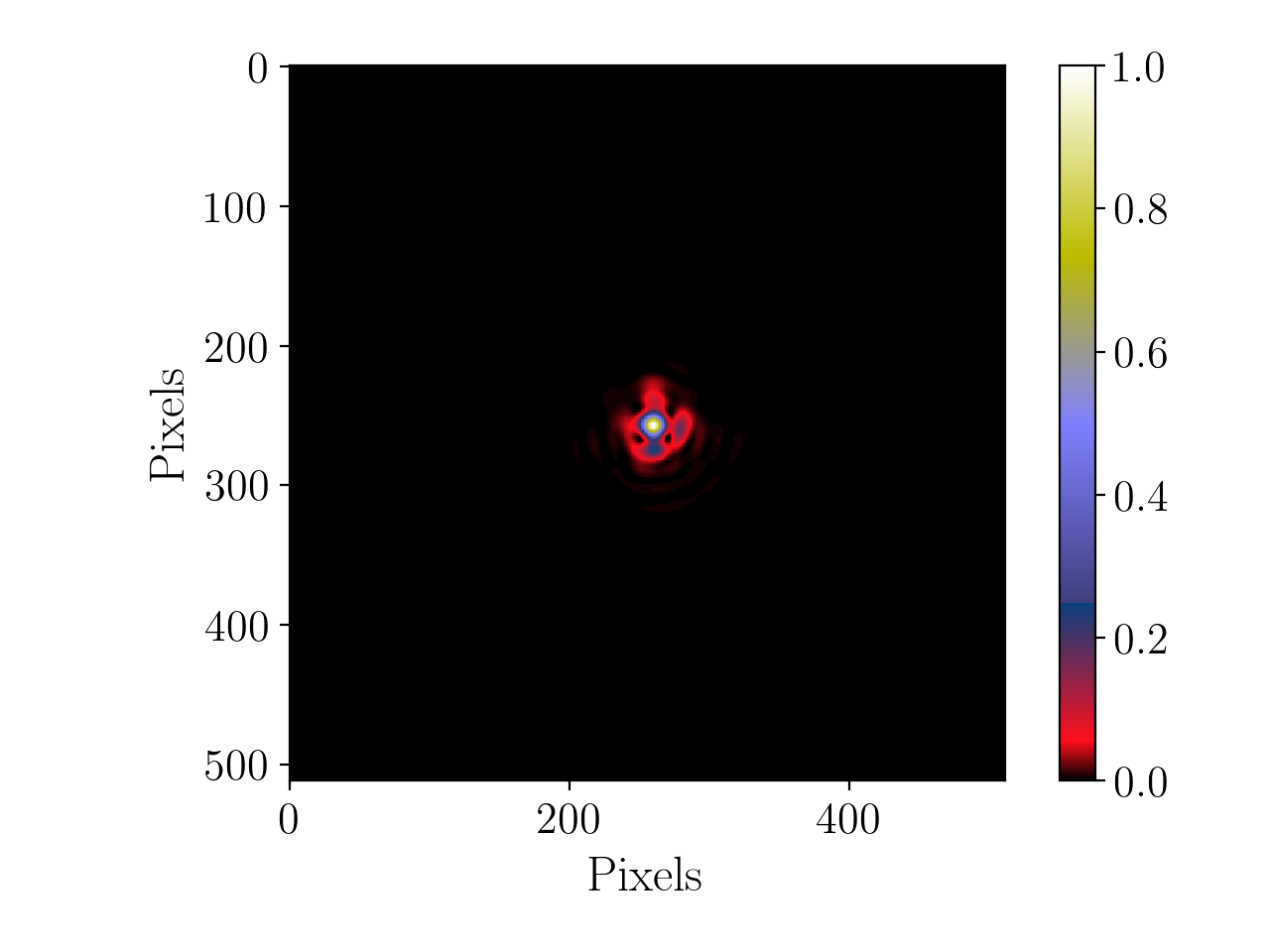}
    \end{subfigure}\\
    \begin{subfigure}{0.49\linewidth}
        \centering
        \includegraphics[width=.8\textwidth]{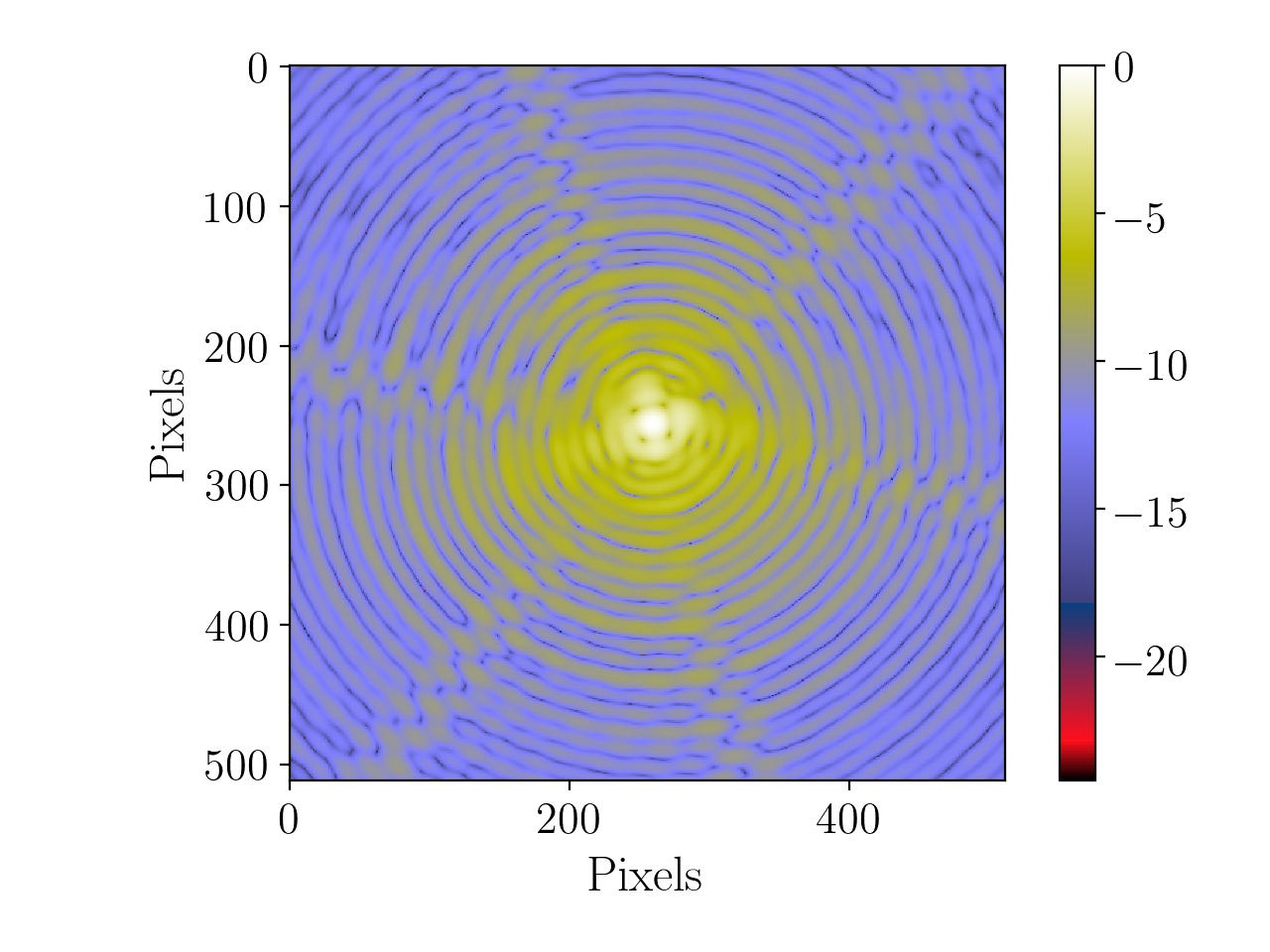}
    \end{subfigure}
    \begin{subfigure}{0.49\linewidth}
        \centering
        \includegraphics[width=.8\textwidth]{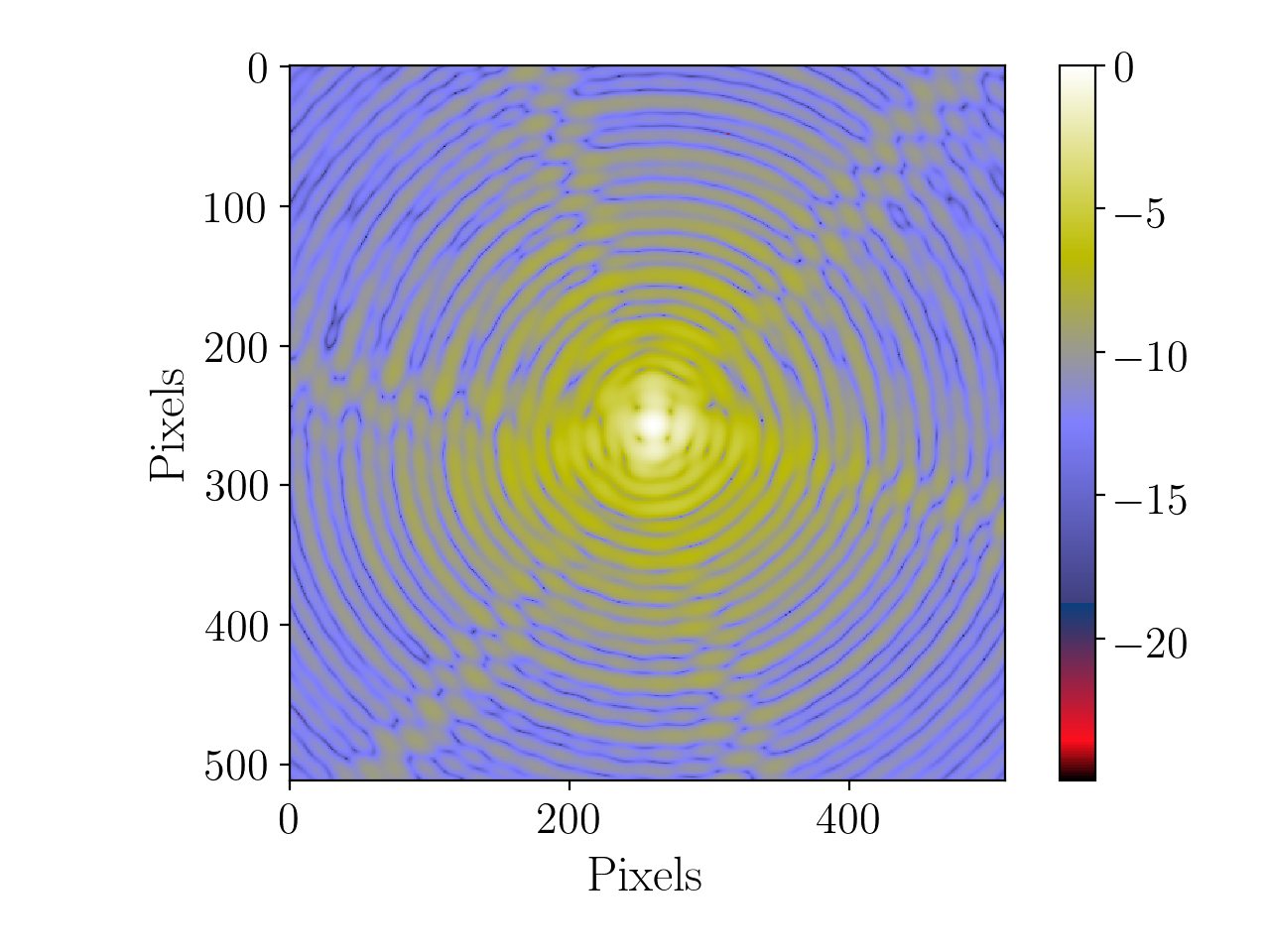}
    \end{subfigure}
    \caption{Visual examples of simulated Euclid PSF in the natural (\textit{top row}) and logarithmic (\textit{bottom row}) domains, at the original pixel sampling of the simulation (about 12 times finer than \Euclid). Each stamp is approximately 4.25\,arcsec across.}
    \label{fig:example_PSF}
\end{figure*}

The PSFs we use are simulations of \Euclid's VIS instrument's PSFs~\cite[as described in][\S 4.1]{ngole2015}, located in the central part of the FOV, sampled at a single wavelength of $600$\,nm. As mentioned in \autoref{sec:intro}, this is a simplification of the true Euclid PSF, since we neglect its chromatic variations, and the detector and guiding effects are absent from the simulations. This data set contains $597$ such PSFs, each consisting of a $512\times 512$ stamp with a pixel size of approximately $0.0083$\,arcsecond, that is to say sampled on a much finer grid than the \Euclid pixel size. See \autoref{fig:example_PSF} for two examples chosen at some of the top-right- and bottom-left-most positions. We note these simulations correspond to a ``toleranced'' realization of the instrument. The amount of variations seen in the instrumental PSF here is rather pessimistic, and the true flight model is likely to exhibit more spatial stability.

As previously discussed, in a real-life observing situation, the only information (in the nonparametric framework of this work) from which we would derive our PSF models would be obtained from stars within the field, which lie at positions different from those where we wish to estimate it. We thus randomly split our sample of PSFs into two parts: a training set of $300$ PSFs, the position of which we refer to as "star positions"; and a test set with the remaining $297$ PSFs (at "galaxy positions").
The number of stars in our training sample is of order $10$ times smaller than the expected average number of usable stars present in a VIS science exposure, though using all the available stars simultaneously would require taking into account the variations of the PSF across different CCDs.
 In \autoref{sec:PSFmodel}, we only use the PSFs at star positions to try and produce estimations of the PSF at the galaxy positions. Conversely, from \autoref{sec:gals} onward, we solely focus on and use objects at the galaxy positions.

\begin{figure*}
        \centering
    \begin{subfigure}{0.24\linewidth}
        \centering
        \includegraphics[width=.8\textwidth]{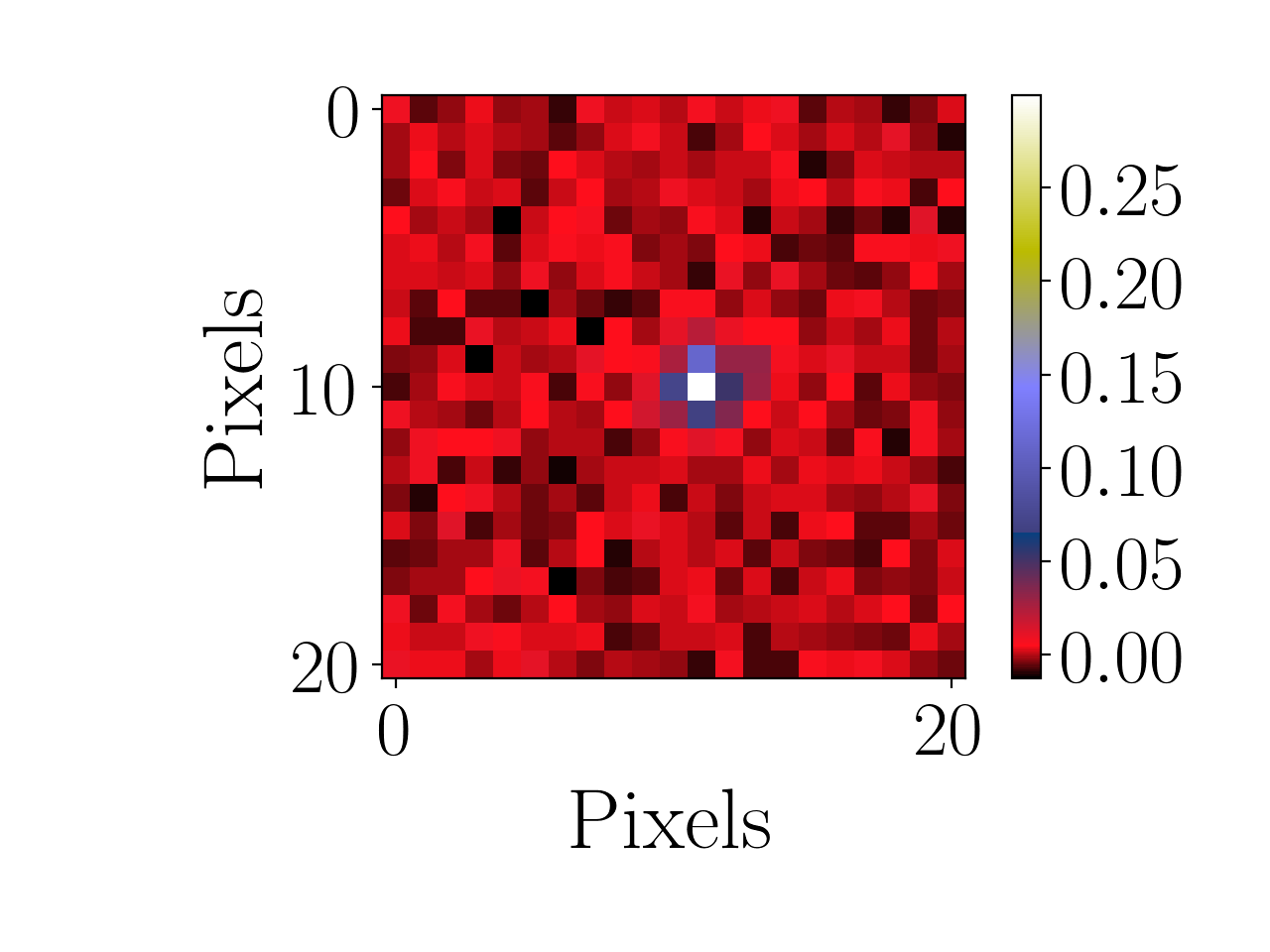}
    \end{subfigure}
    \begin{subfigure}{0.24\linewidth}
        \centering
        \includegraphics[width=.8\textwidth]{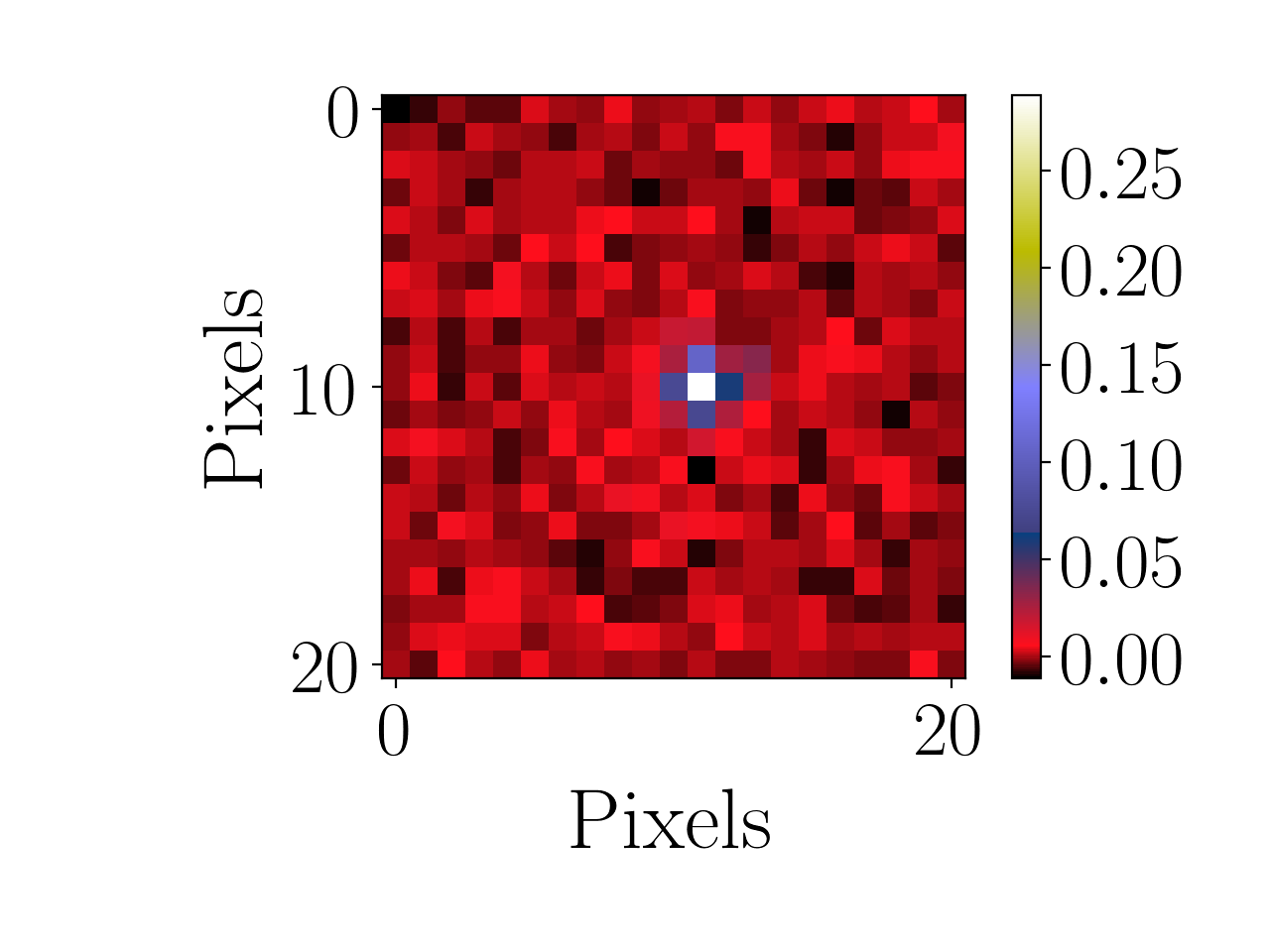}
    \end{subfigure}
    \begin{subfigure}{0.24\linewidth}
        \centering
        \includegraphics[width=.8\textwidth]{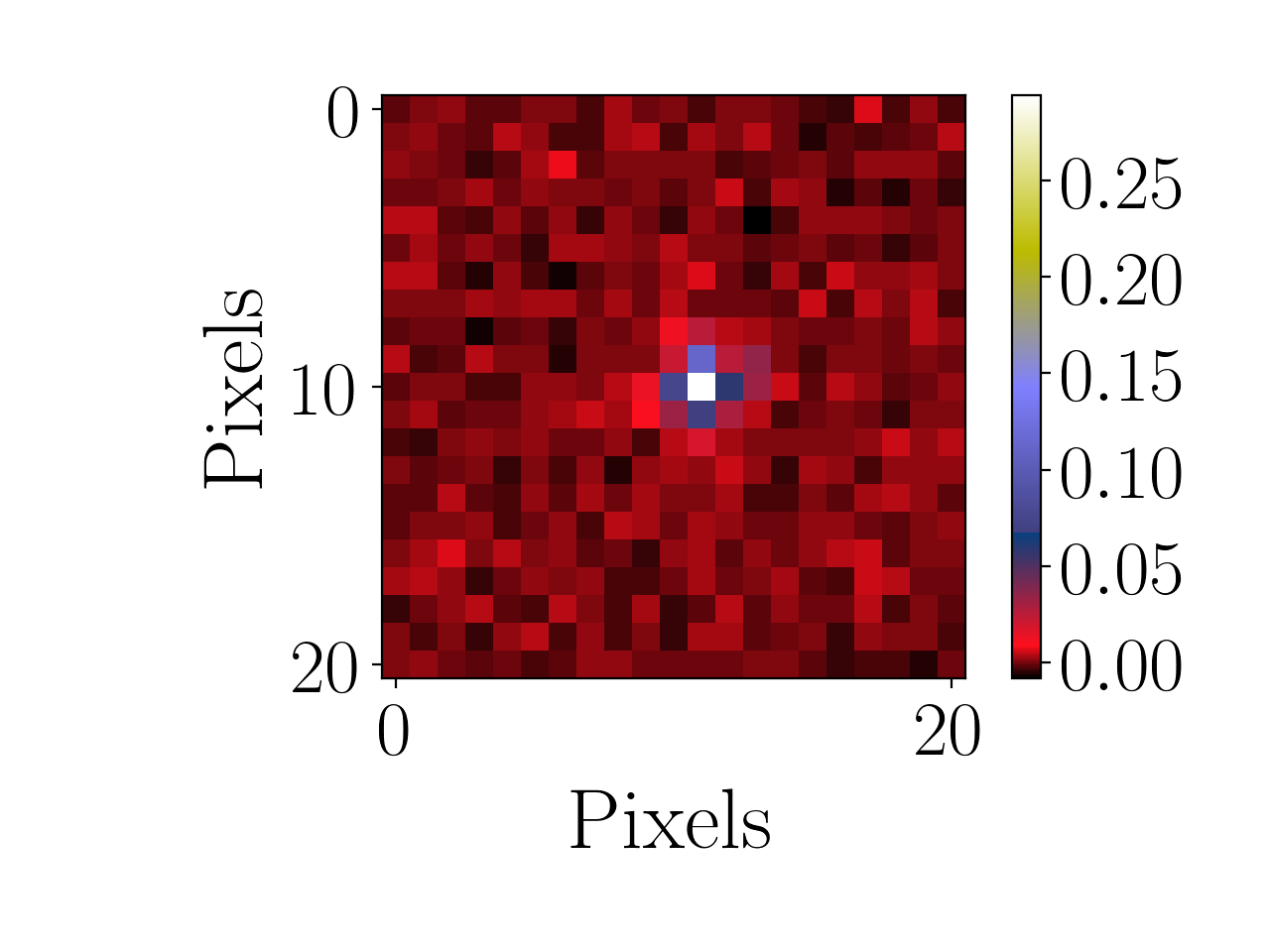}
    \end{subfigure}
    \begin{subfigure}{0.24\linewidth}
        \centering
        \includegraphics[width=.8\textwidth]{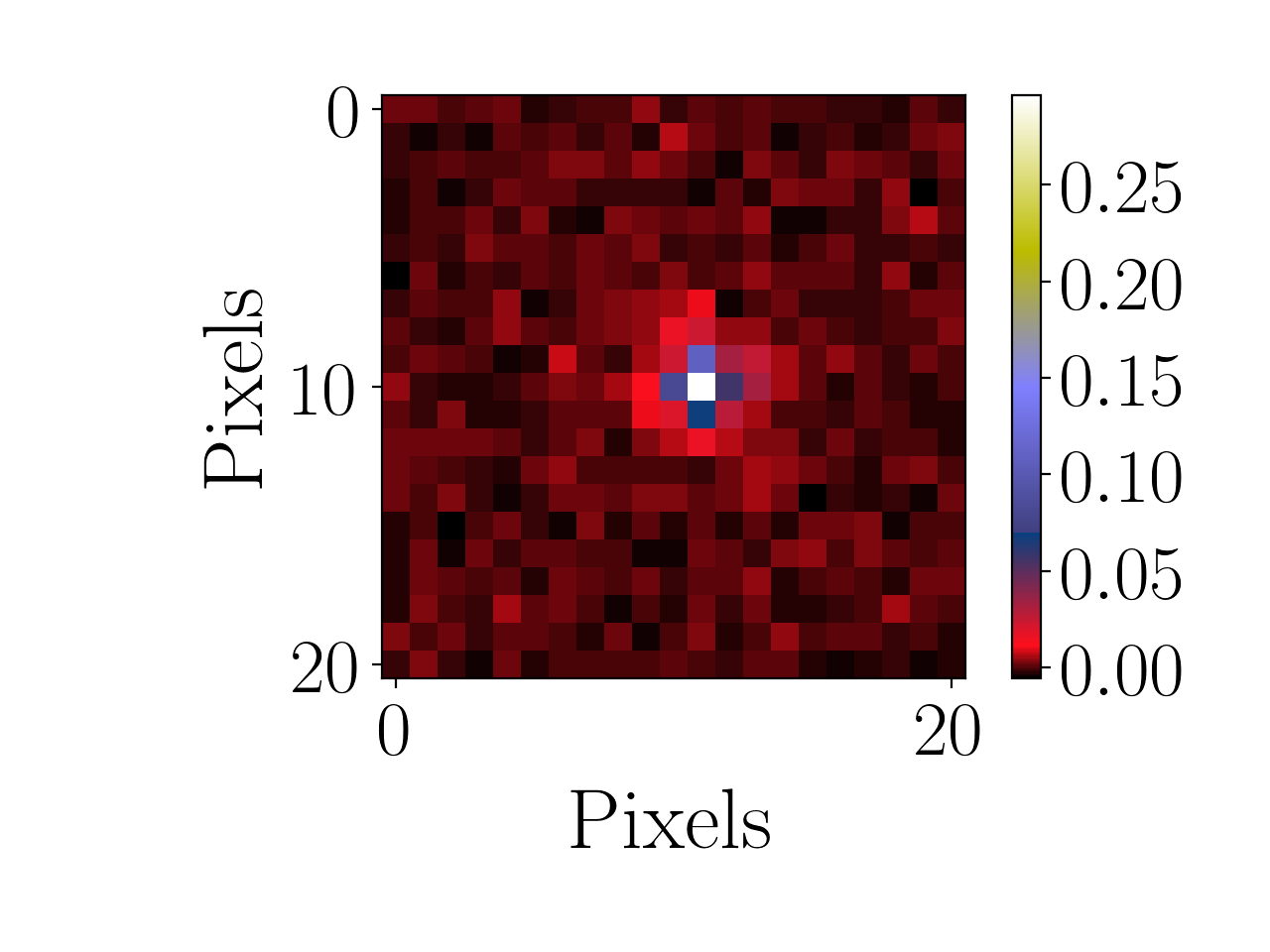}
    \end{subfigure}
    \caption{Example observed (under-sampled) star stamps, at the various SNRs considered (\textit{from left to right}, 10, 20, 35, 50), from which the PSF models will be estimated. Each stamp is approximately 2.1\,arcsec across.}
    \label{fig:ugstars}
\end{figure*}

\Euclid's sampling frequency is at 0.688 of the Nyquist rate, which sets our goal in terms of super resolution at achieving an up-sampling factor of $1/D = 2$~\citep{cropper2013}. To simulate observed stars, we sampled all $300$ PSFs in the training set at the nominal \Euclid pixel scale of $0.1$\,arcsecond. This is achieved by first applying a mean filter (which amounts to the approximation that the VIS pixel response is a perfect top hat), then sampling pixels at the correct rate. We applied a random sub-pixel shift to each resulting image, then truncated the stamps to be of size $21\times 21$ around the pixel closest to the object's centroid. Indeed, in observing situations, our PSF models would likely (definitely, in the cases of both \texttt{PSFEx} and RCA) be fit on image stamps containing a suitable star extracted from the full image. Our models would thus necessarily need to deal with the resulting truncation effects. Lastly, we added various levels of white Gaussian noise with standard deviation $\sigma$, yielding five different sets of observed stars at average signal-to-noise ratios (SNR) of $10, 20, 35,$ and $50$, where SNR is defined as

\begin{align}
        \mathrm{SNR} = \frac{\|x\|^2_2}{\sigma^2 p^2}\;,
\end{align}
for image $x$ of size $p\times p$. An example of the resulting star images is shown in \autoref{fig:ugstars}.

From these, we estimate the PSF at twice the \Euclid pixel sampling and at the galaxy positions in \autoref{sec:PSFmodel}. Because of the up-sampling, the resulting PSFs are stored in stamps of $42\times 42$ pixels. 

For comparison, we also prepared a set of ``known'' PSFs $\hat H ^{\mathrm{kn}}$ at those positions, by sampling the 297 test PSFs at half of \Euclid's pixel size and truncating the resulting images to $42\times 42$ pixels. While not the ideal case (where the continuous PSF image would be perfectly known), this fiducial, unattainable case amounts to the best possible PSF our approaches to super-resolution, denoising and spatial interpolation could possibly achieve. We note that this would require some extra conditions to be met, for example by the population of random shifts undergone by the under-sampled images (which is, under the safe assumption that shifts are randomly distributed, also directly related to the number of observed stars). Using the notations of \autoref{sec:notations}, $\hat H ^{\mathrm{kn}}$ would be the the PSF obtained if only $\mathcal{F}_\mathrm{s}$ remained while $F_\mathrm{d}$ had been perfectly corrected for. In other words, the only effects degrading these PSFs are those of sampling (at our target of half \Euclid's pixel size), and truncation at the best possible stamp size given that of our star images.

\subsection{PSF modeling}\label{sec:PSFmodel}
We first assumed the star images described in \autoref{sec:psf} had already been extracted, and we performed both super-resolution and spatial interpolation using RCA, as described in Sects.~\ref{sec:RCA} and~\ref{sec:frca}. This yielded a set of 297 RCA-estimated $42\times 42$ PSFs, $\hat H ^{\mathrm{RCA}}$, at galaxy positions per SNR level.

\texttt{PSFEx} was designed to run on catalogs extracted using companion software \texttt{Source Extractor}, and thus requires a little more setup. For each SNR level, we first created a fake full image of $12\,000\times 12\,000$ pixels, into which the $300$ stars are placed at their respective positions. We then ran \texttt{Source Extractor} on the resulting images, with parameters selected so that all stars were detected and extracted correctly, and no spurious detections occurred. When run, \texttt{PSFEx} performs a further selection across all objects extracted by \texttt{Source Extractor}, which is usually desirable for fitting the PSF model to appropriate stars. In our case, however, since we already knew our \texttt{Source Extractor} catalog to be perfect, we tuned \texttt{PSFEx}'s selection parameters so that as many stars as possible were used. One was nonetheless rejected at SNR 50. The parameters related to the model are the following:

\begin{verbatim}
PSF_SAMPLING    .5               
PSF_SIZE        42,42           
PSFVAR_KEYS     X_IMAGE,Y_IMAGE 
PSFVAR_GROUPS   1,1            
PSFVAR_DEGREES  2              
\end{verbatim}
Namely, \texttt{PSFEx} learns a set of PSF basis elements $\left(S_{ij}\right)_{i,j}$ so that the PSF at position $x,y$ is estimated as in \autoref{eq:polynomial}, with $d=2$ (repeating the experiments with $d=3$ led to very poor PSF models).
All other \texttt{PSFEx} parameters were left to their default value. Again, this produced one set of estimated PSFs per SNR level, $\hat H^{\mathrm{PSFEx}}$, with the same stamp and pixel sizes as $\hat H^{\mathrm{kn}}$ and $\hat H^{\mathrm{RCA}}$.

\begin{figure*}
        \centering
    \begin{subfigure}{0.49\linewidth}
        \centering
        \includegraphics[width=.8\textwidth]{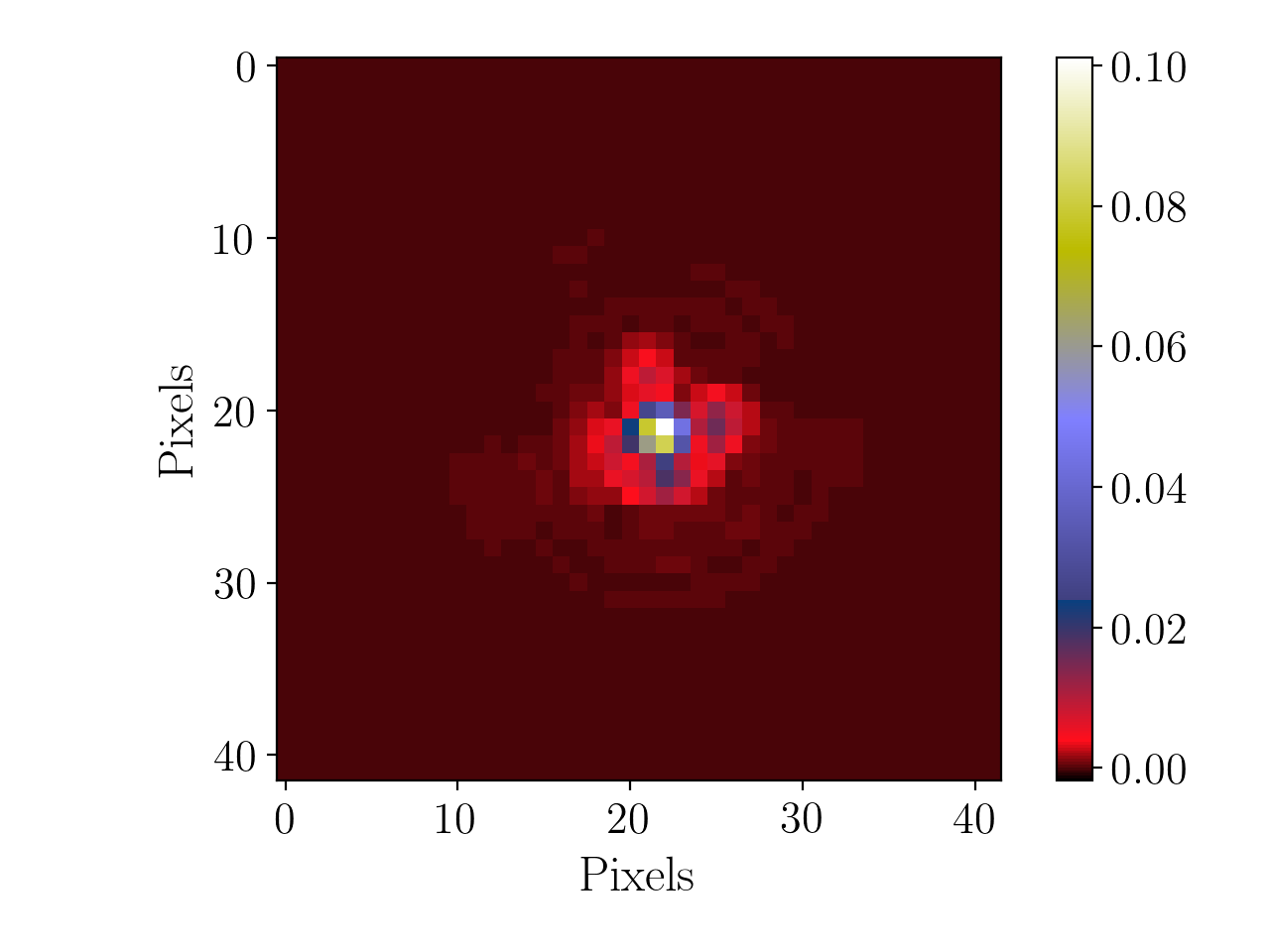}
    \end{subfigure}
    \begin{subfigure}{0.49\linewidth}
        \centering
        \includegraphics[width=.8\textwidth]{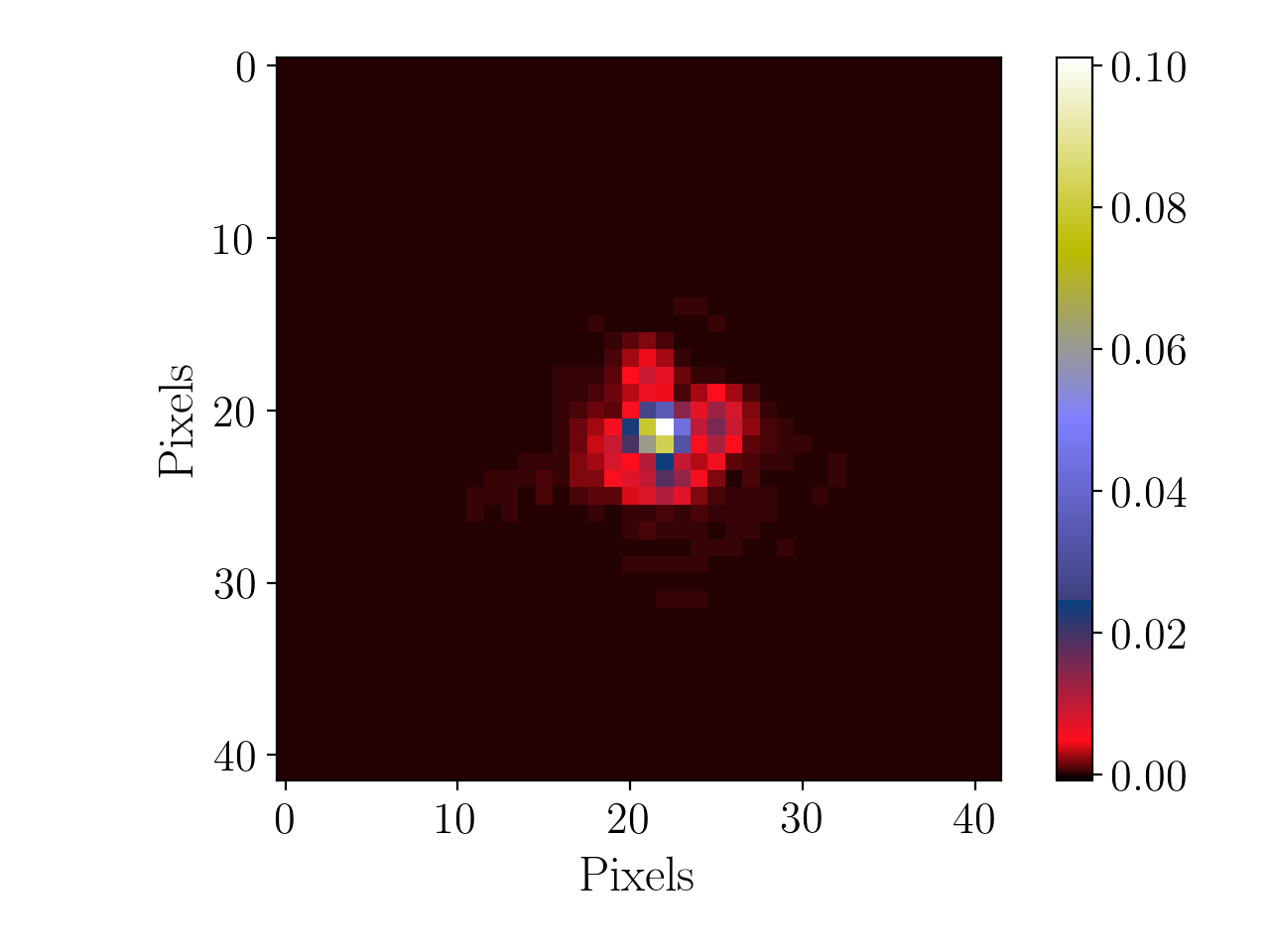}
    \end{subfigure}\\
    \begin{subfigure}{0.49\linewidth}
        \centering
        \includegraphics[width=.8\textwidth]{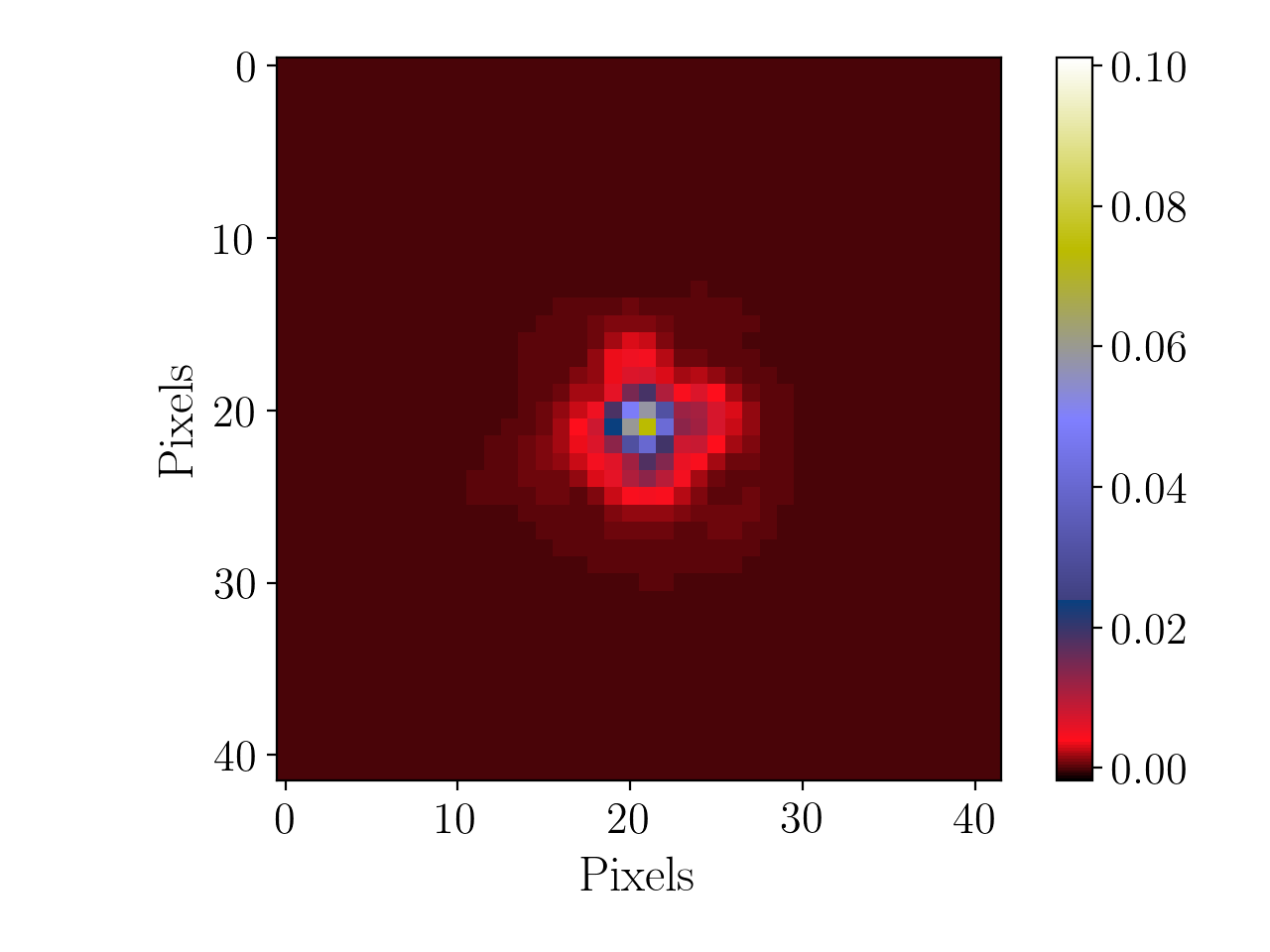}
    \end{subfigure}
    \begin{subfigure}{0.49\linewidth}
        \centering
        \includegraphics[width=.8\textwidth]{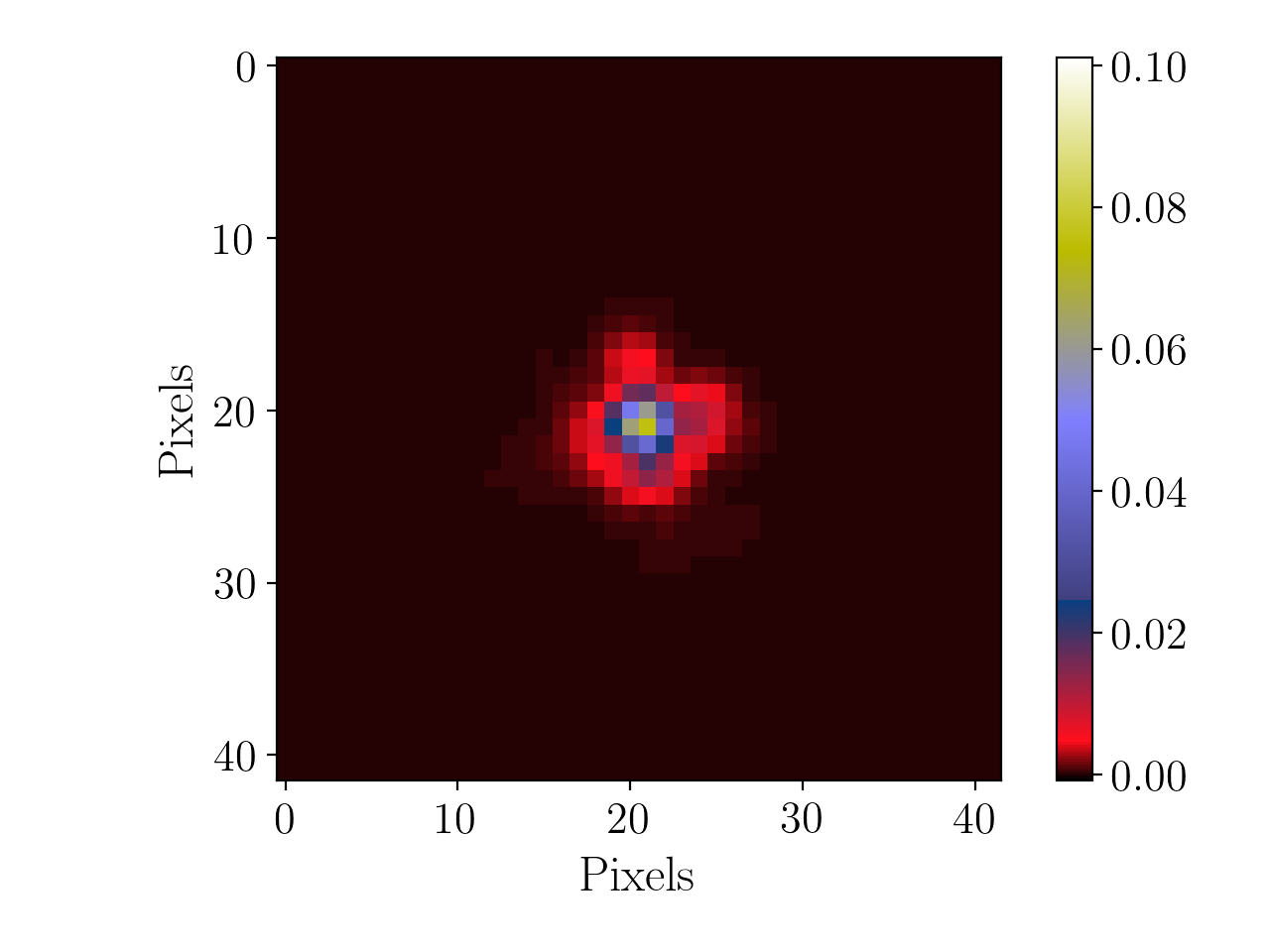}
    \end{subfigure}\\
    \begin{subfigure}{0.49\linewidth}
        \centering
        \includegraphics[width=.8\textwidth]{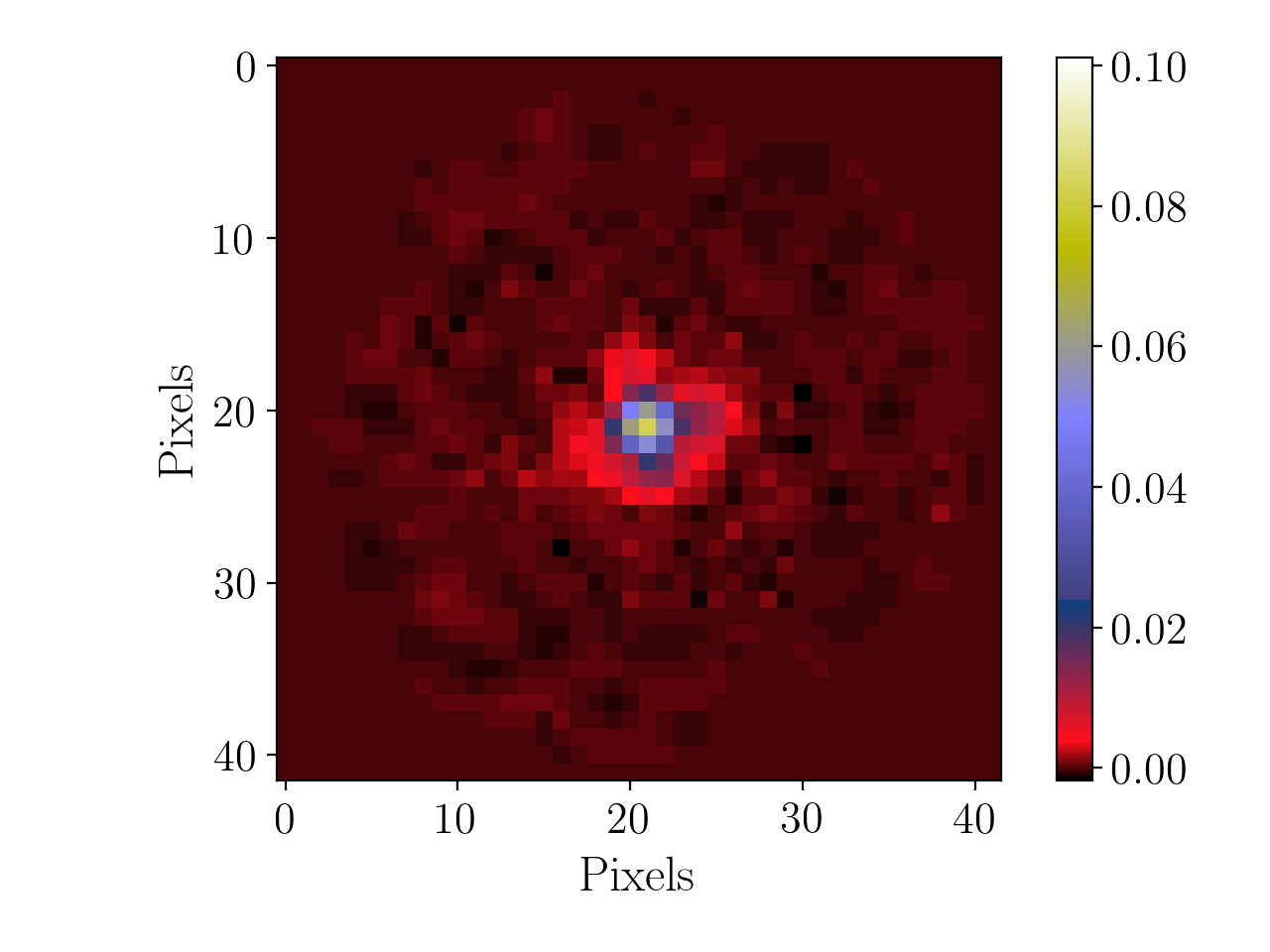}
    \end{subfigure}
    \begin{subfigure}{0.49\linewidth}
        \centering
        \includegraphics[width=.8\textwidth]{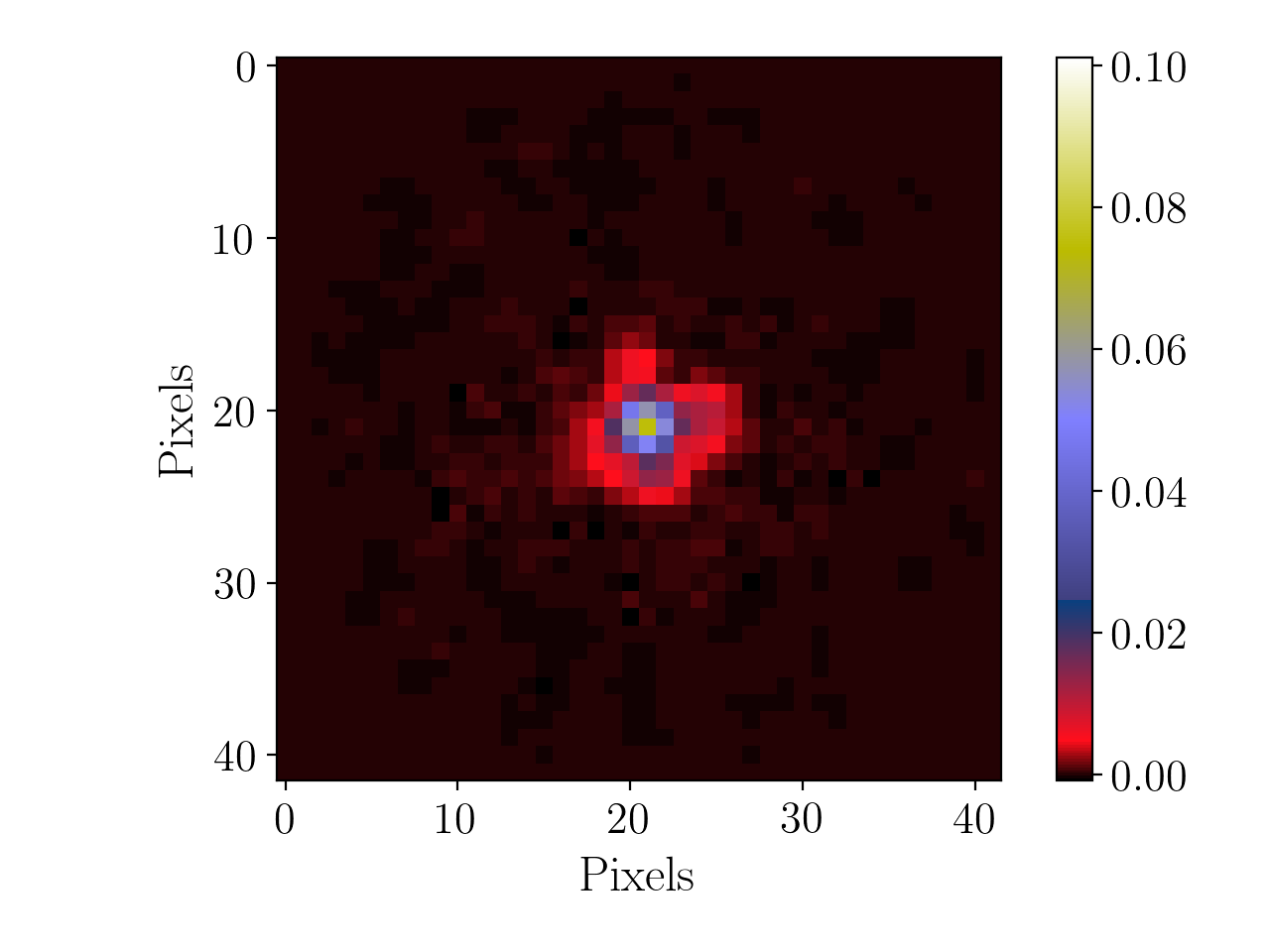}
    \end{subfigure}
    \caption{Examples from three sets of estimated PSFs described in \autoref{sec:psf} (\textit{from top to bottom}: $\hat H^{\mathrm{kn}}$, $\hat H^{\mathrm{RCA}}$, $\hat H^{\mathrm{PSFEx}}$), at SNR of $10$ (\textit{left column}) and $50$ (\textit{right}). Each stamp is approximately 2.1\,arcsec across. We note that the only difference between the two top-most images is the color map, matched to be the same as that of $\hat H^{\mathrm{RCA}}$ and $\hat H^{\mathrm{PSFEx}}$ estimated for each SNR.}
    \label{fig:PSF_models}
\end{figure*}

Examples of $\hat H^{\mathrm{kn}}$, $\hat H^{\mathrm{RCA}}$, and $\hat H^{\mathrm{PSFEx}}$, at the galaxy position corresponding to the simulated PSF on the left-hand side of \autoref{fig:example_PSF}, are given in \autoref{fig:PSF_models} for the worst and best-case noise scenarios.

\subsection{Results}\label{sec:psfcomp}
\cite{paulin2008} established a basis for studying the impact of imperfect PSF models on the shape measurement of galaxies. Working in the framework of unweighted quadrupole moments, they found that the error on the measured galaxy ellipticity is

\begin{align}\label{eq:paulin}
        \hat{e}_i = e_i\left(1 + \frac{\delta(R^2_\mathrm{PSF})}{R^2_\mathrm{gal}}\right) - \left(\frac{R^2_\mathrm{PSF}}{R^2_\mathrm{gal}} \delta e_i^\mathrm{PSF} + \frac{\delta(R^2_\mathrm{PSF})}{R^2_\mathrm{gal}}e_i^\mathrm{PSF}\right)\;,
\end{align}
where $R^2_\mathrm{obj}$ is the size of $\mathrm{obj}$ (which can be either a PSF or a galaxy) defined from quadrupole moments, and $\delta$ indicates the difference between a quantity of the true PSF and that of the model. Following \autoref{eq:paulin}, we can first quantify the quality of our PSF by looking at the errors in size and ellipticity, $\delta(R^2_\mathrm{PSF}), \delta e_i^\mathrm{PSF}$, for both models, on the 297 test PSFs. 

For the former, both models tend to overestimate the size of the PSF, likely because the super-resolution is performed on a small sample of very narrow objects. This size error has a much stronger contribution to the quantities in \eqref{eq:paulin} than the ellipticity error. RCA already reduces this bias in our current set-up, with an improvement of about $24\%$ at all noise levels. This still leads to an RMS on the relative size $\delta(R^2_\mathrm{PSF})/R^2_\mathrm{PSF}$ that is about $10^4$ times too high to match the requirements. Beyond the need to use more stars simultaneously to build the model, which also emerges from every other current shortcoming of our approach, this strong bias will already be greatly reduced in a more realistic \Euclid scenario, since a broadband PSF is necessarily broader than the monochromatic PSF we consider in this work, regardless of the target object's spectral energy distribution (SED).

\begin{figure}
    \centering
    \includegraphics[width=\linewidth]{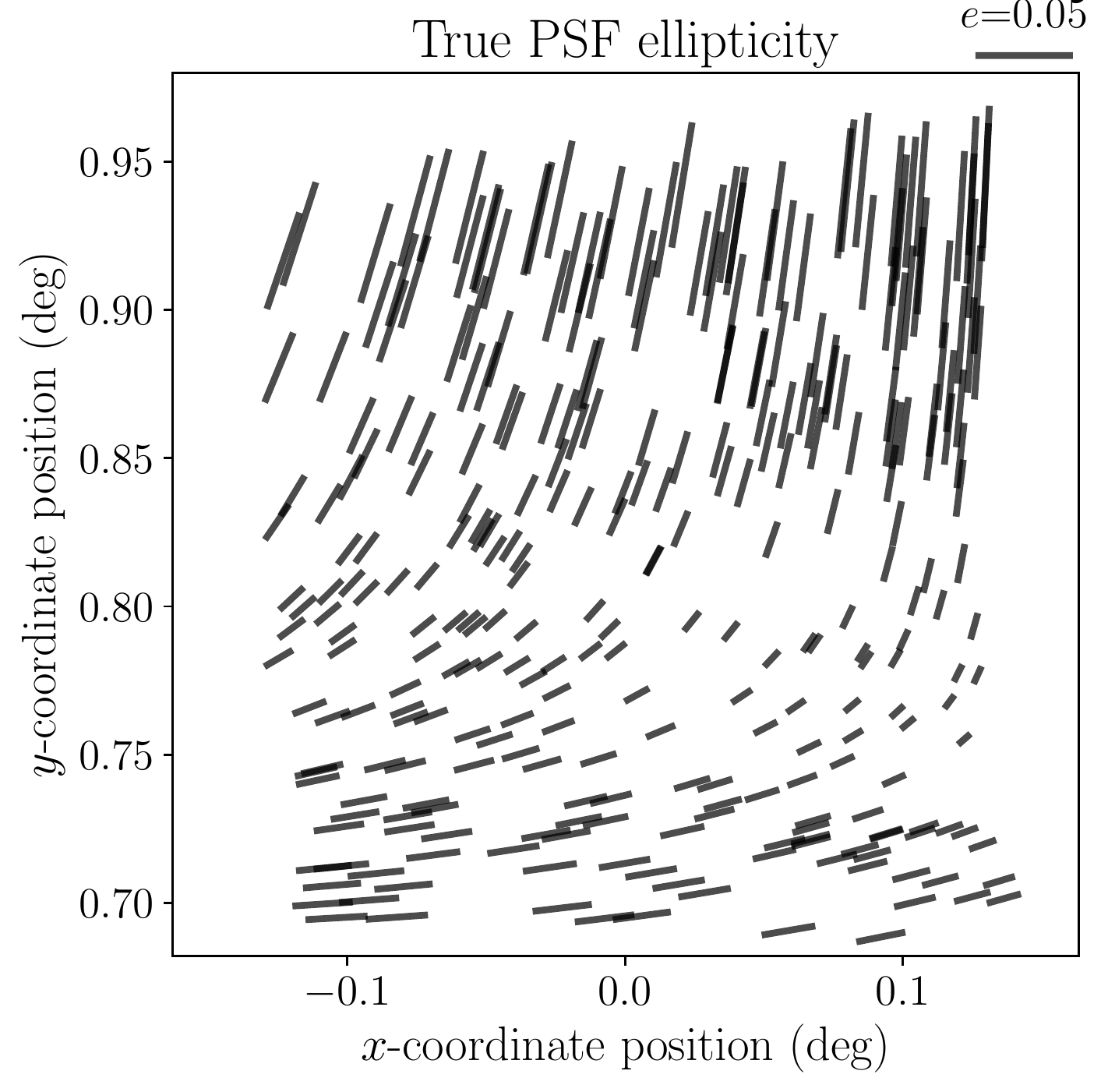}
    \caption{True PSF ellipticity as a function of position. We note these correspond to a pessimistic realization of the instrument, and variations of the true instrument's PSF are likely to be of smaller amplitude.}    
    \label{fig:truellscatter}
\end{figure}

\begin{figure*}
    \centering
    \includegraphics[width=\linewidth]{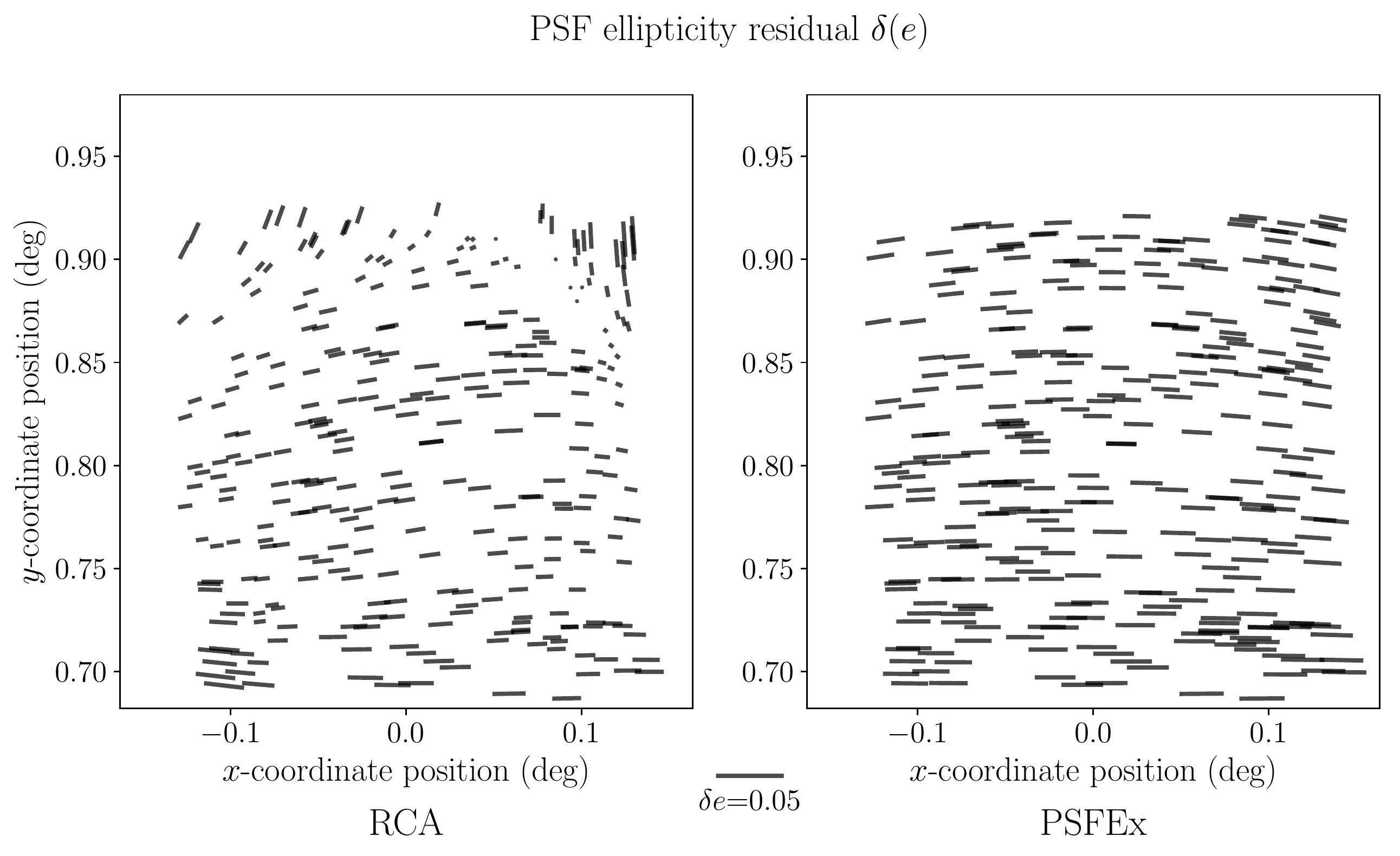}
    \caption{PSF ellipticity residuals as a function of position, for both PSF models. \textit{Left}: RCA; \textit{right}: \texttt{PSFEx}.}    
    \label{fig:ellscatter}
\end{figure*}

\begin{figure}
    \centering
    \begin{subfigure}{\linewidth}
        \centering
        \includegraphics[width=.8\textwidth]{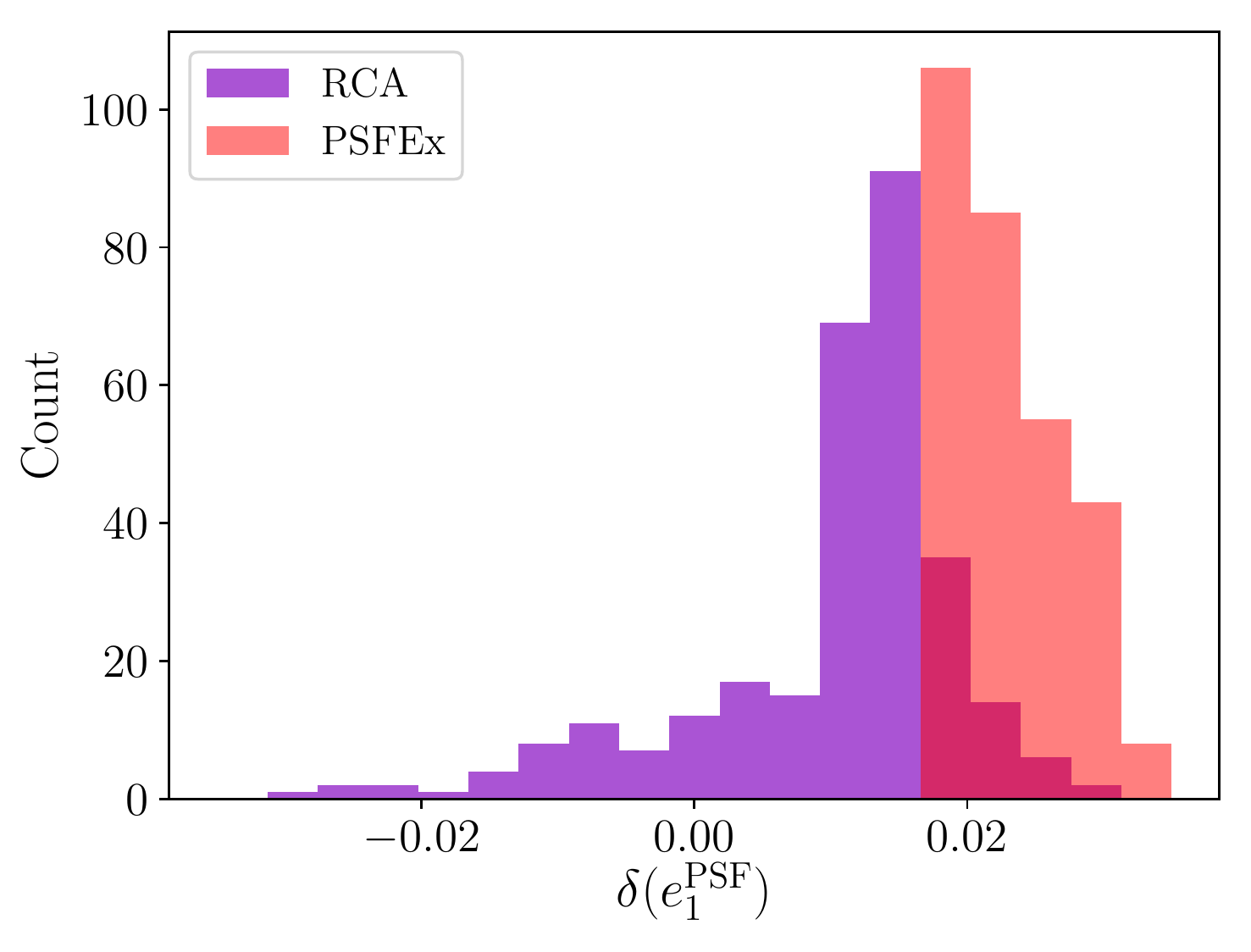}
        \label{fig:deltae1}
    \end{subfigure}\\
    \begin{subfigure}{\linewidth}
        \centering
        \includegraphics[width=.8\textwidth]{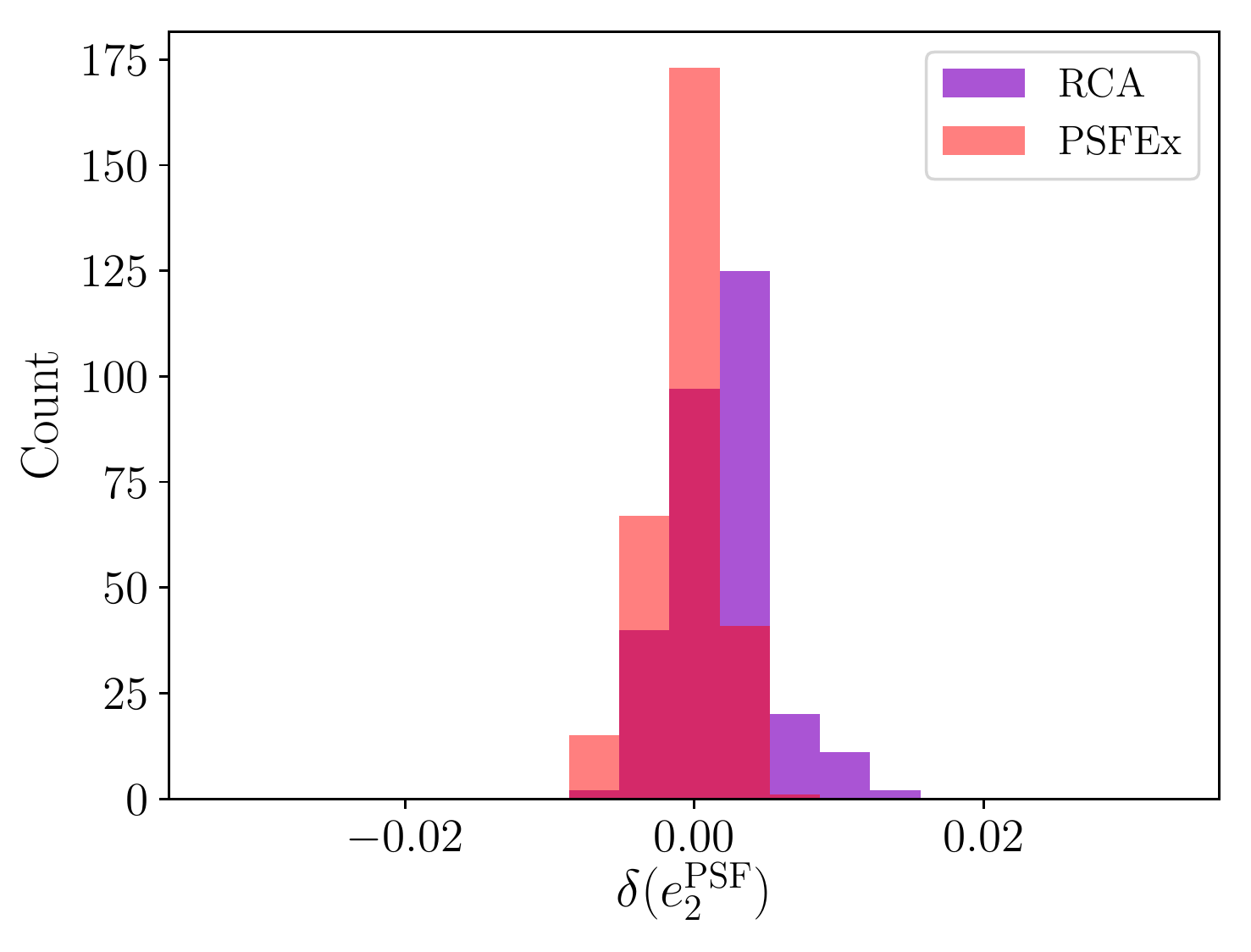}
    \end{subfigure}
    \caption{Distribution of ellipticity residuals for both PSF models. Measurements were made with star SNR 35. \textit{Top}: first ellipticity component; \textit{bottom}: second ellipticity component.}
    \label{fig:deltaehist}
\end{figure}

 The values of the true PSF ellipticity at each galaxy position in our test set is shown in \autoref{fig:truellscatter}. The corresponding ellipticity residuals for each PSF model, $\delta e_i^\mathrm{PSF}$, and their distribution across all positions are shown, respectively, in Figs.~\ref{fig:ellscatter} and~\ref{fig:deltaehist}, when computed on stars with SNR 35. Noticeable residuals are present for both methods, though they are of lower amplitude in the case of RCA. \autoref{fig:truellscatter} shows a strong asymmetry between the two ellipticity components, with most objects showing mostly horizontally or vertically oriented sticks. This indicates the first ellipticity component has both higher values, and much stronger variations across the field than the second (which contributes to diagonal orientations). Residuals in Figs.~\ref{fig:ellscatter} and~\ref{fig:deltaehist} are, in turn, similarly dominated by the first component. This is due to the pixel grid on which the input simulated PSFs were sampled. A simple rotation of the reference frame before sampling would reduce (or invert, through a $\pi/4$-rotation) the difference in amplitudes between the two ellipticity components.

As could already be observed in \autoref{fig:ellscatter}, Fig. 9 shows that \texttt{PSFEx} leads to a strong bias in the first ellipticity component that is systematically overestimated. This occurs at all SNRs and indicates that \texttt{PSFEx}, as it is, cannot capture the variations of the Euclid PSF model from under-sampled stars.

\begin{figure}
        \centering
        \includegraphics[width=\linewidth]{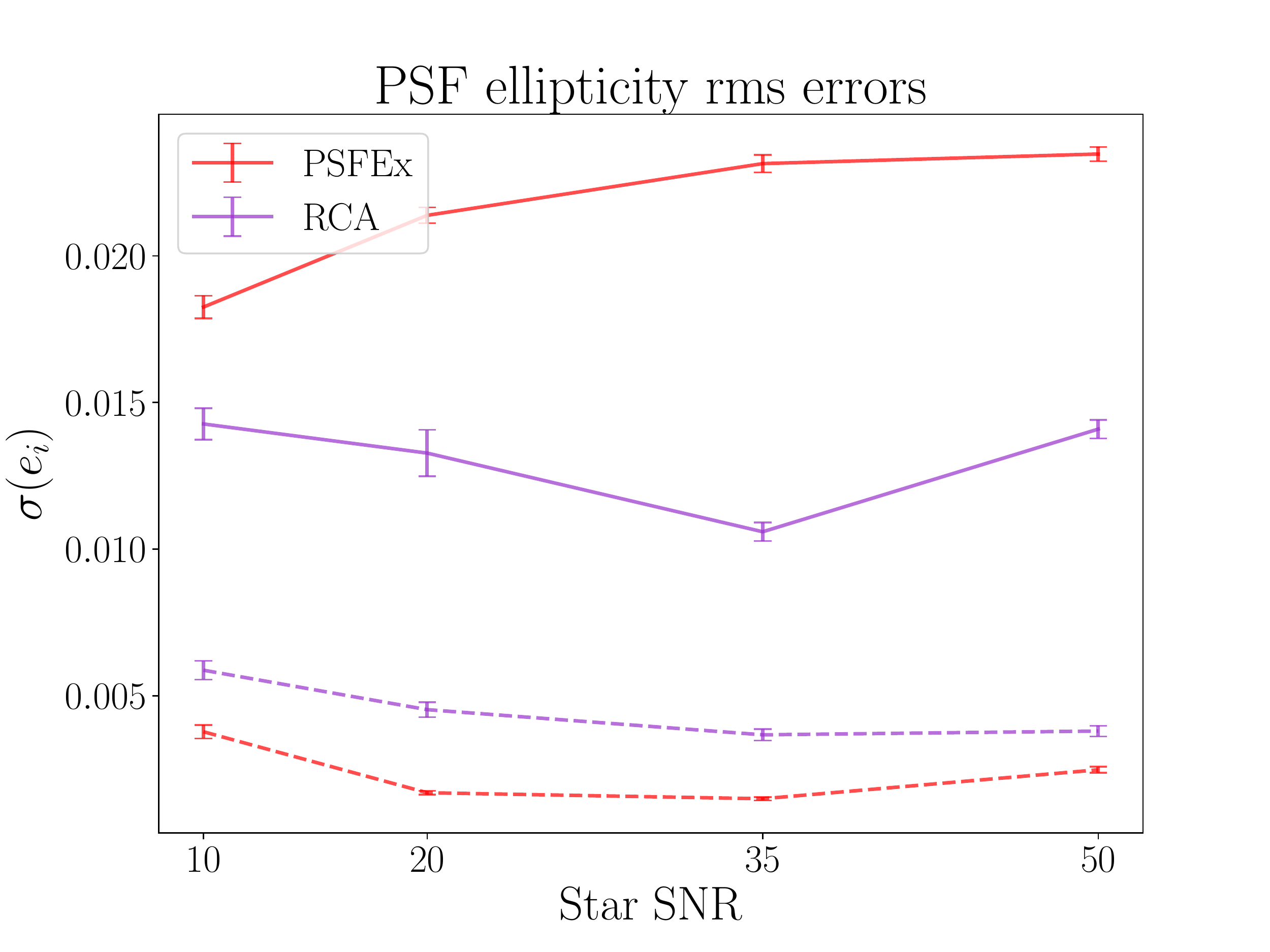}
        \caption{RMS error on each PSF ellipticity component for the two models, as a function of input star SNR. Continuous lines are for the first ellipticity component, dashed for the second. Error bars are computed by jackknife over the $297$ test positions.}    
        \label{fig:PSFrms}
\end{figure}

The RMS error per star SNR level is shown in \autoref{fig:PSFrms}. We observe the same overall behavior of both PSF models at all star SNRs, with RCA performing better at $e_1$ recovery, and worse at capturing the much smaller $e_2$ variations. As mentioned in \autoref{sec:intro}, \Euclid's requirements for weak lensing are that the RMS on both PSF ellipticity components should be lower than $5\times10^{-5}$. As expected, our purely nonparametric approach is far (at a factor of 100-300) from achieving these requirements on its own and with such few stars, though it already yields a significant improvement over \texttt{PSFEx}. 

\begin{figure}
\centering
\includegraphics[width=\linewidth]{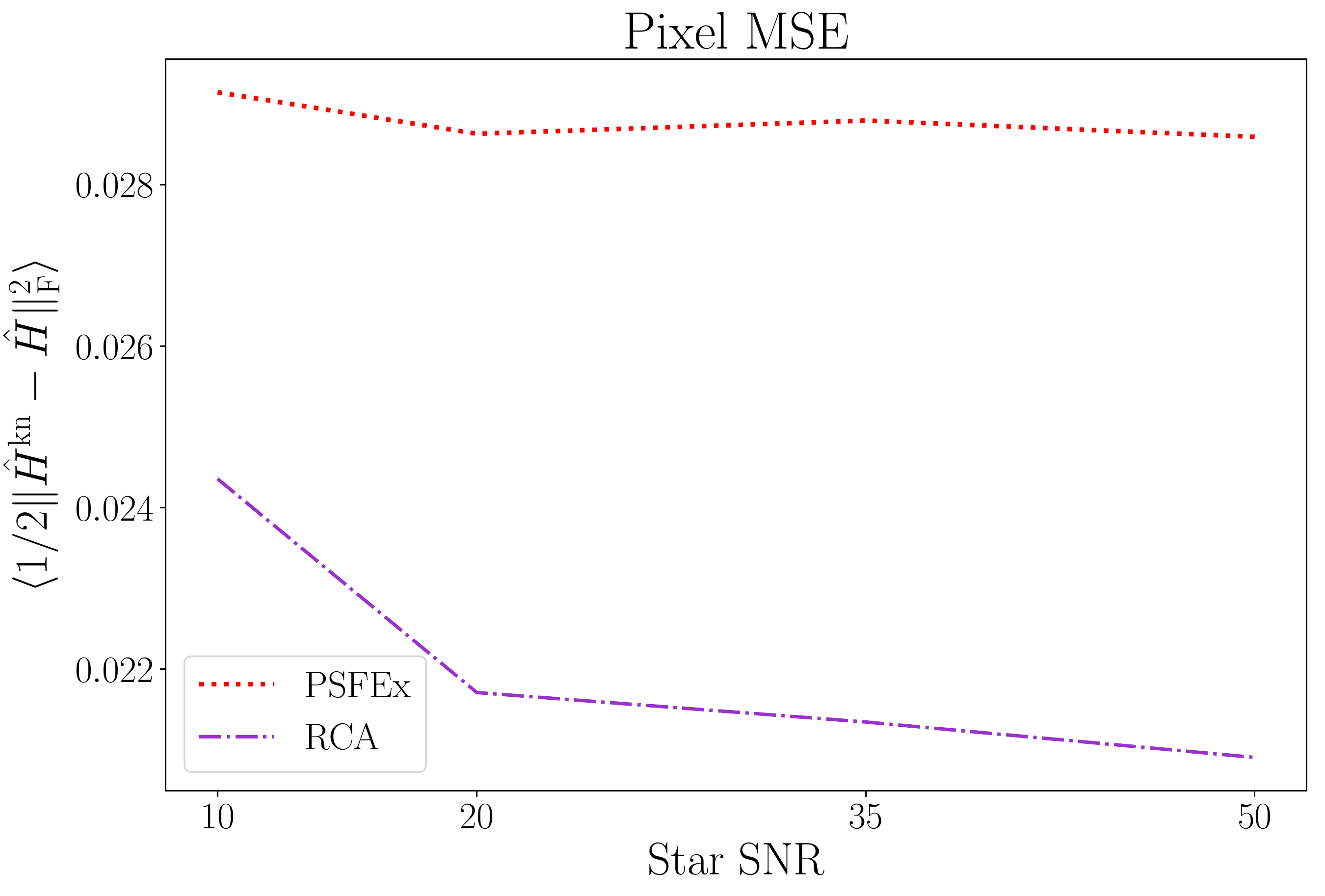}
\caption{Average pixel error as a function of star SNR.}    
\label{fig:pixelMSE}
\end{figure}

The RMS on the first ellipticity component gets increasingly worse for higher SNR values in the case of \texttt{PSFEx}, which might indicate the presence of spurious effects in the model that get attenuated by higher levels of noise. 
It might seem puzzling that the error we observe in the case of RCA is lower for an SNR of 35 than it is for one of 50. We observe the same effect when rerunning RCA on several different realizations of noise at those levels. A natural concern would be that this could indicate that the quality of our PSF model gets worse with decreasing levels of noise; however, the pixel error between our RCA PSFs and the known ones does get smaller, as shown in \autoref{fig:pixelMSE}. These effects illustrate an important point: when building the PSF model, neither RCA nor \texttt{PSFEx} explicitly aim at matching the observed stars' shapes. It is therefore possible that a ``better'' model, as defined from the actual functionals both approaches aim to minimize (in Eqs.~\ref{eq:RCApb} and~\ref{eq:expb}, respectively), leads to a poorer ellipticity component. This is what we observe in \autoref{fig:PSFrms}: the PSF model outputted by RCA, when run on given stars, varies smoothly as a function of their noise level. The overall quality of the model monotonically increases with SNR, as seen in \autoref{fig:pixelMSE}, eventually converging to the model that would be obtained if there were no noise in the input stars. The ellipticity of the model at any arbitrary position also varies smoothly, but there is no guarantee these variations monotonically tend to the true ellipticity. While the effect we observe here is much smaller (and is, in fact, not visually identifiable when comparing the models obtained at SNRs of 35 and 50), as a crude illustration, consider a PSF with two outer rings: the first one having a dampening effect on the full PSF's first component ellipticity $\delta(e_1^\mathrm{inner})<0$, and the second leading to an increase $\delta(e_1^\mathrm{outer})=-\delta(e_1^\mathrm{inner})>0$. For a given number of stars, suppose the best possible error achievable (with no noise) is $\delta(e_1^*) > 0$. Let us assume the quality of the reconstruction of the central part of the PSF is unchanged regardless of input noise levels, and both rings are completely lost to noise at low SNR. As we increase the SNR of input stars, the model would eventually capture the inner ring, while still completely missing the outer one. At this stage, the dampening effect of the first ring would counteract the overestimation of the central part's ellipticity, thus leading to a smaller ellipticity error $0 < \delta(e_1^*) + \delta(e_1^\mathrm{inner}) < \delta(e_1^*)$. If the SNR were to keep increasing, however, the outer ring would eventually be captured by the model, increasing once again the overestimation of the first component ellipticity.

\section{Impact on galaxy shape measurement}\label{sec:gaals}
Equation \eqref{eq:paulin} is exact in the case of unweighted moments. However, the Euclid PSF has divergent second-order moments and a complex profile that leads to strong ellipticity gradients, which further complicates the use of weight functions, as shown by~\cite{hoekstra1998}. Their necessary addition introduces mixing with higher-order moments, as seen, for example, in the DEIMOS~\citep{melchior2011} formalism. \cite{massey2012} extended the study of~\cite{paulin2008} to include these new terms, which result in prefactors:

\begin{align}\label{eq:prefpaulin}
\hat{e}_i = e_i\left(1 + \frac{\delta(R^2_\mathrm{PSF})}{P_RR^2_\mathrm{gal}}\right) - \frac{1}{P_RP_e}\left(\frac{R^2_\mathrm{PSF}}{R^2_\mathrm{gal}} \delta e_i^\mathrm{PSF} + \frac{\delta(R^2_\mathrm{PSF})}{R^2_\mathrm{gal}}e_i^\mathrm{PSF}\right)\;,
\end{align}
where we only introduce the terms that have a direct impact on the contribution of the PSF modeling errors~\citep[][also include those due to non-convolutional detector effects, and those introduced by the shape measurement process]{massey2012}. If the PSF is Gaussian, $P_R$ and $P_e$ are exactly equal to 1. If this holds, or is a good approximation (e.g., for ground-based PSF), \autoref{eq:prefpaulin} reverts to the~\cite{paulin2008} case, i.e. our \autoref{eq:paulin}, and the PSF modeling errors can thus be considered separately from the shape measurement applied. To test whether this remains true in our present case of an \Euclid-like PSF, in this section, we perform image simulations and galaxy shape measurement using each PSF model. In particular, in \autoref{sec:shapemeasurement}, we apply both a moments-based shape measurement method, and one based on model fitting. While each comes with its own method-dependent biases, we would expect the contribution of the PSF modeling errors to be the same in both cases if the assumption that the prefactors in \autoref{eq:prefpaulin} vanish held true.

\subsection{Galaxies and observations}\label{sec:gals}
We performed galaxy-image simulations using the freely available GalSim software\footnote{\url{https://github.com/GalSim-developers/GalSim}}~\citep{rowe2015}. The galaxy parameters are identical to those used in several branches of the GREAT3 challenge~\citep{mandelbaum2014}, themselves based on fitting the COSMOS population. This gives us a population of $2\,040\,000$ galaxies that are either drawn from a single Sersic profile, or composed of both a bulge (following a de Vaucouleurs profile) and a disk (with an exponential profile). We applied 204 different random shear values, each of them to a set of $10\,000$ different galaxies. These sets include the 90-degree rotated counterpart to each galaxy, so as to ensure intrinsic ellipticity truly averages at zero.

The main difference between our image simulations and those used in GREAT3 is, naturally, the PSF used. For our study, we randomly assigned one of the 297 Euclid PSFs (at galaxy positions) to each of the galaxies, imported them in GalSim and performed the convolution with the galaxy profile.

Our observations were then generated by sampling the resulting convolved profile on stamps of $42\times 42$ pixels at half the nominal VIS pixel scale, to match our super-resolved PSFs. We note that in a real-life \Euclid setting, the observed galaxies would also suffer from under-sampling; however, we choose not to take it into account in this work in order to better isolate the effects of imperfect PSF modeling on shape measurement. Similarly, rather than matching the observations' noise levels to those we used for the stars, we instead always added white Gaussian noise with $\sigma=0.01$ (leading to an average SNR of about 50).

\subsection{Shape measurement}\label{sec:shapemeasurement}
With both the estimated PSFs and observed galaxies described in the previous sections, we can now perform the actual shape measurement step. For a given galaxy of intrinsic ellipticity $e^\mathrm{int} = (e^\mathrm{int}_1,e^\mathrm{int}_2)$ and having undergone a shear $(g_1,g_2)$, our shape measurement method yields

\begin{align}\label{eq:measured_ell}
    \hat{e}_i \approx e^\mathrm{int}_i + g_i\;.
\end{align}
The shear itself can then be obtained by averaging over sets of objects:

\begin{align}\label{eq:shear_avg}
    \hat{g} = \langle \hat{e} \rangle &\approx \langle e^\mathrm{int} \rangle + \langle g \rangle = g\;.
\end{align}
In our case, we know $\langle e^\mathrm{int} \rangle$ is exactly 0. Numerous shape measurement methods that yield $\hat{e}$ (and thus $\hat{g}$) exist. However, they are known to be imperfect and introduce bias. Since we are interested in the impact of imperfect PSF models, in order to quantify the amount of error that is introduced by the shape measurement itself, we started by measuring the shape of each observed galaxy using the corresponding known PSF $\hat{H}^\mathrm{kn}$. Then, for each of our star noise levels, we repeated the measurement of the same object, both with the RCA-estimated $\hat{H}^\mathrm{RCA}$ and the \texttt{PSFEx} $\hat{H}^\mathrm{PSFEx}$.

Broadly speaking, shape measurement methods used in weak lensing studies fall into one of two categories: moments-based approaches and model fitting. The former rely on computing estimates of the shape of the object from their second-order quadrupole moments, and PSF correction is typically carried out by also computing the PSF model image's moments and correcting for these. On the other hand, model fitting methods assume some analytical model for the profile of the galaxies. The parameters of the model are then selected on a per-object basis, by fitting the observations with a sampled profile, convolved with the PSF (hence the process sometimes being referred to as ``forward" fitting).

Because of the considerable difference between those two approaches, especially with regard to how the PSF is taken into account, we perform our experiments with one method of each type. The most well-known moments-based approach is the KSB method, first introduced by~\cite{kaiser1995}. Various improvements and implementations of the KSB method have since then been proposed. In the present work, we use its implementation within the HSM~\citep{hirata2003,mandelbaum2005} library of GalSim, where the size of the (circular) Gaussian window function is matched to that of the observed galaxy. For model fitting, we use the freely available \texttt{im3shape} package\footnote{\url{https://bitbucket.org/joezuntz/im3shape-git}} described in~\cite{zuntz2013}. In the results shown in \autoref{sec:results}, \texttt{im3shape} was run with most parameters left to default, except for those related to the images stamp size, noise level, ranges for the estimation of the object's centroid and PSF handling (see \autoref{appdx:addfigs} for a complete list). 

A particular consequence of this is that the model chosen for galaxies is a de Vaucouleurs bulge combined with an exponential disk, which, in turn, is the exact model used for generating some of our observations (though some others are composed of a single Sersic profile with index $n\notin \{1,4\}$). However, \texttt{im3shape} thus configured assumes the bulge and disk to have the exact same ellipticity, orientation and relative size, which is not necessarily the case for our simulated galaxies. Nonetheless, this means the actual galaxy profiles used by \texttt{im3shape} are fairly close to those of the observations, perhaps more so than what could be expected from real data. In other words, our model fitting experiments may not suffer from so-called model bias quite as much as could be expected in a more realistic setting~\citep{voigt2010}. However, our emphasis in the present work is on the effect of PSF modeling errors on galaxy shape measurement, and whether both approaches are similarly affected by them. For a study of the impact of model bias on shape measurement, see~\cite{pujol2017}.

In some cases, the KSB implementation we used fails to compute the shapes of certain objects, or returns ellipticity estimates with an absolute value of more than 1. When this occurs with any of our three PSFs, we remove these objects from the analysis. This leads to about 72\,000 objects being put aside. The exact amount of objects removed per SNR and PSF type are given in \autoref{appdx:addfigs}. We note that the model fitting approach always provides an estimate of the shape, and thus all 2\,040\,000 objects are used.

\subsection{Results}\label{sec:results}
We first describe the results of our shape measurement experiment by themselves. Then, in \autoref{sec:paulin}, we compare them to the widely used analytical prediction of \cite{paulin2008}.

\subsubsection{Ellipticity measurements}\label{sec:galshapes}
We first consider the measured shape of galaxies themselves. Regardless of the PSF model and shape measurement method applied, we obtain an estimate of the overall galaxy shape, which includes both its intrinsic ellipticity and the undergone shear as shown in \autoref{eq:measured_ell}. Despite their differences, both approaches suffer from some form of model bias. 

Quadrupole moments are extremely sensitive to noise effects. To overcome this sensitivity, KSB uses a (matched, in our case) Gaussian window function. This of course induces bias, which has been compared to the effect of model bias in the case of model fitting methods, for instance by~\cite{viola2014}. As discussed at the end of \autoref{sec:shapemeasurement}, in our current setup, \texttt{im3shape} measurements are on the contrary fairly exempt from model bias. Regardless, these potentially strong biases are due solely to the shape measurement methods themselves, and should be independent from the PSF modeling. Since the impact of the latter is our focus here, we study the \textit{relative} ellipticity error of the various combinations of PSF models and shape measurements, that is,

\begin{align}
        \langle(\hat e_i^\mathrm{kn} - \hat e_i^\mathrm{RCA})^2\rangle,\langle(\hat e_i^\mathrm{kn} - \hat e_i^\mathrm{PSFEx})^2\rangle\;,
\end{align}
where the average is taken over all objects. We note that the overall amplitude of errors is still related to the intrinsic biases of each shape measurement method, which could be alleviated by a proper calibration scheme. However, these results can still be used to inform us about the two PSF models and their impact on galaxy shape estimation.

When using KSB, we find a clear improvement of order 50-60\% in the shape error when using the proposed approach over \texttt{PSFEx}. Similarly to results shown on the PSF models themselves in \autoref{sec:psfcomp}, this seems to indicate that both models yield significantly different PSFs, and that our RCA-based approach is more successful at reconstructing the true PSF. 

Interestingly, however, the observed difference is much smaller (of order 10-20\%) between the two PSF models when shapes are measured through model fitting. This would seem to imply that these methods are less sensitive to PSF modeling errors than moments-based methods, which was not an especially expected outcome: for instance,~\cite{pujol2017} found no significant difference in sensitivity on various other potential factors when comparing methods of each type. This difference in behaviour when faced with imperfect PSF models could be related to effects due to mixing with higher order moments discussed at the beginning of \autoref{sec:gaals}. This is further studied in \autoref{sec:paulin}.

The results are shown, for all SNR levels, PSF models and shape measurement approaches, in \autoref{tab:relative_errors} of the appendix.

\subsubsection{Shear bias}\label{sec:shrbias}
A second way to study the impact of PSF models is to look at the actual inferred shear itself. In our case, since we know the intrinsic galaxy ellipticities average to zero (and we can correct for it if not, for example if some objects were tagged as outliers and removed from the analysis), we only have to average across a set of 10\,000 objects with the same applied shear to obtain our shear estimator, as shown in \autoref{eq:shear_avg}.

A common way to parameterize the bias made on shear measurement is to extend it to first order:

\begin{align}\label{eq:shr_bias}
        \hat g_i \approx (1+m_i)g_i + c_i\;,
\end{align}
where $m_i,c_i$ are the multiplicative and additive shear bias, respectively, for shear component $i\in\{1,2\}$. As in the previous section, we compute the value of those two parameters for each combination of PSF and shape measurement technique. Once again, we emphasize that the goal of the present work is not to compare shape measurement methods per se, but rather how PSF model errors impact them. The focus should thus be on the differences of $c$ and $m$ values between different PSF models, rather than on the actual values themselves.

\begin{figure}
        \centering
                \includegraphics[width=\linewidth]{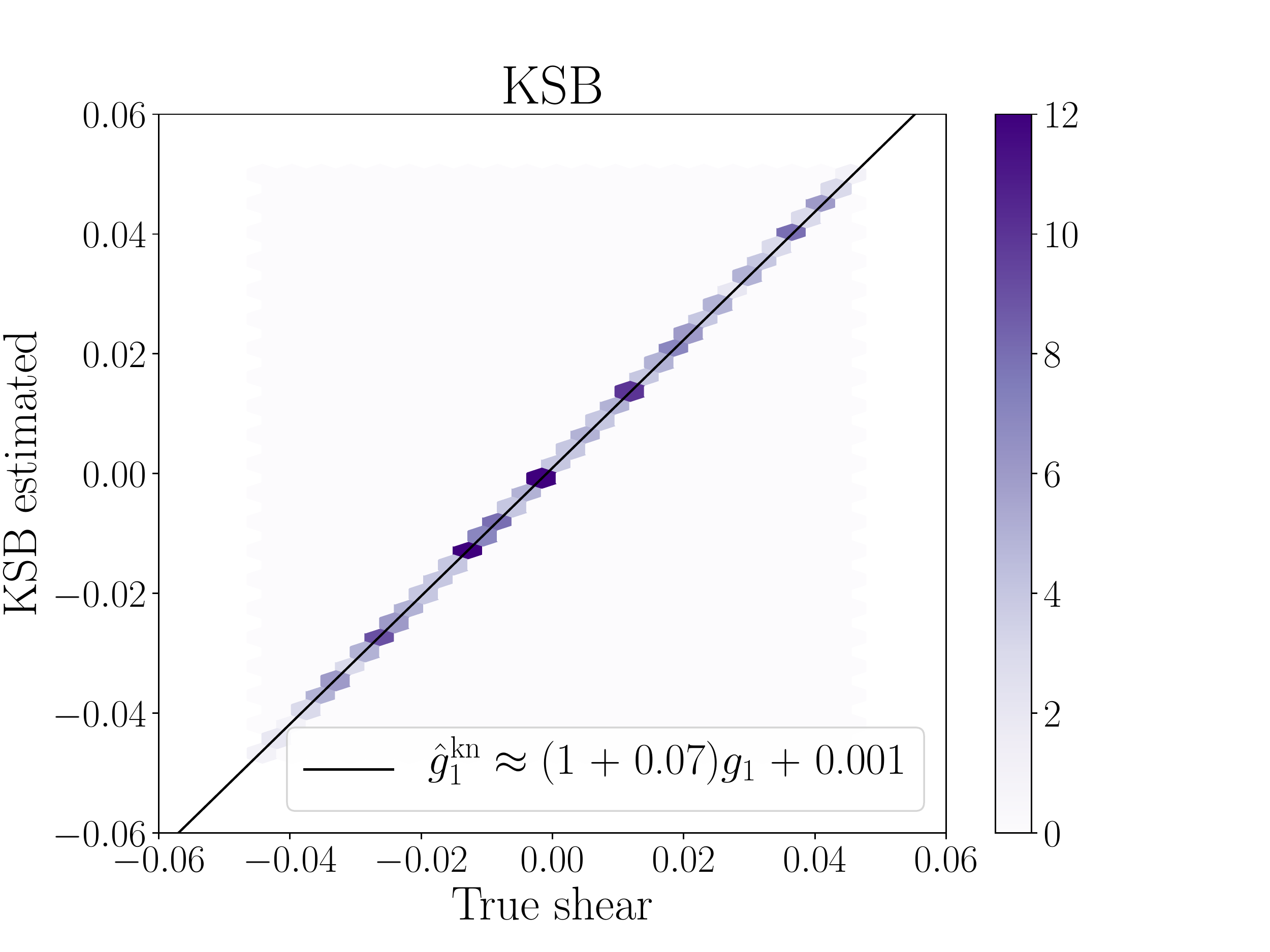}
                \caption{2D density of true and measured shear; the colors correspond to the number of occurrences of measured shear values when using the known PSF, each from approximatively 10,000 galaxies, for the corresponding input shear. The line shows the best-fit linear regression, yielding the bias values. Shapes were measured with KSB.}
        \label{fig:g1kn_ksb}
\end{figure}
        
\begin{figure}
        \begin{subfigure}{\linewidth}
                \includegraphics[width=\linewidth]{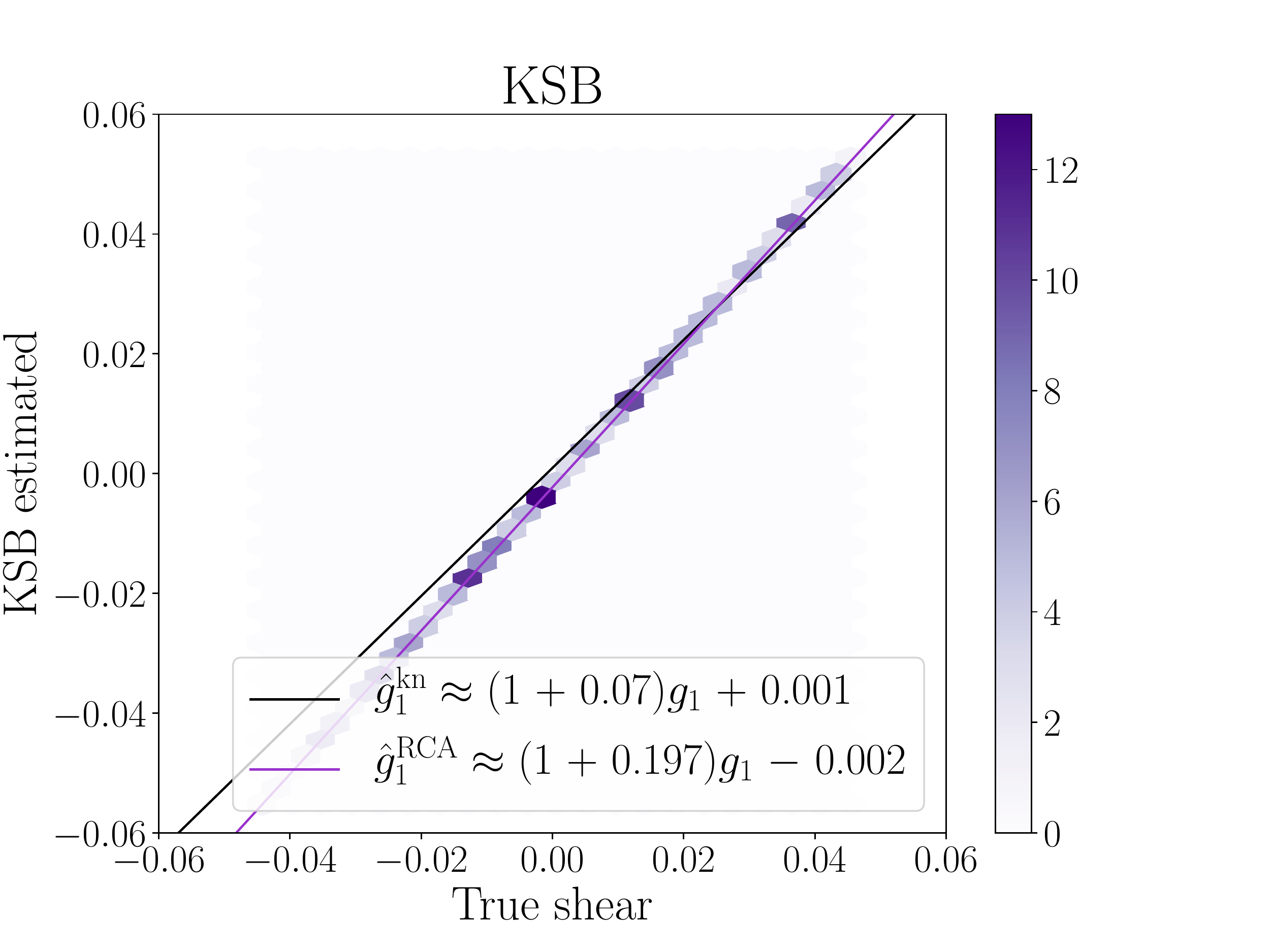}
        \end{subfigure}\\
        \begin{subfigure}{\linewidth}
                \includegraphics[width=\linewidth]{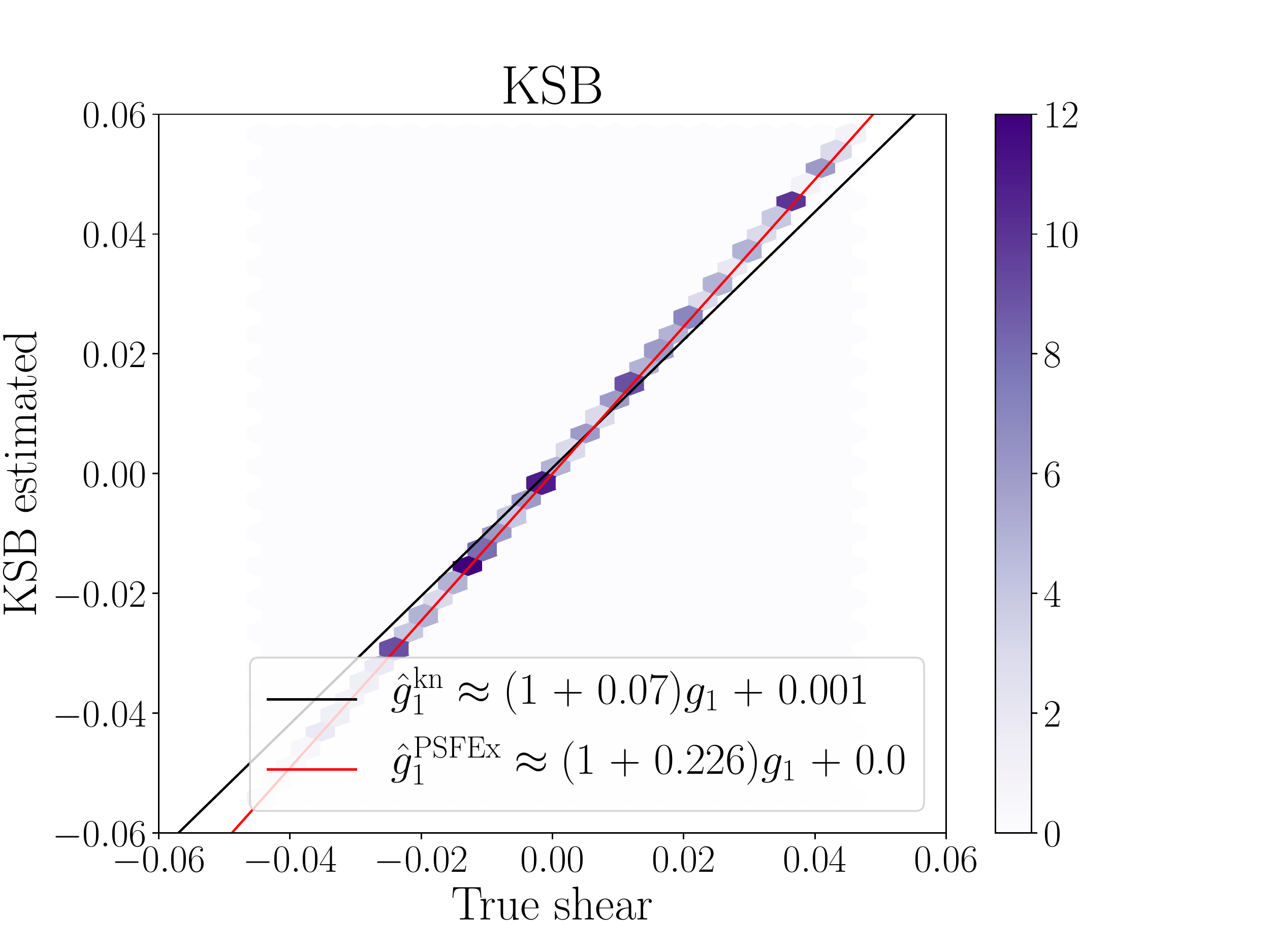}
        \end{subfigure}
        \caption{Similar to \autoref{fig:g1kn_ksb}, 2D density of true and measured shear, using PSF models for the latter. The line corresponding to first-order shear bias is shown for both the PSF models (in color) and the best case scenario (in black). Shape measurement is performed using KSB.}
    \label{fig:g1mod_ksb}
\end{figure}

In the case of KSB, the distributions of the first component of measured and true shears, as well as the best fit linear regression yielding the two bias values, are shown in \autoref{fig:g1kn_ksb} for the known PSF. The same figures for both our models, with star SNR of 35, are shown in \autoref{fig:g1mod_ksb}, with the line corresponding to shear bias in the ideal case shown in black for comparison.

PSF modeling induces a stronger shear bias in both cases, with over a factor of two gain in multiplicative bias compared to the ideal scenario. The RCA-based PSF leads to an improvement, in both components, of the multiplicative bias compared to \texttt{PSFEx}, as seen in \autoref{tab:biasvSNR} (where, similarly to the ellipticities in \autoref{sec:galshapes}, we show the relative biases $\Delta m_i, \Delta c_i$ after subtracting that measured using the known PSF). 

\begin{table*}
    \centering
    \begin{tabular}{c|l|cc|cc|cc|cc}
        \multicolumn{2}{c|}{SNR} & \multicolumn{2}{c|}{10} & \multicolumn{2}{c|}{20} & \multicolumn{2}{c|}{35} & \multicolumn{2}{c}{50}\\
        \multicolumn{2}{c|}{Ellipticity component}&  $1$st & $2$nd & $1$st & $2$nd & $1$st & $2$nd & $1$st& $2$nd \\
        \hline
        \multirow{2}{*}{$\Delta m_i$} & RCA & 0.14 & 0.12 & 0.13 & 0.11 & 0.13 & 0.12 & 0.13 & 0.12\\
        & \texttt{PSFEx} & 0.16 & 0.14 & 0.16 & 0.15 & 0.16 & 0.14 & 0.16 & 0.14\\
        \hline  
        \multirow{2}{*}{$\Delta c_i \times 10^3$} & RCA & -4.70 & 0.48 & -6.37 & 0.24 & -3.23 & 0.89 & -3.30 & 1.47\\
        & \texttt{PSFEx} & -2.88 & 1.22 & -3.05 & -0.19 & -0.96 & 0.13 & -1.10 & 0.31
    \end{tabular}
    \caption{KSB-induced shear biases $\Delta m_i, \Delta c_i$ as a function of the SNR of stars on which the PSF models were fitted.}
    \label{tab:biasvSNR}
\end{table*}
Conversely, the lower half of \autoref{tab:biasvSNR} indicates our RCA-based PSFs lead to a higher additive bias than the \texttt{PSFEx} ones. This additive bias is present at every star SNR, though it undergoes strong variations. This higher RCA additive bias is especially noticeable on the first of the two shear components, despite both the bias and RMS error on the first component PSF ellipticity being smaller, as is shown in \autoref{sec:psfcomp}. A common way to investigate the relationship between PSF and additive bias is to reparameterize \autoref{eq:shr_bias} like so:

\begin{align}
        \hat g_i \approx (1+m_i)g_i + c'_i + \alpha e_i^\mathrm{PSF}\;,
\end{align}
where $\alpha$ then quantifies the amount of PSF leakage. We note, however, that this quantity can contain both PSF effects that were not fully captured by the shape measurement step, and effects emanating from errors in the PSF model itself. It would therefore not be informative in our present case, where the additive bias appears stronger for the PSF model with the smallest errors despite the same shape measurement being applied in both cases.

\begin{figure}
        \begin{subfigure}{\linewidth}
                \includegraphics[width=\linewidth]{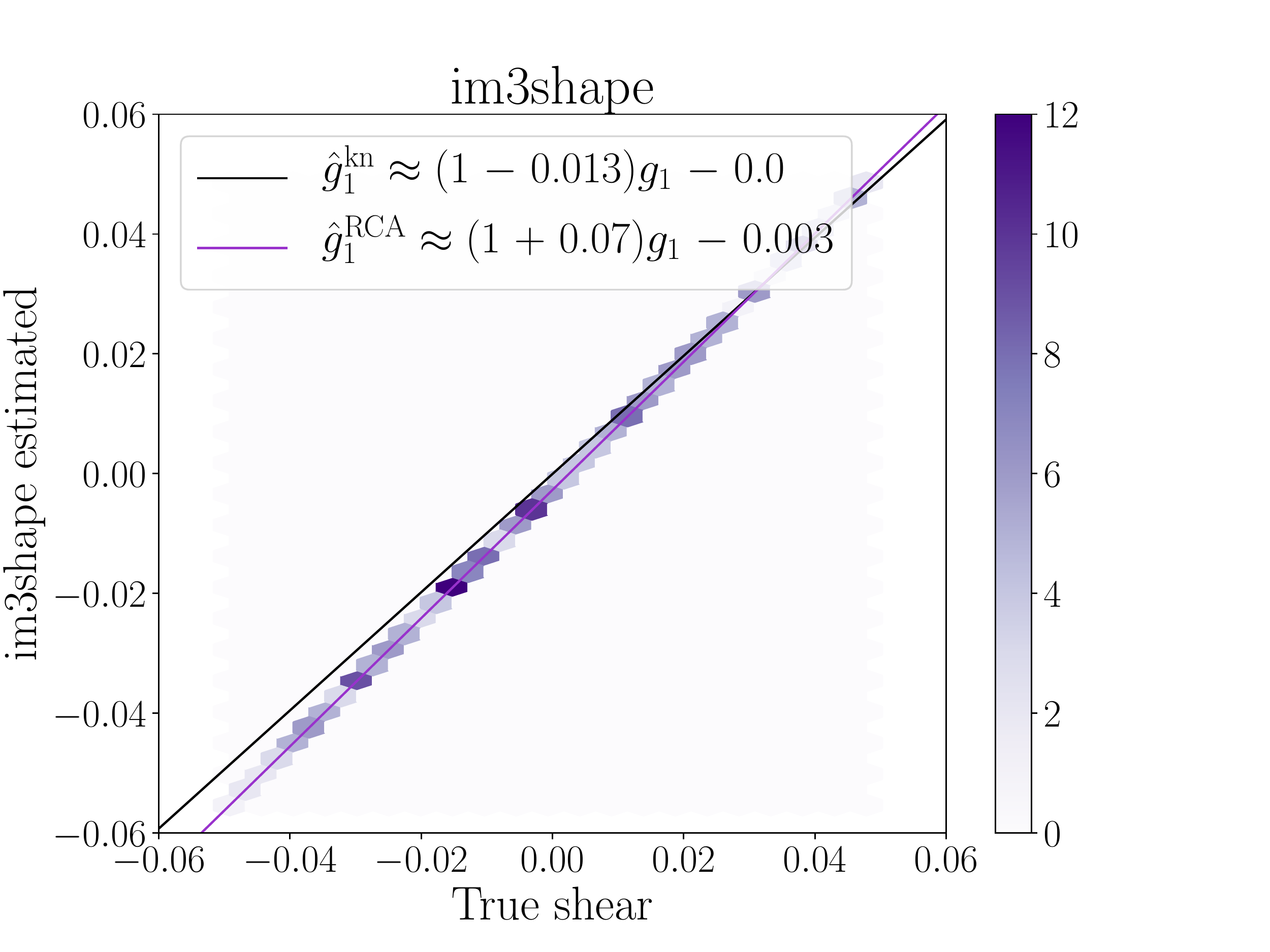}
        \end{subfigure}\\
        \begin{subfigure}{\linewidth}
                \includegraphics[width=\linewidth]{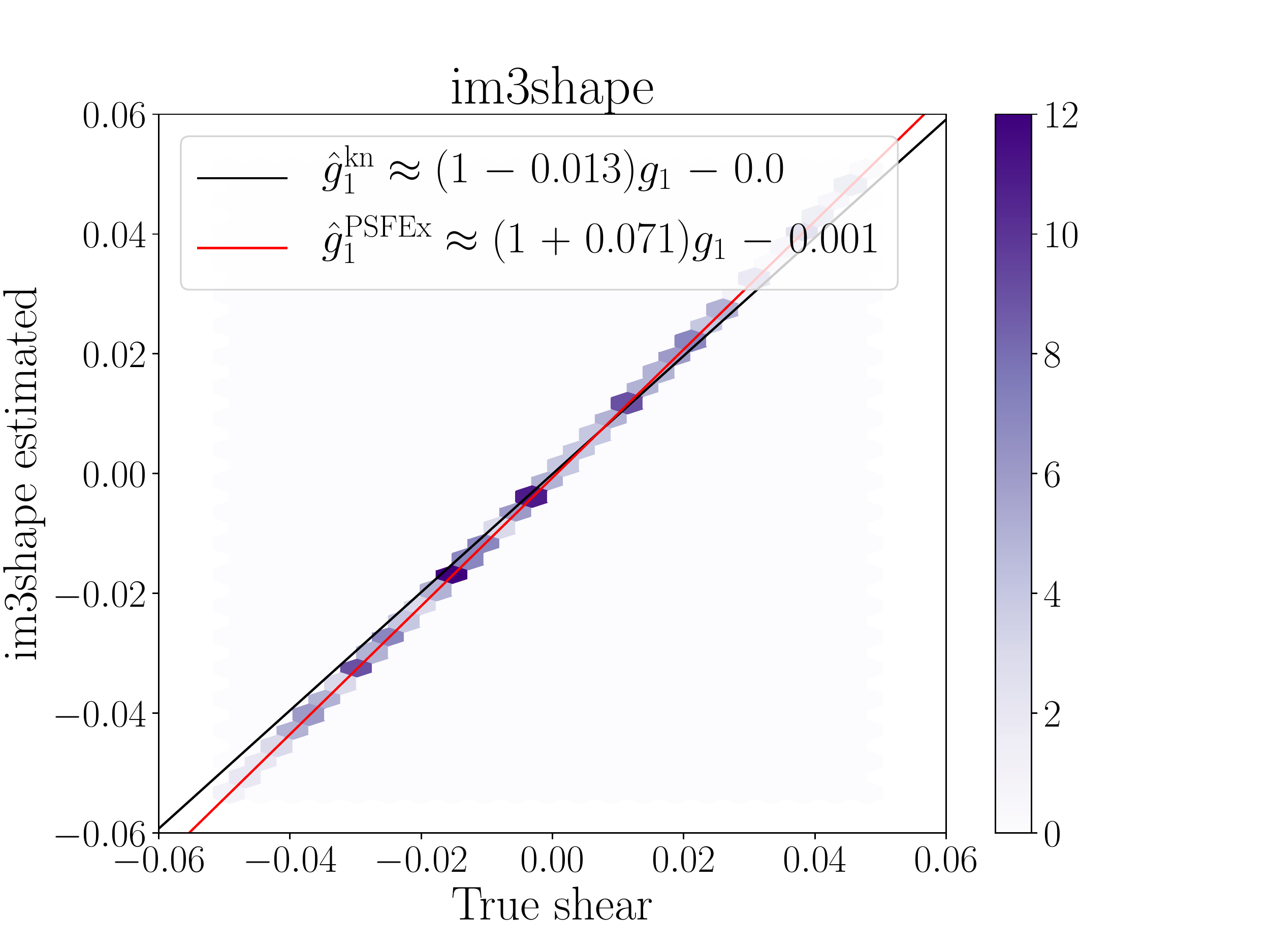}
        \end{subfigure}
        \caption{Same as \autoref{fig:g1mod_ksb}, when shape measurement is performed using model fitting.}
    \label{fig:g1mod_im3}
\end{figure}

A study of the shear biases obtained with our different PSF models when using \texttt{im3shape} also seems to indicate the presence of a slight additive bias when using the RCA PSF. This is illustrated in \autoref{fig:g1mod_im3}, which features the same shear 2D densities and linear fit as \autoref{fig:g1mod_ksb}, also at star SNR 35, when the shape measurement is performed by model fitting. In terms of multiplicative bias, the difference between the known and modeled PSFs is much smaller than it was with KSB, and insignificant in-between models, which once again seems to indicate a lower sensitivity to PSF modeling errors of model fitting methods.

\subsubsection{Comparison to analytical predictions}\label{sec:paulin}
\begin{figure}
    \centering
    \includegraphics[width=\linewidth]{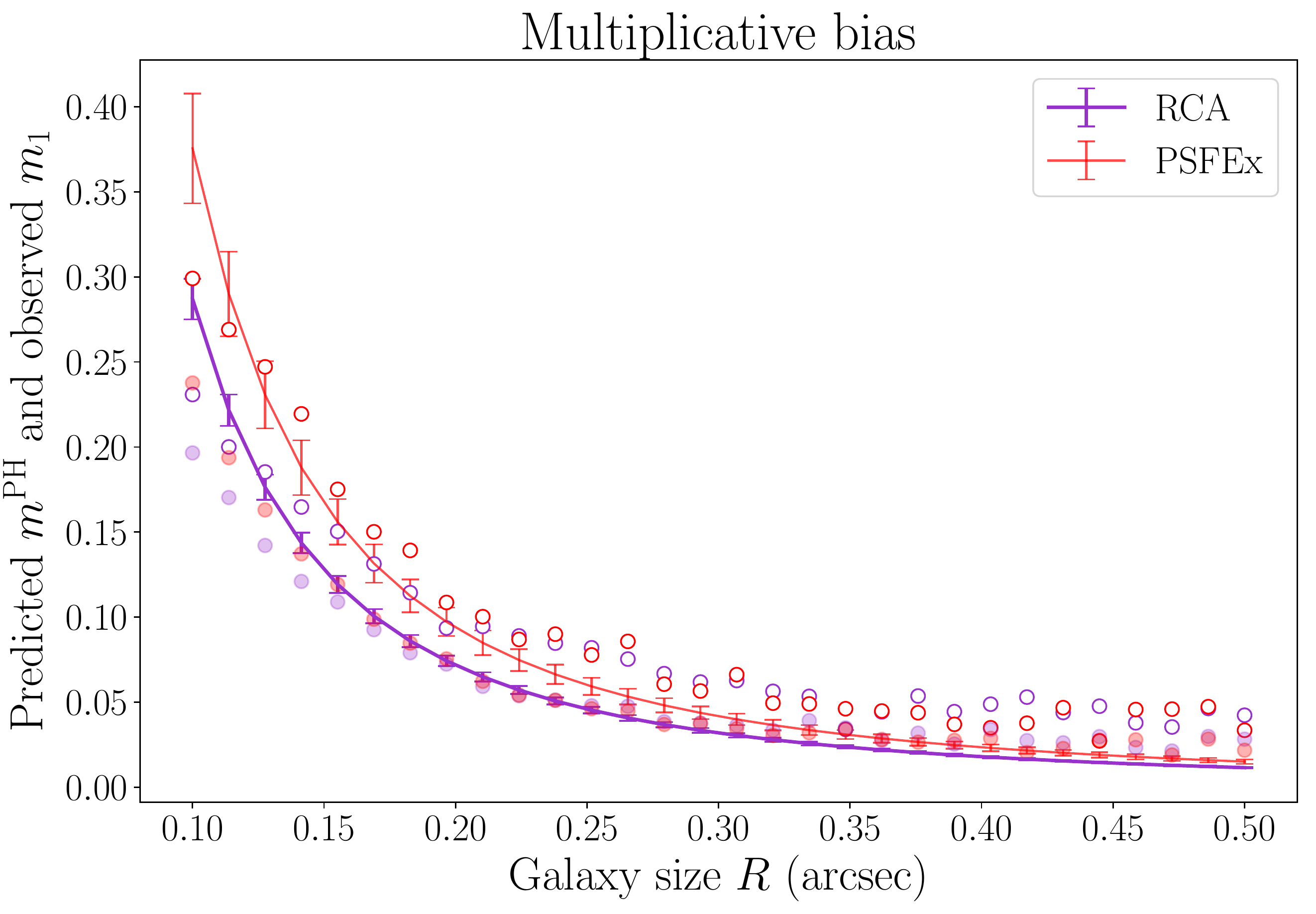}
    \caption{Multiplicative bias induced by the PSF models, as predicted from \autoref{eq:phm} (continuous line and error bars) and observed when measuring galaxy shapes with KSB (empty points) or \texttt{im3shape} (filled points).}
    \label{fig:mpaulin}
\end{figure}

\begin{figure}
    \begin{subfigure}{\linewidth}
        \includegraphics[width=\linewidth]{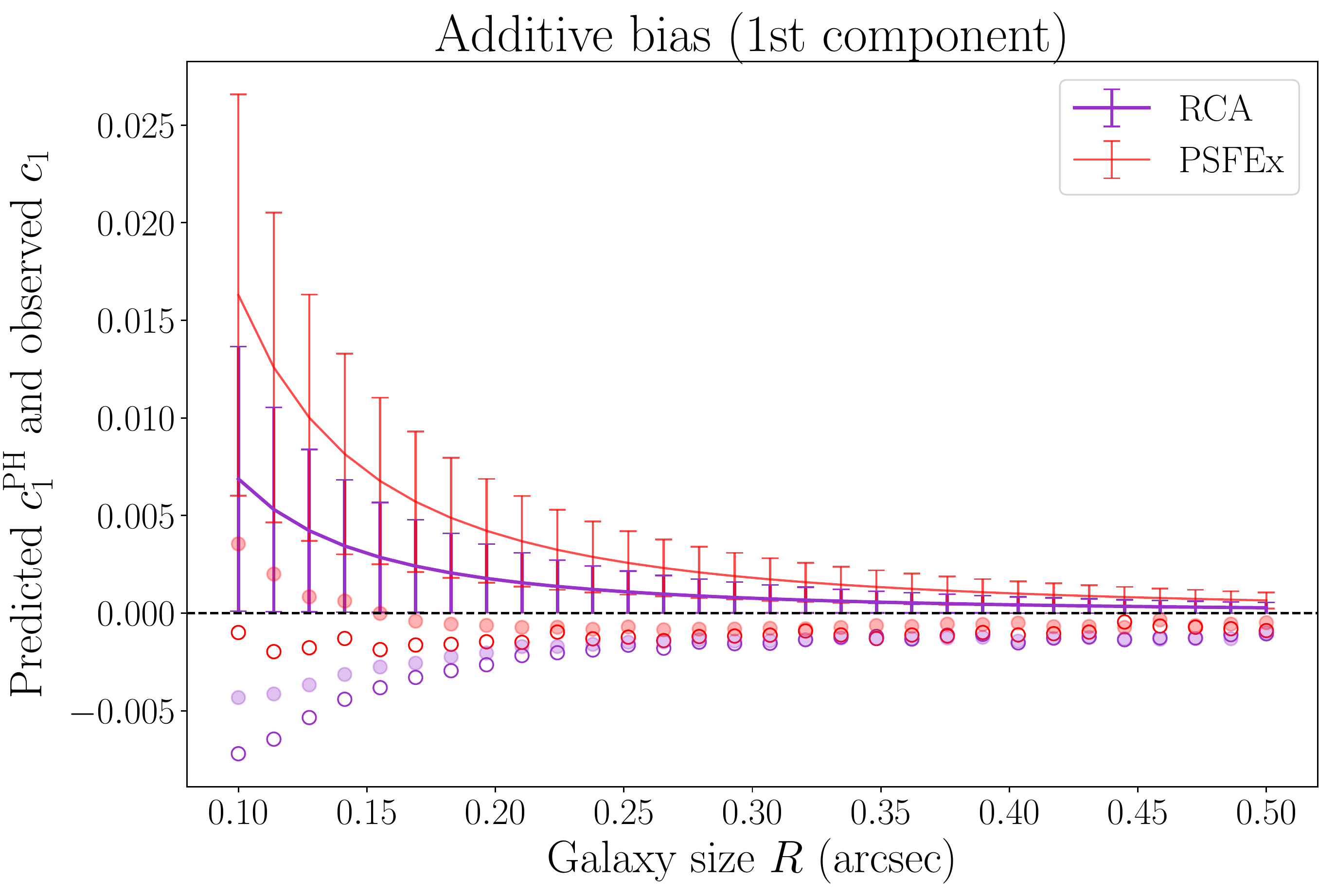}
    \end{subfigure}\\
    \begin{subfigure}{\linewidth}
        \includegraphics[width=\linewidth]{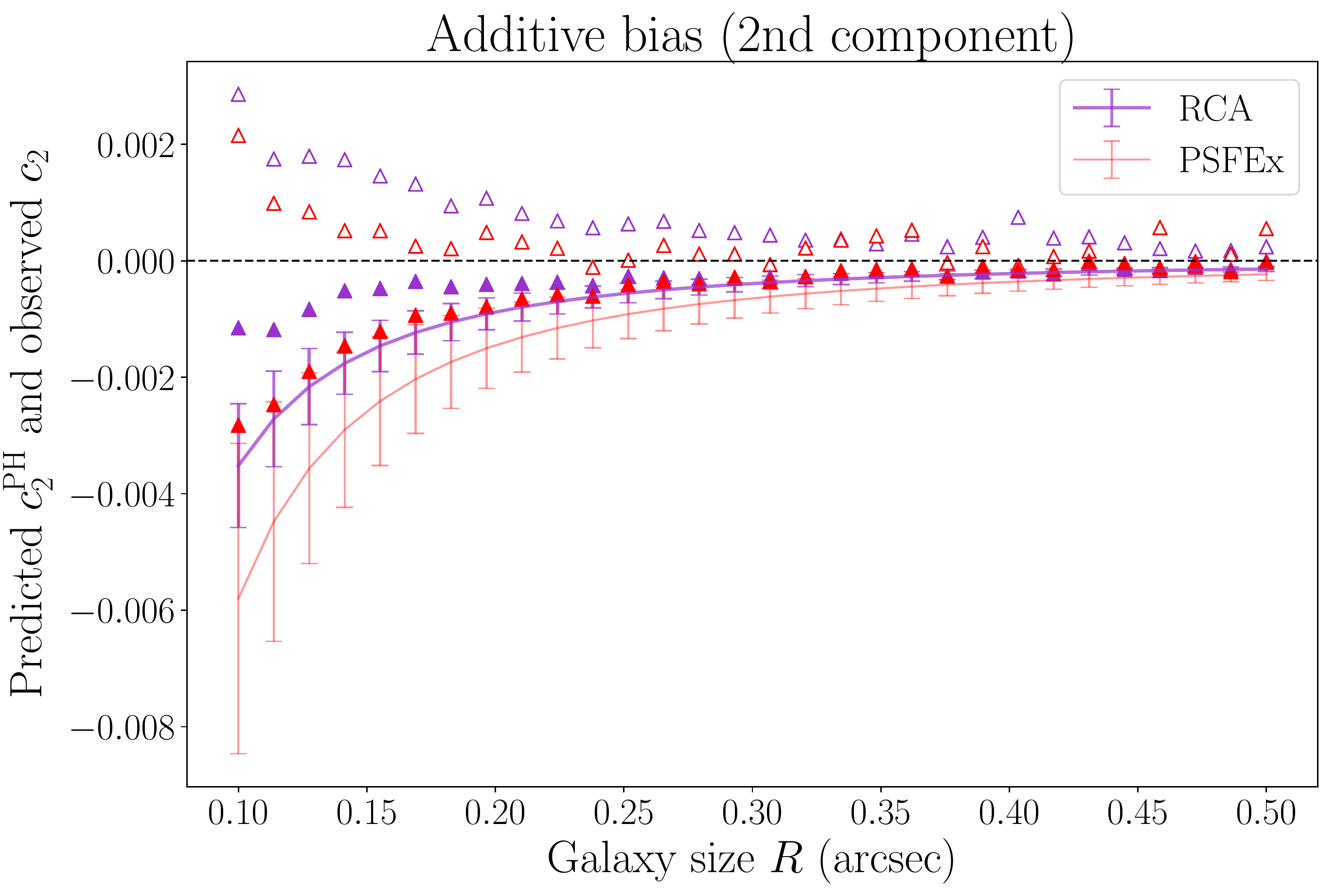}
    \end{subfigure}\\
    \caption{Similar to \autoref{fig:mpaulin}, for the additive biases predicted from \autoref{eq:phc} (continuous line and error bars), and those observed with KSB (empty points) and \texttt{im3shape} (filled points). Note that in this case, the analytical predictions are different for each ellipticity component (\textit{top}: first; \textit{bottom}: second) because of the left-hand term in \autoref{eq:phc}.}
    \label{fig:cpaulin}
\end{figure}

The results shown in \autoref{sec:shrbias} are already at odds with those predicted by \autoref{eq:paulin}, since we observe different relative biases introduced by the same PSF model errors depending on the shape measurement used. To better illustrate this discrepancy, following from \autoref{eq:paulin}, for any set of galaxy, true PSF, and PSF model, we define the expected contribution to multiplicative bias, 

\begin{align}\label{eq:phm}
m^\mathrm{PH} \eqdef \frac{\delta(R^2_\mathrm{PSF})}{R^2_\mathrm{gal}}\;,
\end{align} 
which is the same for both ellipticity components, and the contribution to each component of the additive bias
\begin{align}\label{eq:phc}
c_i^\mathrm{PH} \eqdef -\left(\frac{R^2_\mathrm{PSF}}{R^2_\mathrm{gal}} \delta e_i^\mathrm{PSF} + \frac{\delta(R^2_\mathrm{PSF})}{R^2_\mathrm{gal}}e_i^\mathrm{PSF}\right)\;.
\end{align}
These are shown in Figs.~\ref{fig:mpaulin} and~\ref{fig:cpaulin}, respectively, for a range of galaxy sizes. Here and throughout this section, we use PSFs modeled at star SNR 35. The error bars correspond to the variations across our 297 estimated PSFs. 
Starting from our full experiment, we separated the pure-Sersic galaxies, split them by size, and recomputed the shear biases we observed per galaxy size bin. Similar to \autoref{sec:shrbias}, we then computed the relative biases, $\Delta m, \Delta c_i$, by removing the bias measured with the known PSFs to those of both PSF models. These values are then over plotted for each galaxy size bin in Figs.~\ref{fig:mpaulin} and~\ref{fig:cpaulin}, and show strong deviations from the analytical predictions. 

For instance, we showed in \autoref{sec:psfcomp} that the RCA model led to smaller errors in both the first PSF ellipticity component, $\delta e_1^\mathrm{PSF}$, and its size, $\delta\left(R^2_\mathrm{PSF}\right)$. It follows that we expect a lower relative (positive) additive bias when using the RCA PSFs, which is the opposite of what we observe with our full experiment. The worse performance in $e_2^\mathrm{PSF}$ recovery is compensated by RCA's smaller $\delta(R^2_\mathrm{PSF})$, which, as previously mentioned, largely outweighs the contribution of the ellipticity error term here. This leads to smaller $c_2^\mathrm{PH}$ values when compared with the prediction for \texttt{PSFEx}. However, we see that the biases we observe in practice are strongly dependent on the shape measurement method. With \texttt{im3shape}, the $c_2$ contribution is indeed smaller for RCA, though they were overestimated by the analytical prediction for both PSF models. With KSB, it is higher for RCA than it is for \texttt{PSFEx}, and while $c_2^\mathrm{PH}<0$ for both PSF models, they lead to a positive contribution when propagated to KSB-measured shapes.

As discussed at the beginning of \autoref{sec:gaals}, we know the analytical predictions are exact when the prefactors in \autoref{eq:prefpaulin} vanish. In order to test whether these are the reason for the differences we observed, we generated a new set of simulations. The galaxies have the same properties (size, shape, applied shear) as those described in \autoref{sec:gals}, but are drawn from a 2D Gaussian distribution. Similarly, the PSF applied have identical shape properties as our Euclid PSFs, but are also Gaussian. Lastly, we recreated a set of ``RCA'' and a set of ``\texttt{PSFEx}'' PSFs, Gaussian as well, but with the same shape errors $\delta e_i^\mathrm{PSF}, \delta\left(R^2_\mathrm{PSF}\right)$ as those measured on our actual star-fit models. While we have access to the true galaxy sizes from the GREAT3 input catalogs, the PSF shapes have to be measured in all three cases. We once again used the HSM library of GalSim, which matches the size of the weighting function to the object being measured. We chose the size of our Gaussian PSFs to be the same as that of the matched window, which leads to a constant factor of two in their unweighted $R^2_\mathrm{PSF}$.

\begin{figure}
    \centering
    \includegraphics[width=\linewidth]{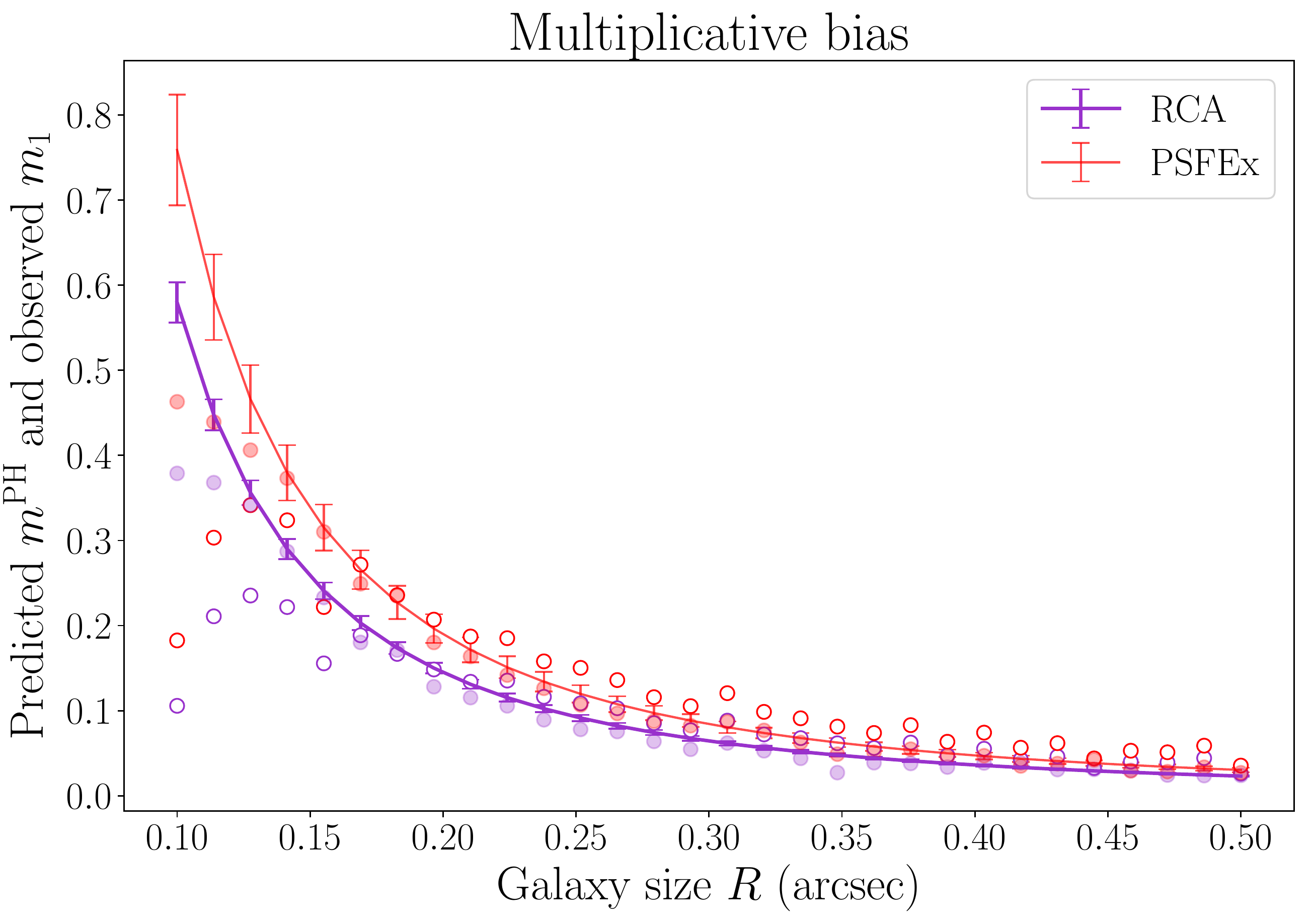}
    \caption{Same as \autoref{fig:mpaulin}, when PSFs and galaxies are Gaussian.}
    \label{fig:mgauss}
\end{figure}

\begin{figure}
    \begin{subfigure}{\linewidth}
        \includegraphics[width=\linewidth]{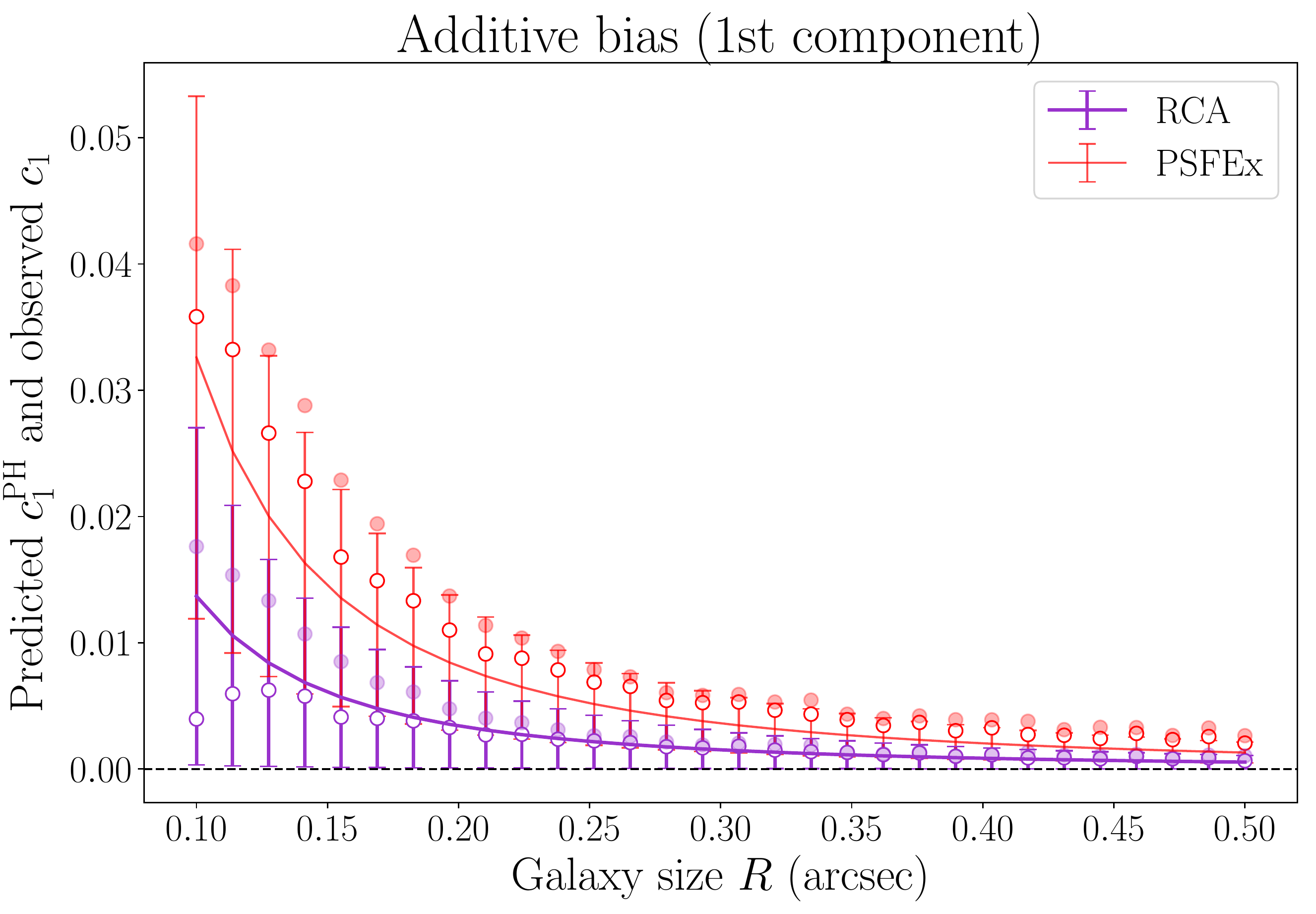}
    \end{subfigure}\\
    \begin{subfigure}{\linewidth}
        \includegraphics[width=\linewidth]{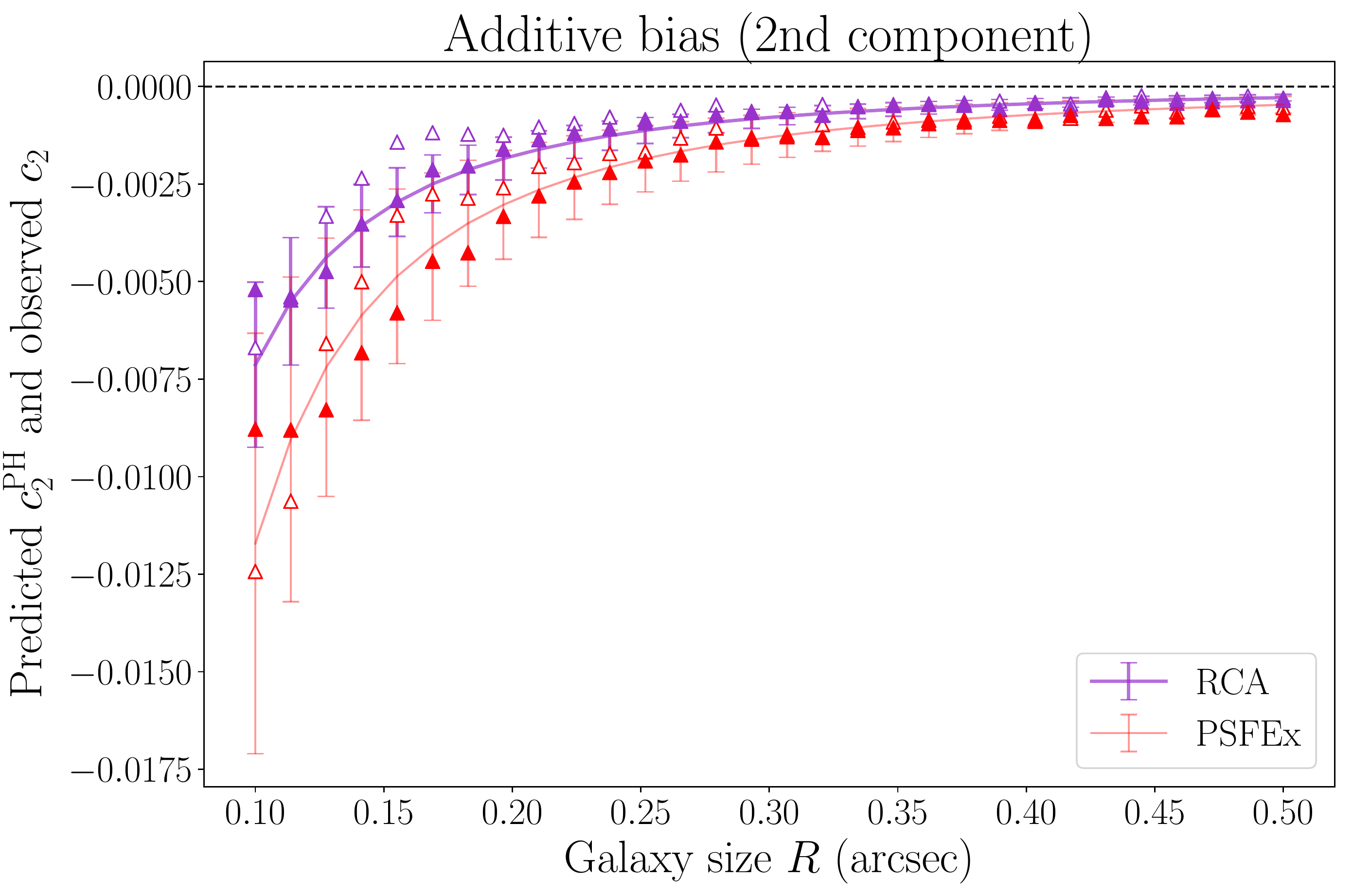}
    \end{subfigure}\\
    \caption{Same as \autoref{fig:cpaulin}, when PSFs and galaxies are Gaussian.}
    \label{fig:cgauss}
\end{figure}

The results are shown in Figs.~\ref{fig:mgauss} and~\ref{fig:cgauss} for the multiplicative and additive components, respectively, and show good agreement with the predicted values. The first few galaxy size bins lead to smaller measured multiplicative biases, though these are only due to the small number of galaxies at these sizes.

\section{Conclusion}\label{sec:conclusion}
   In this work, we extended a previously proposed approach for PSF estimation, taking necessary steps toward a fully nonparametric approach applicable in the context of the upcoming \Euclid survey. A study of the PSF models and their residuals shows our model outperforms the proven and widely used \texttt{PSFEx}. This could indicate a better handling of the super resolution, as hints at potential issues with \texttt{PSFEx}'s super-resolution mode were recently observed in HSC~\citep{bosch2017}.
   Our method is still, however, far from achieving the \Euclid requirements. As a nonparametric approach, its main limitation lies in the number of available stars, and a natural path of improvement is thus simultaneous use of stars from different exposures, that is, taking into account the temporal variability of the PSF. Another approach is that of a parametric PSF model, which is under development for \Euclid (Duncan et al., in prep.) and should allow us to reach the requirements. Ultimately, the combination of both approaches will likely do better than each taken separately, which warrants further study of nonparametric models, how to improve them, and make them capable of handling the specificities of \Euclid's PSF.

  As mentioned in the introduction, in the present work, we make many simplifying assumptions regarding the VIS PSF. In particular, we considered a single monochromatic PSF. This will no longer be an acceptable assumption in the case of \Euclid~\citep{eriksen2018}, and aside from the need to use a greater number of stars, a necessary improvement to the presented method and focus for future work will be to take those chromatic variations into account. Dealing with this issue in the nonparametric framework is of considerable difficulty, since the observed stars give measurements of the PSF integrated with their own SED. However, we recently showed~\citep{schmitz2018} that optimal transport tools were particularly well suited to represent the chromatic variations undergone by the VIS PSF. Introducing these tools into our nonparametric model could allow us to break the additional degeneracy caused by the chromatic variations of the PSF being integrated with the stars' SEDs, and to extract monochromatic components that could then be recombined with the galaxies' SEDs.
    
   Despite the improvement in model quality, the use of our approach as the PSF in galaxy shape measurement unexpectedly leads to stronger additive shear biases than when using \texttt{PSFEx}. Following this observation, as well as other observed discrepancies, this paper also shows that, in the case of \Euclid, the way the PSF modeling errors impact shear measurement can be more complicated than previously thought, and method dependent. In particular, the~\cite{paulin2008} formalism no longer holds. Our experiments show this is likely coming from additional terms arising from the necessary addition of a window function to compute quadrupole moments. Similar effects could thus occur for any diffraction-limited telescope.

\begin{acknowledgements}
      The Euclid PSFs used in this work were provided by Koryo Okumura, Samuel Ronayette and J\'er\^ome Amiaux. The authors are grateful to Lance Miller, Henk Hoekstra and Christopher Duncan for their comments on an earlier draft of this paper, to the anonymous, external referee for their comments and suggestions, to Arnau Pujol and Florent Sureau for numerous helpful discussions and advice on shape measurement, and to Imane El Hamzaoui for her help with some of the figures in this paper. This work was supported by the Centre National d'Etudes Spatiales and the European Community through the grant DEDALE (contract no. 665044) within the H2020 Framework Program. JB acknowledges support by Funda{\c c}{\~a}o para a Ci{\^e}ncia e a
Tecnologia through national funds (UID/FIS/04434/2013) and Investigador FCT
contract IF/01654/2014/CP1215/CT0003., and by FEDER through COMPETE2020 (POCI-01-0145-FEDER-007672). \AckECon
\end{acknowledgements}

\bibliographystyle{aa}
\bibliography{PSF_prop}

\begin{thebibliography}{53}
\expandafter\ifx\csname natexlab\endcsname\relax\def\natexlab#1{#1}\fi

\bibitem[{{Bertin}(2011)}]{bertin2011}
{Bertin}, E. 2011, in Astronomical Society of the Pacific Conference Series,
  Vol. 442, Astronomical Data Analysis Software and Systems XX, ed. I.~N.
  {Evans}, A.~{Accomazzi}, D.~J. {Mink}, \& A.~H. {Rots}, 435

\bibitem[{Bertin \& Arnouts(1996)}]{bertin1996}
Bertin, E. \& Arnouts, S. 1996, A\&AS, 117, 393

\bibitem[{Bosch {et~al.}(2017)Bosch, Armstrong, Bickerton, Furusawa, Ikeda,
  Koike, Lupton, Mineo, Price, Takata, {et~al.}}]{bosch2017}
Bosch, J., Armstrong, R., Bickerton, S., {et~al.} 2017, PASJ, 70, S5

\bibitem[{Buhmann(2003)}]{buhmann2003}
Buhmann, M.~D. 2003, Radial basis functions: theory and implementations,
  Vol.~12 (Cambridge University Press)

\bibitem[{Condat(2013)}]{condat2013}
Condat, L. 2013, Journal of Optimization Theory and Applications, 158, 460

\bibitem[{Coulton {et~al.}(2018)Coulton, Armstrong, Smith, Lupton, \&
  Spergel}]{coulton2018}
Coulton, W.~R., Armstrong, R., Smith, K.~M., Lupton, R.~H., \& Spergel, D.~N.
  2018, ApJ, 155, 258

\bibitem[{Cropper {et~al.}(2013)Cropper, Hoekstra, Kitching, Massey, Amiaux,
  Miller, Mellier, Rhodes, Rowe, Pires, {et~al.}}]{cropper2013}
Cropper, M., Hoekstra, H., Kitching, T., {et~al.} 2013, MNRAS, 431, 3103

\bibitem[{Cypriano {et~al.}(2010)Cypriano, Amara, Voigt, Bridle, Abdalla,
  Refregier, Seiffert, \& Rhodes}]{cypriano2010}
Cypriano, E., Amara, A., Voigt, L., {et~al.} 2010, MNRAS, 405, 494

\bibitem[{Eriksen \& Hoekstra(2018)}]{eriksen2018}
Eriksen, M. \& Hoekstra, H. 2018, MNRAS, 477, 3433

\bibitem[{Gentile {et~al.}(2013)Gentile, Courbin, \& Meylan}]{gentile2013}
Gentile, M., Courbin, F., \& Meylan, G. 2013, A\&A, 549, A1

\bibitem[{Green {et~al.}(2012)Green, Schechter, Baltay, Bean, Bennett, Brown,
  Conselice, Donahue, Fan, Gaudi, {et~al.}}]{green2012}
Green, J., Schechter, P., Baltay, C., {et~al.} 2012, arXiv preprint
  arXiv:1208.4012

\bibitem[{Hammond {et~al.}(2011)Hammond, Vandergheynst, \&
  Gribonval}]{hammond2011}
Hammond, D.~K., Vandergheynst, P., \& Gribonval, R. 2011, Applied and
  Computational Harmonic Analysis, 30, 129

\bibitem[{Herbel {et~al.}(2018)Herbel, Kacprzak, Amara, Refregier, \&
  Lucchi}]{herbel2018}
Herbel, J., Kacprzak, T., Amara, A., Refregier, A., \& Lucchi, A. 2018, ArXiv
  e-prints [\eprint[arXiv]{1801.07615}]

\bibitem[{Hirata \& Seljak(2003)}]{hirata2003}
Hirata, C. \& Seljak, U. 2003, MNRAS, 343, 459

\bibitem[{Hoekstra(2004)}]{hoekstra2004}
Hoekstra, H. 2004, MNRAS, 347, 1337

\bibitem[{Hoekstra {et~al.}(1998)Hoekstra, Franx, Kuijken, \&
  Squires}]{hoekstra1998}
Hoekstra, H., Franx, M., Kuijken, K., \& Squires, G. 1998, ApJ, 504, 636

\bibitem[{Hoekstra {et~al.}(2015)Hoekstra, Herbonnet, Muzzin, Babul, Mahdavi,
  Viola, \& Cacciato}]{hoekstra2015}
Hoekstra, H., Herbonnet, R., Muzzin, A., {et~al.} 2015, MNRAS, 449, 685

\bibitem[{Hoekstra {et~al.}(2017)Hoekstra, Viola, \& Herbonnet}]{hoekstra2017}
Hoekstra, H., Viola, M., \& Herbonnet, R. 2017, MNRAS, 468, 3295

\bibitem[{Jarvis {et~al.}(2016)Jarvis, Sheldon, Zuntz, Kacprzak, Bridle, Amara,
  Armstrong, Becker, Bernstein, Bonnett, {et~al.}}]{jarvis2016}
Jarvis, M., Sheldon, E., Zuntz, J., {et~al.} 2016, MNRAS, 460, 2245

\bibitem[{Kaiser {et~al.}(1995)Kaiser, Squires, \& Broadhurst}]{kaiser1995}
Kaiser, N., Squires, G., \& Broadhurst, T. 1995, ApJ, 449, 460

\bibitem[{Kilbinger(2015)}]{kilbinger2015}
Kilbinger, M. 2015, Reports on Progress in Physics, 78, 086901

\bibitem[{{Krist}(1995)}]{krist1995}
{Krist}, J. 1995, in Astronomical Society of the Pacific Conference Series,
  Vol.~77, Astronomical Data Analysis Software and Systems IV, ed. R.~A.
  {Shaw}, H.~E. {Payne}, \& J.~J.~E. {Hayes}, 349

\bibitem[{Kuijken {et~al.}(2015)Kuijken, Heymans, Hildebrandt, Nakajima, Erben,
  de~Jong, Viola, Choi, Hoekstra, Miller, {et~al.}}]{kuijken2015}
Kuijken, K., Heymans, C., Hildebrandt, H., {et~al.} 2015, MNRAS, 454, 3500

\bibitem[{Kuntzer \& Courbin(2017)}]{kuntzer2017}
Kuntzer, T. \& Courbin, F. 2017, A\&A, 606, A119

\bibitem[{Kuntzer {et~al.}(2016)Kuntzer, Tewes, \& Courbin}]{kuntzer2016}
Kuntzer, T., Tewes, M., \& Courbin, F. 2016, A\&A, 591, A54

\bibitem[{Kuntzer(2018)}]{kuntzer2018thesis}
Kuntzer, T.~A. 2018, PhD thesis, EPFL

\bibitem[{Laureijs {et~al.}(2011)Laureijs, Amiaux, Arduini, Augueres,
  Brinchmann, Cole, Cropper, Dabin, Duvet, Ealet, {et~al.}}]{laureijs2011}
Laureijs, R., Amiaux, J., Arduini, S., {et~al.} 2011, ArXiv e-prints
  [\eprint[arXiv]{1110.3193}]

\bibitem[{{LSST Science Collaboration} {et~al.}(2009){LSST Science
  Collaboration}, {Abell}, {Allison}, {Anderson}, {Andrew}, {Angel}, {Armus},
  {Arnett}, {Asztalos}, {Axelrod}, \& et~al.}]{LSST2009}
{LSST Science Collaboration}, {Abell}, P.~A., {Allison}, J., {et~al.} 2009,
  arXiv e-prints [\eprint[arXiv]{0912.0201}]

\bibitem[{Mandelbaum {et~al.}(2005)Mandelbaum, Hirata, Seljak, Guzik,
  Padmanabhan, Blake, Blanton, Lupton, \& Brinkmann}]{mandelbaum2005}
Mandelbaum, R., Hirata, C.~M., Seljak, U., {et~al.} 2005, MNRAS, 361, 1287

\bibitem[{Mandelbaum {et~al.}(2017)Mandelbaum, Miyatake, Hamana, Oguri, Simet,
  Armstrong, Bosch, Murata, Lanusse, Leauthaud, {et~al.}}]{mandelbaum2018}
Mandelbaum, R., Miyatake, H., Hamana, T., {et~al.} 2017, PASJ, 70, S25

\bibitem[{Mandelbaum {et~al.}(2015)Mandelbaum, Rowe, Armstrong, Bard, Bertin,
  Bosch, Boutigny, Courbin, Dawson, Donnarumma, {et~al.}}]{mandelbaum2015}
Mandelbaum, R., Rowe, B., Armstrong, R., {et~al.} 2015, MNRAS, 450, 2963

\bibitem[{Mandelbaum {et~al.}(2014)Mandelbaum, Rowe, Bosch, Chang, Courbin,
  Gill, Jarvis, Kannawadi, Kacprzak, Lackner, {et~al.}}]{mandelbaum2014}
Mandelbaum, R., Rowe, B., Bosch, J., {et~al.} 2014, ApJS, 212, 5

\bibitem[{Massey {et~al.}(2012)Massey, Hoekstra, Kitching, Rhodes, Cropper,
  Amiaux, Harvey, Mellier, Meneghetti, Miller, {et~al.}}]{massey2012}
Massey, R., Hoekstra, H., Kitching, T., {et~al.} 2012, MNRAS, 429, 661

\bibitem[{Melchior {et~al.}(2011)Melchior, Viola, Sch{\"a}fer, \&
  Bartelmann}]{melchior2011}
Melchior, P., Viola, M., Sch{\"a}fer, B.~M., \& Bartelmann, M. 2011, MNRAS,
  412, 1552

\bibitem[{Miller {et~al.}(2013)Miller, Heymans, Kitching, Van~Waerbeke, Erben,
  Hildebrandt, Hoekstra, Mellier, Rowe, Coupon, {et~al.}}]{miller2013}
Miller, L., Heymans, C., Kitching, T., {et~al.} 2013, MNRAS, 429, 2858

\bibitem[{Ngol{\`e} \& Starck(2017)}]{ngole2017}
Ngol{\`e}, F. \& Starck, J.-L. 2017, SIAM Journal on Imaging Sciences, 10, 1549

\bibitem[{Ngol{\`e} {et~al.}(2016)Ngol{\`e}, Starck, Okumura, Amiaux, \&
  Hudelot}]{ngole2016}
Ngol{\`e}, F., Starck, J.-L., Okumura, K., Amiaux, J., \& Hudelot, P. 2016,
  Inverse Problems, 32, 124001

\bibitem[{Ngol{\`e} {et~al.}(2015)Ngol{\`e}, Starck, Ronayette, Okumura, \&
  Amiaux}]{ngole2015}
Ngol{\`e}, F., Starck, J.-L., Ronayette, S., Okumura, K., \& Amiaux, J. 2015,
  A\&A, 575, A86

\bibitem[{Paulin-Henriksson {et~al.}(2008)Paulin-Henriksson, Amara, Voigt,
  Refregier, \& Bridle}]{paulin2008}
Paulin-Henriksson, S., Amara, A., Voigt, L., Refregier, A., \& Bridle, S. 2008,
  A\&A, 484, 67

\bibitem[{Paulin-Henriksson {et~al.}(2009)Paulin-Henriksson, Refregier, \&
  Amara}]{paulin2009}
Paulin-Henriksson, S., Refregier, A., \& Amara, A. 2009, A\&A, 500, 647

\bibitem[{Pujol {et~al.}(2017)Pujol, Sureau, Bobin, Courbin, Gentile, \&
  Kilbinger}]{pujol2017}
Pujol, A., Sureau, F., Bobin, J., {et~al.} 2017, ArXiv e-prints
  [\eprint[arXiv]{1707.01285}]

\bibitem[{Raguet {et~al.}(2013)Raguet, Fadili, \& Peyr{\'e}}]{raguet2013}
Raguet, H., Fadili, J., \& Peyr{\'e}, G. 2013, SIAM Journal on Imaging
  Sciences, 6, 1199

\bibitem[{Rowe(2010)}]{rowe2010}
Rowe, B. 2010, MNRAS, 404, 350

\bibitem[{Rowe {et~al.}(2011)Rowe, Hirata, \& Rhodes}]{rowe2011}
Rowe, B., Hirata, C., \& Rhodes, J. 2011, ApJ, 741, 46

\bibitem[{Rowe {et~al.}(2015)Rowe, Jarvis, Mandelbaum, Bernstein, Bosch, Simet,
  Meyers, Kacprzak, Nakajima, Zuntz, {et~al.}}]{rowe2015}
Rowe, B., Jarvis, M., Mandelbaum, R., {et~al.} 2015, Astronomy and Computing,
  10, 121

\bibitem[{Schmitz {et~al.}(2018)Schmitz, Heitz, Bonneel, Ngole, Coeurjolly,
  Cuturi, Peyr{\'e}, \& Starck}]{schmitz2018}
Schmitz, M.~A., Heitz, M., Bonneel, N., {et~al.} 2018, SIAM Journal on Imaging
  Sciences, 11, 643

\bibitem[{Semboloni {et~al.}(2013)Semboloni, Hoekstra, Huang, Cardone, Cropper,
  Joachimi, Kitching, Kuijken, Lombardi, Maoli, {et~al.}}]{semboloni2013}
Semboloni, E., Hoekstra, H., Huang, Z., {et~al.} 2013, MNRAS, 432, 2385

\bibitem[{Starck {et~al.}(2011)Starck, Murtagh, \& Bertero}]{starck2011}
Starck, J.-L., Murtagh, F., \& Bertero, M. 2011, in Handbook of Mathematical
  Methods in Imaging (Springer), 1489--1531

\bibitem[{Starck {et~al.}(2015)Starck, Murtagh, \& Fadili}]{starck2015}
Starck, J.-L., Murtagh, F., \& Fadili, J. 2015, Sparse image and signal
  processing: Wavelets and related geometric multiscale analysis (Cambridge
  University Press)

\bibitem[{Viola {et~al.}(2014)Viola, Kitching, \& Joachimi}]{viola2014}
Viola, M., Kitching, T., \& Joachimi, B. 2014, MNRAS, 439, 1909

\bibitem[{Voigt \& Bridle(2010)}]{voigt2010}
Voigt, L. \& Bridle, S. 2010, MNRAS, 404, 458

\bibitem[{Zuntz {et~al.}(2013)Zuntz, Kacprzak, Voigt, Hirsch, Rowe, \&
  Bridle}]{zuntz2013}
Zuntz, J., Kacprzak, T., Voigt, L., {et~al.} 2013, MNRAS, 434, 1604

\bibitem[{Zuntz {et~al.}(2018)Zuntz, Sheldon, Samuroff, Troxel, Jarvis,
  MacCrann, Gruen, Prat, S{\'a}nchez, Choi, {et~al.}}]{zuntz2018}
Zuntz, J., Sheldon, E., Samuroff, S., {et~al.} 2018, MNRAS, 481, 1149

\end{thebibliography}

\begin{appendix}
\section{A primer on graph theory}\label{appdx:graphprimer}
\begin{figure}
    \centering
    \includegraphics[width=\linewidth]{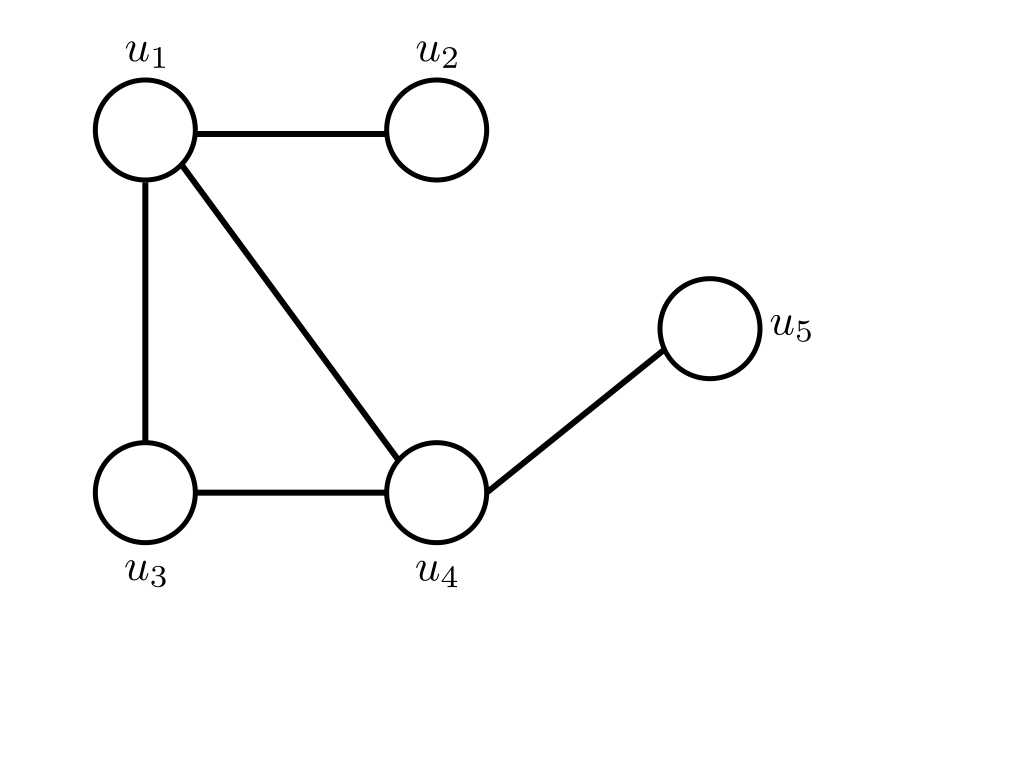}
    \caption{Example unweighted graph.}
    \label{fig:unweightedgraph}
\end{figure}
In this appendix, we give a very brief introduction to some graph theory concepts relevant to our method. Let us first define the Laplacian matrix. Consider an unweighted graph such as that shown in \autoref{fig:unweightedgraph} (where each vertex $i$ is identified through $u_i$ in order to keep the notations consistent, though there is no notion of position here). We define the \textit{degree}, $d(i)$, of vertex $i$ as the number of edges connected to it. In our example, we have $d(1)=d(4)=3, d(2)=d(5)=1,$ and $d(3)=2$. The degree matrix $D$ is simply the diagonal matrix with $D_{ii} = d(i)$. It contains information about the graph's connectivity, but nothing about its actual structure (as we cannot know which vertices contribute to another's degree). This is contained in the \textit{adjacency} matrix $A$, which, in the unweighted case, is simply defined as

\begin{align}
     A_{ij} = 
    \begin{cases}
       1 &\mbox{if there is an edge between vertices $i$ and $j$},\\
       0 &\mbox{otherwise}.
    \end{cases}
\end{align}
We can now define the \textit{Laplacian} matrix of a graph as

\begin{align}\label{eq:lapdef}
    L = D-A\;.
\end{align}
In the example of \autoref{fig:unweightedgraph}, the Laplacian would be

\begin{align}
L = \begin{pmatrix}
    3 & -1 & -1 & -1 & 0\\
    -1 & 1 & 0 & 0 & 0\\
    -1 & 0 & 2 & -1 & 0\\
    -1 & 0 & -1 & 3 & -1\\
    0 & 0 & 0 & -1 & 1
\end{pmatrix}\;.
\end{align}
The Laplacian matrix is a tool central to graph theory. In our case, the edges are weighted by a function of the distance between the stars corresponding to the vertices. Generalizing the definition of $L$ to the weighted case is an intuitive procedure. While the adjacency matrix entries up to now only contained a binary information (either two vertices are connected, or they are not), in the weighted case, we replace that with the weight on the corresponding edge: the entries of $A$ now tell us quantitatively \textit{how} connected two vertices are. Similarly, for the degree matrix to quantify the amount of connectivity of a given vertex rather than just count the number of edges, we define the degree of node $i$ as $d(i) = \sum_{j} A_{ij}$, that is, the sum of the weights carried by all edges connected to vertex $i$. From that definition, we immediately infer that our matrices $P_{e_k,a_k}$ defined in \autoref{eq:graphlap} are precisely the Laplacian matrices of a fully connected graph with the edge between $i$ and $j$ weighted by $1/\|u_i-u_j\|_2^{e_k}$, multiplied (entry-wise) by a matrix $\tilde{L_\mathrm{D}}$, defined as

\begin{align}\label{eq:ld}
    \tilde {L_\mathrm{D}} \eqdef a_k\mathrm{Id} - \begin{pmatrix}
    0 & -1 & -1 & \dots &-1\\
    -1 & 0 & -1 & \dots & -1\\
    \vdots & & \ddots & & \vdots \\
    -1 & \dots & -1 & 0 & -1\\
     -1 & \dots & -1 & -1 & 0
    \end{pmatrix}\;,
\end{align}
where $\mathrm{Id}$ is the identity matrix. 

The role of $e_k$ in associating each of our graphs to a certain spatial frequency is straightforward: the higher its value, the stronger the decay in edge weight as the distance between two vertices increases, leading to the graph capturing lower spatial frequencies. Comparing \autoref{eq:ld} to \eqref{eq:lapdef} gives an intuitive (though not rigorous) interpretation as to the role of $a_k$: it amounts to multiplying the degree matrix of our graph by $a_k$, in turn affecting its overall connectivity.

As a pathway toward defining wavelets on graphs,~\cite{hammond2011} introduced, by analogy with the usual transform, the Fourier transform on graphs. For a graph $G$ with Laplacian $L$, $\left(V_l\right)_l$ denotes its eigen vectors. For any function $f$ defined on the vertices of $G$ (like, in our case, each row $A_k$ containing the coefficients of each star for a particular eigenPSF), we define its Fourier transform as

\begin{align}
    \hat{f}(l) \eqdef \langle V_l, f \rangle\;.
\end{align}
The matrix $V$ introduced in \autoref{sec:rcagraph} is nothing but the concatenation of the eigen vectors associated to each eigen PSF's graph. Factorizing $A$ by $V^\top$ and imposing the rows of the resulting matrix $\alpha$ to be sparse thus simply amounts to imposing the coefficients associated to our eigen PSFs to be sparse in the Fourier domain of each associated graph (themselves capturing, by construction, a particular spatial frequency).

\section{Galaxy shape measurement experiment}\label{appdx:addfigs}
\begin{table*}
    \centering
    \begin{tabular}{c|l|cc|cc|cc|cc}
        \multicolumn{2}{c|}{SNR} & \multicolumn{2}{c|}{10} & \multicolumn{2}{c|}{20} & \multicolumn{2}{c|}{35} & \multicolumn{2}{c}{50}\\
        \multicolumn{2}{c|}{Ellipticity component}&  $1$st & $2$nd & $1$st & $2$nd & $1$st & $2$nd & $1$st& $2$nd \\
        \hline
        \multirow{2}{*}{KSB} & RCA & 1.19 & 0.76 & 1.04 & 0.73 & 1.02 & 0.76 & 1.00 & 0.77\\
        & \texttt{PSFEx} & 2.34 & 1.88 & 2.29 & 1.95 & 2.26 & 1.91 & 2.28 & 1.93\\
        \hline  
        \multirow{2}{*}{\texttt{im3shape}} & RCA & 0.73 & 0.42 & 0.67 & 0.40 & 0.64 & 0.41 & 0.62 & 0.40\\
        & \texttt{PSFEx} & 0.81 & 0.48 & 0.80 & 0.50 & 0.79 & 0.50 & 0.79 & 0.49
    \end{tabular}
    \caption{Relative ellipticity error on the measured galaxy shapes, $\langle (\hat e_i^\mathrm{kn} - \hat e_i)^2\rangle \times 10^3$, for both PSF models and both shape measurement methods.}
    \label{tab:relative_errors}
\end{table*}
This appendix contains additional details on the galaxy shape experiment described in \autoref{sec:gaals}.  \autoref{tab:relative_errors} shows the relative errors on measured galaxy shapes, for all combinations of noise level, PSF model, and shape measurement approach, while \autoref{tab:outliers} contains the number of objects removed from the analysis when using KSB.

Below is the configuration file used when running \texttt{im3shape} in our experiments:
        
    \begin{minipage}{\linewidth}
        \begin{verbatim}
                noise_sigma = 0.01 
                background_subtract = NO
                psf_truncation_pixels = 50.0
                stamp_size = 42
                
                sersics_x0_start = 21.0 
                sersics_y0_start = 21.0        
                sersics_x0_min = 18.0        
                sersics_y0_min = 18.0        
                sersics_x0_max = 24.0        
                sersics_y0_max = 24.0    
                                 
                psf_input = psf_image_cube            
                perform_pixel_integration = NO 
                upsampling = 1
                central_pixel_upsampling = NO
                padding = 0
        \end{verbatim}
    \end{minipage}
We note that default values are used for all parameters not specified in this config file.

\begin{table}
        \centering
    \begin{tabular}{l|cccc}
        SNR & 10 & 20 & 35 & 50\\
        \hline
        Known & 26\,741 & 26\,702 & 26\,787 & 26\,798\\
        \texttt{PSFEx} & 63\,541 & 63\,356 & 63\,236 & 63\,111\\
        RCA & 54\,345 & 51\,664 & 52\,100 & 51\,739\\
        \textbf{Total} & \textbf{72\,951} & \textbf{71\,551} & \textbf{71\,128} & \textbf{70\,902}
    \end{tabular}
    \caption{Number of objects where the HSM implementation of KSB fails to compute the PSF-corrected shapes per star SNR level. The total amount corresponds to the union of the 3 PSF-specific sets of such outliers.}
    \label{tab:outliers}
\end{table}

\end{appendix}

\end{document}